%% file: qedDisorderDraft3.tex
\documentclass[aps,pre,preprint,onecolumn,citeautoscript,superscriptaddress,floatfix]{revtex4-1}  
\DeclareMathAlphabet{\mathpzc}{OT1}{pzc}{m}{it}

\input{headersDraft2}

\usepackage[textsize=tiny]{todonotes}
\presetkeys{todonotes}{color=green!40}{}

\begin{document}

\title{QED$_3$ with quenched disorder:\\ quantum critical states with interactions and disorder}
 \author{Alex Thomson}
 \affiliation{Department of Physics, Harvard University, Cambridge, Massachusetts, 02138, USA}
 \author{Subir Sachdev}
 \affiliation{Department of Physics, Harvard University, Cambridge, Massachusetts, 02138, USA}
 \affiliation{Perimeter Institute for Theoretical Physics, Waterloo, Ontario N2L 2Y5, Canada}
 \date{\today}
 \vspace{0.6in}
\begin{abstract} 
Quantum electrodynamics in 2+1-dimensions (QED$_3$) 
is a strongly coupled conformal field theory (CFT) of a U(1) gauge field coupled to $2N$ two-component massless fermions. 
The $N=2$ CFT has been proposed as a ground state of the spin-1/2 kagome Heisenberg antiferromagnet.
We study QED$_3$ in the presence of weak quenched disorder in its two spatial directions.
When the disorder explicitly breaks the fermion flavor symmetry from SU($2N$)$\rightarrow$U(1)$\times$SU($N$) but preserves 
time-reversal symmetry, we find that the theory flows to a non-trivial fixed line at non-zero disorder with a continuously varying dynamical critical exponent $z>1$.
We determine the zero-temperature flavor (spin) conductivity along the critical line.
Our calculations are performed in the large-$N$ limit, and the disorder is handled using the replica method.
\end{abstract}

\maketitle

\section{Introduction}
\label{sec:intro}

While our understanding of magnetic systems and spin liquids in particular has made great progress in the last two decades, most systems have been studied in the clean limit with translational symmetry present. 
In this paper, we explore the behavior of a critical spin liquid described by a conformal field theory (CFT) when perturbed by weak quenched disorder. 

\begin{figure}
\centering
\includegraphics[scale=0.4]{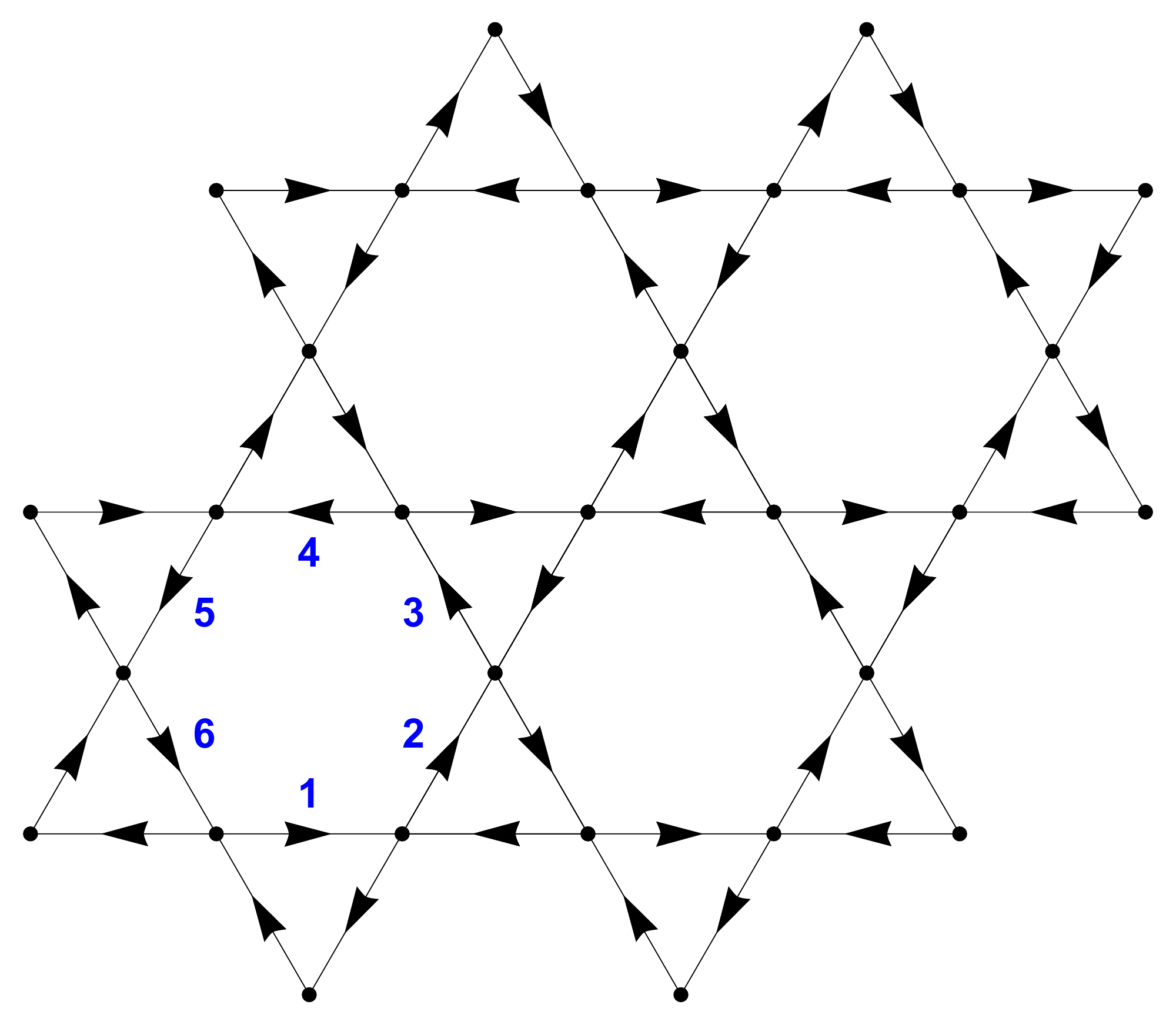}
\caption{The kagome lattice. The arrows indicate the convention chosen for the bond directions of the spin chirality operator, $\v{S}_i\times\v{S}_j$, where $i$ and $j$ label nearest-neighbour sites. The order of the cross product is taken such that first spin sits at the lattice site pointing towards the site of the second spin. Later, we will use the same ordering convention to define nearest-neighbour bond operators $\v{S}_i\cdot\v{S}_j$.}
\label{fig:bondOrientation}
\end{figure}

The CFT we consider is 2+1 dimensional quantum electrodynamics (QED$_3$), a strongly coupled theory of a U(1) gauge field coupled to $2N$ massless two-component 
fermions \cite{Rantner2001,Rantner2002}.
This CFT is one of the proposed ground states of the spin-1/2 kagome Heisenberg antiferromagnet, $H_\mathrm{H}=J\sum_{\Braket{ij}}\v{S}_i\cdot\v{S}_j$,
where $J>0$ and $\Braket{ij}$ labels nearest-neighbour sites on a kagome lattice (shown in Fig.~\ref{fig:bondOrientation}) \cite{Hastings2000,Ran2007,Hermele2008}. (We note that other proposed ground states are gapped $\mathbb{Z}_2$
spin liquids \cite{SSkagome}, and the choice between the CFT and the $\mathbb{Z}_2$ spin liquids remains a matter of continuing
debate \cite{2016arXiv160609639M,2016arXiv161002024J,2016arXiv161106238H,2016arXiv161106990L,Iqbal1,Iqbal2}.)
In addition, QED$_3$ may also describe certain deconfined critical points \cite{Senthil2004a,Senthil2004b} between topological phases \cite{Grover2013,Barkeshli2013}. 

The QED$_3$ action is written 
\eq{\label{eqn:qedAction}
S_\mathrm{qed}\[\psi,\bpsi,A\]&=-\int d^2x\,d\t\,\bpsi_\a \g^\m\(\ptl_\m-{iA_\m\o\sqrt{2N}}\)\psi_\a+{1\o4e^2(2N)}\int d^2x\,d\t\,\(\ptl_\m A_\n-\ptl_\n A_\m\)^2
}
where $\alpha$ labels the $2N$ fermion flavors, and we have denoted the Euclidean spacetime coordinates as $r=(x,\t)$. 
The $\psi_\a$'s are 2-component spinors, with $\bpsi_\a=\psi^\dag_\a\t^z$ and $\g^\m=\(\t^z,\t^y,-\t^x\)$ where the $\t^a$'s are Pauli matrices.
The dimension of the charge is $\[e^2\]=+1$ and so under the renormalization group (RG) flow we expect $e^2\rightarrow\infty$; this will be discussed in greater detail in Sec.~\ref{sec:pureQED3}.
This theory possesses an explicit global SU($2N$) symmetry under which the fermions flavors are rotated into one another. 

The action in Eq.~\eqref{eqn:qedAction} specifically describes non-compact QED$_3$ {\it i.e.\/}
there are no monopoles operators in the action, and flux conservation is a global symmetry: $\ptl_\m J^\m_\mathrm{top}=0$, where $J^\m_\mathrm{top}=\ep^{\m\n\r}\ptl_\n A_\r$. 
Because $S_\mathrm{qed}$ arises in condensed matter as the low-energy description of a lattice model, monopole events must be allowed in the ultraviolet (UV). However, Berry phases from the underlying lattice spins
can lead to destructive interference between monopole tunneling events \cite{PhysRevLett.62.1694,PhysRevB.42.4568,Senthil2004a,Senthil2004b}, and it could well be the case
that monopoles carrying the smallest magnetic charge are prohibited for the clean kagome antiferromagnet;
the minimal magnetic charge for allowed monopoles in the kagome antiferromagnet is unknown, and its determination
remains an important open problem.
In order for non-compact QED$_3$ to be the correct low-energy description, the smallest allowed monopole operators 
must be irrelevant perturbations.
When the number of fermion flavors is low, this is not the case and the monopoles to proliferate, confining the theory \cite{Polyakov1977,Polyakov1987}.
As matter is added to the system, the scaling dimension of the monopoles increases and they eventually become irrelevant \cite{MURTHY1990557,Hermele2004,Borokhov2002,Pufu14,Pufu15}.
The number of fermion flavours required before this occurs is currently unknown, but estimates place it around $2N_\mathrm{monopole}^c\lesssim 12$ for the smallest monopole charge \cite{Pufu15}. 
In this paper, we work in the large-$N$ limit, where all possible monopole operators are strongly  
irrelevant \cite{MURTHY1990557}. 
There is an additional critical fermion flavour number beneath which QED$_3$ spontaneously generates a chiral mass. 
The exact value of this number is also unknown but is expected to be $2N_\mathrm{chiral}^c\approx 3$ \cite{Teber1,Teber2}.

The kagome antiferromagnet corresponds to the case $N=2$: the four flavors of fermions arise as a result of spin degrees of freedom, as well as an additional two-fold valley degeneracy.
Nonetheless, when we specify to this case, we will operate under the assumption that the large $N$ results 
also apply to the $N=2$ case.

Since some degree of disorder is present in all physical systems, it is important to understand the behavior of these theories under this type of perturbation.
The primary result of this paper is that when time reversal and a global U(1)$\times$SU$(N)$ flavour symmetry are respected microscopically, there exists a critical line with both non-zero disorder and interactions.
This is obtained by coupling the theory to quenched disorder of the form
\eq{\label{Sdis1}
S_{\mathrm{dis},z}\[\psi,\bpsi\]&=\int d^2x\,d\t
\[M_z(x)\bpsi\s^z\psi(x,\t)++i\mathcal{A}_{j z}(x)\bpsi\s^z\g^j\psi(x,\t)\].
}
Here, ${M}_z$ and $\A_{jz}$ are random fields with zero mean. 
Both fields are independent of time: although QED$_3$ is a relativistic theory, disorder explicits breaks this symmetry. 
This should be contrasted with classical disordered field theories  where the random fields are functions of all of the coordinates in the action.
$M_z$ and $\A_{jz}$ are both Gaussian and entirely determined by their disorder averages:
\eq{
\overline{M_z(x)M_z(x')}&={g_{t,z}\o2}\d^2\(x-x'\),
&
\overline{\A_{iz}(x)\A_{jz}(x')}&=\d_{ij}{g_{\A,z}\o2}\d^2\(x-x'\),
&
\overline{M_{z}(x)\A_{jz}(x')}&=0.
}
The variances $g_{t,z}$ and $g_{\A,z}$ control the strength of the disorder, and, naturally, they must be positive.
Performing a diagrammatic expansion to $\O(g_{\xi}^2,g_\xi/2N)$ with $\xi=(t,z),(\A,z)$, we find a critical line with $g_{t,z}=-8g_{\A,z}+64/(3\pi^2 N)$. Provided the flavor symmetry is not broken further, we expect at least a fixed point to exist at sufficiently large $N$: higher
order corrections could convert the line to a fixed point but are not expected to lead to runaway flows to strong disorder.

In the context of the kagome antiferromagnet, the bilinear $\bpsi \s^z\psi$ can be associated with the $z$-component of the Dzyaloshinskii-Moriya (DM) interaction operator:
\eq{
\sum_{\Braket{ij}\in \mathrm{hex}(\vx)}\hat{\v{z}}\cdot\(\v{S}_{i}\times\v{S}_{j}\),
}
where hex$(\vx)$ labels the hexagon at point $\vx$ and the bonds $\braket{ij}$ are summed in the fashion shown in Fig.~\ref{fig:bondOrientation}. 
Similarly, $i\bpsi\s^z\g^{x,y}\psi$ correspond to spin currents in the microscopic theory.
It follows that the fixed line could be relevant to kagome magnets with randomly varying DM fields.

We also study the RG flow when disorder couples to the more general set of operators:
\eq{\label{eqn:OtherDisorder}
N_s&=\bpsi\psi,
&
N^a&=\bpsi\s^a\psi,
&
J^a_\m&=i\bpsi\s^a\g_\m\psi,
&
J_{\mathrm{top},\m}&=\ep_{\m\n\r}\ptl^\n A^\r
}
where $\s^a=\(\s^x,\s^y,\s^z\)$. 
We find that the U(1)$\times$SU($N$) symmetric critical line is unstable to disorder coupling to either $N^{x,y}$, $J^{x,y}_j$, and $J^z_0$. These theories flow to strong disorder and cannot be accessed with the perturbative methods used here. 
Disorder coupling to the topological current is marginal to $\O(1/(2N)^2)$; however, upon including higher order contributions, the  $J_{\mathrm{top},0}$ disorder strength becomes relevant. 

In Sec.~\ref{sec:MicroOperators} we will see that if the Pauli matrices in the operators of Eq.~\eqref{eqn:OtherDisorder} act on the valley indices of the emergent Dirac fermions, then the mass-like terms $N^a$ should be associated with different valence bond ordering patterns on the kagome lattice \cite{Hermele2008}.
Our analysis therefore indicates that the QED$_3$ phase is unstable to random bond disorder in the kagome antiferromagnet. 

There have been earlier studies of massless Dirac fermions coupled to disorder. A comprehensive analysis for
free Dirac fermions was presented by Ludwig {\it et al.} \cite{PhysRevB.50.7526}. An important ingredient in their
analysis was the coupling of the disorder to components of the current operator $J^\m(r)=i\bpsi\g^\m\psi(r)$.
For the free theory, $J^\m$ has scaling dimension 2 like any other globally conserved current; consequently, the disorder
coupling to $J^\m$ turns out to be marginal at the clean free fixed point, and this has important consequences for the
disordered system. For the QED$_3$ case considered here, the situation 
is dramatically different: because of the presence of the 
gauge field, $J^\m$ is no longer a globally conserved current, 
and its scaling dimension at the CFT fixed point is 3 \cite{Hermele2005}. The corresponding disorder is strongly 
irrelevant, and this is the reason it was not included in Eqs.~(\ref{Sdis1}) and~\eqref{eqn:OtherDisorder}.

Other earlier works with Dirac fermions studied the influence of disorder 
and the $1/r$ Coulomb interactions between the Dirac fermions
\cite{1998PhRvL..80.5409Y,PhysRevB.60.8290}, and were motivated by the study of transitions between quantum Hall states.
Today, they can be applied to graphene. As in our work, 
they found fixed lines at non-zero disorder and interactions.

Our paper begins in Sec.~\ref{sec:DisQED} by discussing our model in more detail.
We start by reviewing some important properties of QED$_3$ in Sec.~\ref{sec:pureQED3}, before presenting the types of disorder under consideration in Sec.~\ref{sec:disType}. 
The renormalization procedure and resulting $\b$-functions are described in Sec.~\ref{sec:RenormAction}.
The remainder of the section discusses the flows which result upon enforcing different symmetries, including the U(1)$\times$SU($N$) symmetric critical line mentioned above (Sec.~\ref{sec:FixedPt}). 
Sec.~\ref{sec:MicroOperators} focuses on applications to the kagome antiferromagnet and translates the fermion bilinears and topological current of the CFT to the microscopic observables of the spin model. 
Finally, in Sec.~\ref{sec:SpinConductivity} the flavor conductivity along the critical line is calculated.
We review out results and conclude in Sec.~\ref{sec:Conclusion}.

\section{Disordered QED$_3$}\label{sec:DisQED}
\subsection{Pure QED$_3$}\label{sec:pureQED3}
\begin{figure}
\centering
\input{effPhotonProp.tex}
\caption{Diagrammatic expression for the effective photon propagator in the large-$N$ limit. The dotted lines indicate the bare photon propagator, $D^0_{\m\n}(p)$, while the fermion bubbles are equal to $\Pi_{\m\n}(q)$. As indicated in the text, only the full photon propagator will be used. }
\label{fig:effectivePhotProp}
\end{figure}
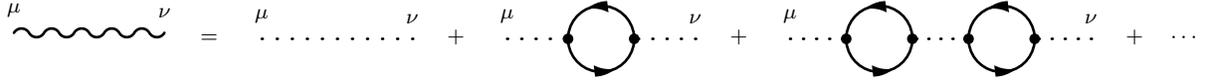

The Euclidean signature action for QED$_3$ is given in Eq.~\eqref{eqn:qedAction}.
In the IR limit, for $N$ large enough, this theory flows to a strongly coupled CFT at $e^2=\infty$. 
All loop contributions to the fermion propagator are suppressed by $1/2N$ and so we will work with the free propagator
\eq{
G(p)&=\d_{\a\b}{ip_\m\g^\m\o p^2}
}
where $\a$ and $\b$ are flavor indices. 
The same is not true of the photon propagator. 
Instead, the $N=\infty$ Green's function must include a summation over the bubble diagrams shown in Fig.~\ref{fig:effectivePhotProp}. 
The effective propagator is determined most simply by adding a non-local gauge fixing term to the action \cite{pufu16}
\eq{
S_{\text{gauge-fixing}}&={1\o32\(\zeta-1\)}\int {d^3p\o(2\pi)^3}{p_\m p_\n\o\abs{p}}A^\m(p)A^\n(-p),
}
where $\zeta$ is an arbitrary parameter which cannot enter into any physical observable.
The resulting free photon propagator is
\eq{
D^{0}_{\m\n}(p)&={2Ne^2\o p^2}\(\d_{\m\n}-{p_\m p_\n\o p^2}\)+{16\(\zeta-1\)\o\abs{p}}{p_\m p_\n\o p^2}\cdot
}
The polarization bubble in Fig.~\ref{fig:effectivePhotProp} can be evaluated (see Appendix~\ref{app:PolarizationLoop}) and gives
\eq{
\Pi^{\m\n}(p)&={\abs{p}\o16}\(\d_{\m\n}-{p_\m p_\n\o p^2}\)\cdot
}
Therefore, the $N=\infty$ propagator is 
\eq{
D^\mathrm{eff}_{\m\n}(p)&=\bigg(\[D^0_{\m\n}(p)\]^{-1}+\Pi_{\m\n}(p)\bigg)^{-1}
={ 16\o \abs{p}}\(\d_{\m\n}-\zeta{p_\m p_\n\o p^2}\)+\O\(p^2\o e^2\).
}
Here, we have used the fact that, because the dimension of $e^2$ is 1, in the infrared limit, $p\rightarrow0$, all terms of $\O(p^2/e^2)$ are suppressed. Provided we use the effective photon propagator and organize our perturbation theory such that no fermion bubbles of the type summed in Fig.~\ref{fig:effectivePhotProp} are repeated, the limit $e^2\rightarrow\infty$ can be taken directly. We will further simplify by working in the $\zeta=0$ gauge.

Since we will regulate the disordered theory using dimensional regularization, we write 
\eq{
S_\mathrm{qed}\[ \psi,\bpsi,A\]&=-\int d^dx\,d\t\,\bpsi_\a\(\sd{\ptl}+{i\m^{-\ep/2} \mathpzc{g}\o\sqrt{2N}}\sd{A}\)\psi_\a,
}
where $d=2+\ep$, $\m$ is an arbitrary scale, and the photon propagator is understood to be $D^\mathrm{eff}_{\m\n}(p)$.
We will often write $D=d+1$ and denote spacetime coordinates by $r=(x,\t)$.
By making the coupling dimensionful, we are taking the engineering dimension of $A_\m$ to be $d/2$. Gauge invariance guarantees that $\mathpzc{g}$ will not be renormalized, and it will be set to unity at the end of the calculation. This is discussed further in Sec.~\ref{sec:RenormAction}.

We now discuss the symmetries and operator content of the theory. 
QED$_3$ has a $\mathrm{SU}(2N$) symmetry under which the flavors rotate into one another:
\eq{
\psi_\a\rightarrow \[\exp\(i\theta_{ab}\s^a T^b\)\]_{\a\b}\psi_\b\,\cdot
}
Here, we have expressed the $(2N)^2-1$ generators of SU($2N$) as
\eq{
\s^a \,T^b&,
&
\s^a,&
&
T^b,
}
where $\s^a$, $a=x,y,z$, are the 2$\times$2 Pauli matrices and $T^a$, $a=1,\dots,N^2-1$, are $N\times N$ traceless, Hermitian matrices normalized such that $\tr\(T^aT^b\)=\d_{ab}/2$. 
Associated with each generator of this symmetry is a conserved current, 
\eq{\label{eqn:ConservedCurr}
J^{ab}_\m(r)&=i\bpsi \s^a T^b\g_\m\psi(r),
&
J^{a0}_\m(r)&=i\bpsi \s^a \g_\m\psi(r),
&
J^{0b}_\m(r)&=i\bpsi T^b\g_\m\psi(r).
}
To all orders in $1/(2N)$, these operators have scaling dimension $\Delta_J=2$. 
When we discuss the symmetry of the theory in the remainder of the paper, we will be referring to the flavour symmetry unless explicitly stated otherwise.

As we remarked in Sec.~\ref{sec:intro}, the irrelevance of monopoles results in an emergent U(1)$_\mathrm{top}$ symmetry associated with a conserved gauge flux current, 
\eq{\label{eqn:topCurrentDef}
J^\m_\mathrm{top}&=\ep^{\m\n\r}\ptl_\n A_\r.
}
Like the SU($2N$) currents, the scaling dimension of $J^\m_\mathrm{top}$ is exactly 2.
In the limit we consider, monopole scaling dimensions are much greater than 2, though, as $N$ descreases, this may cease to be the case.

The global U(1) transformation, $\psi\rightarrow e^{i\theta}\psi$, also has a conserved current, $J^\m(r)=i\bpsi\g^\m\psi(r)$. 
However, because the U(1) phase rotation is also a local symmetry, its current is quite different from the SU($2N$) and U(1)$_\mathrm{top}$ currents.
This is evident upon considering the equations of motion:
\eq{
J^\m&={1\o e^2\sqrt{2N}}\ptl_\n F^{\n\m}={1\o e^2\sqrt{2N}}\,\ep^{\m\n\r}\ptl_\n J_{\mathrm{top},\r}\cdot
} 
Taken as an operator identity, this implies that the global U(1) current is actually a descendent of the gauge field, and, consequently, its scaling dimension is 3 instead of 2 \cite{Hermele2005}.

In addition to the currents, there are $(2N)^2-1$ ``mass" operators which can be constructed from the SU($2N$) generators,
\eq{\label{eqn:MassOps}
N^{ab}(r)&=\bpsi \s^a T^b\psi(r),
&
N^{a0}(r)&=\bpsi \s^a\psi(r),
&
N^{0b}(r)&=\bpsi T^b\psi(r),
}
as well as the usual 2+1 dimensional Dirac mass term:
\eq{\label{eqn:MassOps2}
N_s(r)&={1\o\sqrt{2N}}\;\bpsi\psi(r)\cdot
}
Unlike the currents, at finite $N$ these operators have nontrivial anomalous scaling dimensions \cite{Hermele2005,Hermele2005err,pufu16}.
In particular, since $N_s$ allows for ``photon decay" processes, it becomes less relevant, with a scaling dimension of 
\eq{\label{eqn:massDim}
\Delta_s&=2+{128\o3\pi^2(2N)}+\O\(1\o N^2\)\cdot
}
Conversely, the SU($2N$) masses become more relevant:
\eq{\label{eqn:SU2NmassDim}
\Delta_1&=2-{64\o3\pi^2(2N)}+\O\(1\o N^2\)\cdot
}

\subsection{Disorder}\label{sec:disType}

We are interested in perturbing the QED$_3$ CFT with disorder. A simple scaling argument shows that there are a limited number of operators which can give interesting results upon coupling to disorder. 
We begin by considering disorder coupling to an arbitrary, gauge-invariant operator $\O$ with scaling dimension $\Delta_\O$:
\eq{\label{eqn:DisActionGeneral}
S_{\mathrm{dis},\O}\[\O\]&=\int d^dx\,d\t\,M_\O(x)\O(x,\t)
}
where $M_\O(x)$ is a Gaussian random variable with zero average and with correlations given by
\eq{\label{eqn:DisAverageGeneral}
\overline{M_\O(x)M_\O(x')}&={g_\O\o2}\d^d(x-x')\cdot
}
$g_\O$ is the variance of $M_\O$ and controls the strength of the disorder. 
To allow for a well-controlled perturbative expansion, we assume that $g_\O$ is of the same order as $1/(2N)$; this implies that the bare disorder strength and the electromagnetic interaction are of the same magnitude.

Since $S_\mathrm{dis}$ explicitly breaks Lorentz symmetry, time and space need no longer scale in the same way.  
We express this by 
allowing time to scale as $-z$: $[\tau ]=-z$. 
``$z$" is referred to as the dynamic critical exponent. 
While our assumption that $g_\O\sim\O(1/N)$ ensures that $z-1\sim \O(1/N)$ as well, the possibility that $z\neq1$ at higher orders has several effects which will be important later. 
First, the dimensions of conserved currents are no longer all fixed precisely at 2. 
The scaling dimension of the time component remains 2, but spatial components have dimension $\Delta_{J,xy}=1+z$.
Second, having a dynamic critical exponent different from unity also changes the dimensional analysis of the disorder strength $g_\O$.
Eq.~\eqref{eqn:DisActionGeneral} establishes that $\[M_\O\]=d+z-\Delta_\O$, and with Eq.~\eqref{eqn:DisAverageGeneral}, this indicates that $\[g_\O\]=d+2z-2\Delta_\O$.
It follows that the critical dimension is $1+z$. 
This is the quantum version of the Harris criterion \cite{Harris1974}.

At tree level, $z=1$, so the Harris criterion indicates that in 2$d$ disorder coupling to operators with $\Delta_\O>2$ is irrelevant: at low energies, the system is described by the clean theory. 
Conversely, operators with scaling dimensions less than or equal to 2 are either relevant or marginal perturbations when coupled to disorder.

Referring to the previous section, to leading order in $N$, there are no relevant perturbations and only the global topological current, the SU($2N$) currents, and the mass terms, $N_s$ and $N_{ab}$, are marginal.
However, as mentioned above, at finite $N$, it's possible that the scaling dimension of an allowed monopole operator is less than 2, making it relevant. We
will not examine this possibility in our present large $N$ expansion.
As discussed in Sec.~\ref{sec:intro}, 
the global U(1) current, $J^\m=i\bpsi\g^\m\psi$, is irrelevant because its scaling dimension is 3.

Keeping in mind that in order to compare with the kagome antiferromagnet we must set $N=2$, we couple disorder to operators which break the SU($2N$) symmetry down to SU($N$): 
\eq{\label{eqn:DisorderAction}
S_\mathrm{dis}[\psi,\bpsi]&=\int d^dx\,d\t\bigg[
M_s(x)\bpsi\psi(x,\t)
+M_{t,a}(x)\bpsi\s^a\psi(x,\t)
\nt&\quad
+i\mathcal{A}_{j a}(x)\bpsi\s^a\g^j\psi(x,\t)
+V_a(x)\bpsi\s^a\g^0\psi(x,\t)
\nt&\quad
+i\mathcal{E}_j(x) J^j_\mathrm{top}(x,\t)+\mathcal{B}(x)J^0_\mathrm{top}(x,\t)
\bigg]
}
where $M_s$, $M_{t,a}$, $\mathcal{A}_{ja}$, $V_{a}$, $\mathcal{E}_j$, and $\mathcal{B}$ are Gaussian random variables with vanishing mean.
Here and throughout the paper we use the convention that, when contracting vectors and $\g$-matrices, Roman letters $i,j,\ell,$ etc. indicate that the sum is only over the spatial coordinates $x$, while Greek letters $\m,\n,\s$, etc. include time as well.
We note that since the quenched disorder is classical, the random fields have been expressed in real time. 
That is, the time component of all classical gauge potentials picks up a factor of ``$i$".
Averaging over disorder, we have
\eq{
\overline{M_s(x)M_s(x')}&={g_s\o2}\d^d(x-x'),
&
\overline{V_a(x)V_b(x')}&={g_{v,a}\o2}\d_{ab}\d^d(x-x')
\nt
\overline{M_{t,a}(x)M_{t,b}(x')}&={g_{t,a}\o2}\d_{ab}\d^d(x-x'),
&
\overline{\mathcal{E}_i(x)\mathcal{E}_j(x')}&={g_{\mathcal{E}}\o2}\d_{ij}\d^d(x-x')
\nt
\overline{\mathcal{A}_{i a}(x)\mathcal{A}_{j b}(x')}&={g_{\A,a}\o2}\d_{ab}\d_{ij}\d^d(x-x'),
&
\overline{\mathcal{B}(x)\mathcal{B}(x')}&={g_{\mathcal{B}}\o2}\d^d(x-x')
}
with all other two-points vanishing.  
As in the general case considered above, we assume that the variances, $\{g_s,g_{t,a},g_{\A,a},g_{v,a},g_\E,g_\B\}$, are small and of the same order as $1/(2N)$.

When we interpret these operators in the context of the kagome antiferromagnet, the $\s^a$ matrices will act on spin. 
By recalling that the Dirac mass, $\bpsi\psi$, is odd under time reversal in 2+1 dimensions, we deduce that
the SU(2) mass operators, $i\bpsi\s^a\psi$, should be even. The same logic asserts that the scalar potential operators, $i\bpsi\g^0\s^a\psi$, are odd under time reversal while the vector potential operators, $i\bpsi\g^j\s^a\psi$, are even.
Similarly, the fact that $J^0_\mathrm{top}$ and $J^j_\mathrm{top}$ are the emergent magnetic field and electric fields respectively reveals that they are odd and even under time reversal. 
Therefore, while the zero mean of the quenched disorder fields implies that $S_\mathrm{dis}[\psi,\bpsi]$ preserves time reversal on average, it is only a good symmetry everywhere within the system when $M_s$, $V_a$ and $\B$ are not present (equivalently, $g_s=g_{v,a}=g_\B=0$).
In Sec.~\ref{sec:MicroOperators} we will discuss the microscopic meaning of $S_\mathrm{dis}[\psi,\bpsi]$ in the kagome antiferromagnet more thoroughly.

We will use dimensional regularization with $d=2+\ep$ so that the dimension of the variances is  shifted to $\[g_\xi\]=-\ep$, where $\xi=s$, $(t,a)$, $(\A,a)$, $\(v,a\)$, $\E$, or $\B$.
For convenience, we make the couplings dimensionless by taking $g_\xi\rightarrow \m^{-\ep}g_\xi$ where $\m$ is an arbitrary momentum scale. 
When we perform the renormalization group study, the couplings are restricted to non-negative values because they physically correspond to variances.

The disorder breaks translational symmetry and makes calculating quantities for a given realization of disorder completely intractable.
Instead, the fundamental quantity of interest is the disorder-averaged free energy:
\eq{
\overline{F}&=-\overline{\log Z}
\nt
&=-\log\Bigg[\int DM(x)\,DM_a(x)\,DM_{\m a}(x) \,D\A_{\m a}(x)\,DV_a(x)\,e^{-S_\mathrm{qed}-S_\mathrm{dis}}\,e^{-{\m^\ep\o 2g_{s}^2}\int d^dx\,M_s(x)^2}\,e^{-{\m^\ep\o2g_{\B}^2}\int d^dx \mathcal{B}(x)^2}
\nt
&\quad\quad\times
e^{-{1\o2g_{\E}^2}\int d^dx \mathcal{E}_j(x)\E^j(x)}
\prod_{a=x,y,z}\,e^{-{\m^\ep\o 2g_{t,a}^2}\int d^dx\,M_{t,a}(x)^2}\,
e^{-{\m^\ep\o 2g_{\A,a}^2}\int d^dx\,\mathcal{A}_{j a}(x)A^{ja}(x)}\,
e^{-{\m^\ep\o 2g_{v,a}^2}\int d^dx\,V_a(x)^2}\Bigg].
} 
To solve perturbatively, we employ the replica trick. Using the identity
\eq{
\log Z&=\lim_{n\rightarrow0}{Z^n-1\o n},
}
we instead calculate
\eq{
Z_n&\equiv\overline{Z^n}
=\N\int \prod_{\substack{\a=1,\dots,2N\\\ell=1,\dots,n}}D\psi_{\a\ell}D\bpsi_{\a\ell}DA_\ell
	\,e^{-S_n[\psi_{\a\ell},\bpsi_{\a\ell}]}
}
where $\mathcal{N}$ is a normalization constant and
\eq{
S_n\[\psi,\bpsi,A\]&=
-\sum_\ell\int d^dx\,d\t \,\bpsi_{\ell}(x,\t)\(\sd{\ptl}+{i\mu^{-\ep/2}\mathpzc{g}\o\sqrt{2N}}\sd{A}_\ell\)\psi_{\ell}(x,\t)
\nt
&\quad
+{\m^{-\ep}\o2}\sum_{\ell,m}\int d^dx\,d\t\,d\t'\bigg\{
-g_s\bpsi_{\ell}\psi_{\ell}(x,\t)\bpsi_{ m}\psi_{ m}(x,\t')
-\sum_a g_{t,a}\bpsi_{\ell}\s^a\psi_{\ell}(x,\t)\bpsi_{m}\s_a\psi_{m}(x,\t')
\nt
&\quad-\sum_ag_{\A,a}\,\bpsi_{\ell}i\g^j\s^a\psi_{\ell}(x,\t)\bpsi_{m}i\g_j\s_a\psi_{m}(x,\t')
+\sum_a{g_{v,a}}\,\bpsi_{\ell}i\g^0\s^a\psi_{\ell}(x,\t)\bpsi_{m}i\g^0\s_a\psi_{m}(x,\t')
\nt
&\quad
-{g_{\B}} J^{\ell,0}_{\mathrm{top}}(x,\t)J^{m,0}_{\mathrm{top}}(x,\t')
+{g_{\E}}\, J^{\ell,j}_\mathrm{top}(x,\t)J^m_{\mathrm{top},j}(x,\t')
\bigg\}\cdot
}
In addition to the physical flavor symmetry, the fermions and photon now carry a replica index denoted by $\ell$ and $m$. 
We have suppressed the summation over the flavour indices and will continue to do so in what follows. 
Likewise, the replica indices will often be left implicit.
The Feynman rules corresponding to $S_n\[\psi,\bpsi,A\]$ are provided in Fig.~\ref{fig:FeynRulesAll}.
\begin{figure}

\input{feynRules.tex}
\caption{Feynman rules associated with the replicated action, $S_n\[\psi,\bpsi,A\]$. The diagrams on the first and second rows are diagonal with respect to the replica and flavor indices. In the four-point diagrams, $\ell$ and $m$ are replica indices while $\a,\b,\s,\r$ label the $2N$ fermion flavors.}
\label{fig:FeynRulesAll}
\end{figure}

\section{Renormalization group analysis}\label{sec:RGanalysis}

\subsection{Renormalized action}\label{sec:RenormAction}

\begin{figure}
\centering
\input{vanishingBubble.tex}
\caption{Example of a diagram which vanishes in the replica limit, $n\rightarrow0$. The internal fermion loop involves a sum over all replica indices, and multiplies the diagram by an overall factor of $n$.}
\label{fig:vanishingBubble}
\end{figure}
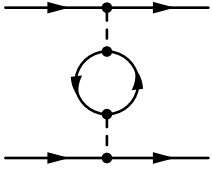

The low energy properties of $S_n\[\psi,\bpsi,A\]$ can be studied with the same renormalization techniques used in many-body systems provided the number of replicas, $n$, is taken to zero at the end of the calculation.
This implies that diagrams which sum over all replicas must be neglected. For instance, Fig.~\ref{fig:vanishingBubble} is proportional to $n$ and should not be included.

We will use renormalized perturbation theory \cite{Amit2005}, making use of a counter term action:
\eq{
S^\mathrm{CT}_n\[\psi,\bpsi\]&=-\sum_\ell\int d^dx\,d\t \,\bpsi_{\ell}
\( i\d_1\g^0{\ptl\o\ptl\t}+ i\d_2\g^j{\ptl\o\ptl x_j} +{i\mu^{-\ep/2}\mathpzc{g}\d_1'\o\sqrt{2N}}A_{0}^{\ell}\g^0+{i\mu^{-\ep/2}\mathpzc{g}\d_2'\o\sqrt{2N}}A_{j}^{\ell}\g^j\)\psi_{\ell}(x,\t)
\nt
&\quad
+{\m^{-\ep}\o2}\sum_{\ell,m}\int d^dx\,d\t\,d\t'\bigg\{
-\d_s\,\bpsi_{\ell}\psi_{\ell}(x,\t)\bpsi_{ m}\psi_{ m}(x,\t')
-\sum_a \d_{t,a}\,\bpsi_{\ell}\s^a\psi_{\ell}(x,\t)\bpsi_{m}\s_a\psi_{m}(x,\t')
\nt
&\quad-\sum_a\d_{\A,a}\,\bpsi_{\ell}i\g^j\s^a\psi_{\ell}(x,\t)\bpsi_{m}i\g_j\s_a\psi_{m}(x,\t')
+\sum_a{\d_{v,a}}\,\bpsi_{\ell}i\g^0\s^a\psi_{\ell}(x,\t)\bpsi_{m}i\g^0\s_a\psi_{m}(x,\t')
\nt
&\quad
-{\d_{\B}}\,J^{\ell,0}_{\mathrm{top}}(x,\t)J^{m,0}_{\mathrm{top}}(x,\t')
+{\d_{\E}}\, J^{\ell,j}_\mathrm{top}(x,\t)J^m_{\mathrm{top},j}(x,\t')
\bigg\}\cdot
}
The counter terms, $\{\d_{1,2},\d_{1,2}',\d_s,\d_{t,a},\d_{v,a},\d_{\A,a},\d_\B,\d_\E\}$, are determined by requiring that all physical observables are finite in a dimensional regularization scheme.
While relativistic invariance is explicitly broken, there is no need track the relative flow of the fermion and photon velocities since the low-energy behaviour of the photon propagator descends entirely from its interaction with the fermions.

The bare action is the sum of $S_n$ and $S^\mathrm{CT}_n$:
\eq{\label{sec:bareAction}
S^B_n\[\psi,\bpsi,A\]&=-\sum_\ell\int d^dx_B\,d\t_B \,\bpsi_{\ell,B}\(i\g^0{\ptl\o\ptl \t_B}+i\g^j{\ptl\o\ptl x_{j,B}}+{i\mathpzc{g}_B\g^0\o\sqrt{2N}}A_{0,B}^\ell+{i\mathpzc{g}_B\g^j\o\sqrt{2N}}A_{j,B}^\ell\)\psi_{\ell,B}(x_B,\t_B)
\nt
&\quad
+{1\o2}\sum_{\ell,m}\int d^dx_B\,d\t_B\,d\t'_B\bigg\{
g_s^B\,\bpsi_{\ell,B}\psi_{\ell,B}(x_B,\t_B)\bpsi_{ m,B}\psi_{ m,B}(x_B,\t'_B)
\nt
&\quad
-\sum_a{g_{t,a}^B}\,\bpsi_{\ell,B}\s^a\psi_{\ell,B}(x_B,\t_B)\bpsi_{m,B}\s_a\psi_{m,B}(x_B,\t'_B)
\nt
&\quad
-\sum_a{g_{\A,a}^B}\,\bpsi_{\ell,B}i\g^j\s^a\psi_{\ell,B}(x_B,\t_B)\bpsi_{m,B}i\g_j\s_a\psi_{m,B}(x_B,\t'_B)
\nt
&\quad
+\sum_a{g_{v,a}^B}\,\bpsi_{\ell,B}i\g^0\s^a\psi_{\ell,B}(x_B,\t_B)\bpsi_{m,B}i\g^0\s_a\psi_{m,B}(x_B,\t'_B)
\nt
&\quad
-{g^B_{\B}}\,J^{\ell,0}_{\mathrm{top},B}(x_B,\t_B)J^{m,0}_{\mathrm{top},B}(x_B,\t'_B)
+{g^B_{\E}}\, J^{\ell,j}_{\mathrm{top},B}(x_B,\t_B)J^m_{\mathrm{top},B,j}(x_B,\t'_B)
\bigg\}
}
where the bare fields and coordinates are
\eq{\label{eqn:BareFieldDef}
\psi_B(x_B,\t_B)&=Z_1^{1/2}\psi(x,\t),
\nt
A_{0, B}(x_B,\t_B)&=Z_{\g,0}^{1/2}A_0(x,\t),
&
A_{j B}(x_B,\t_B)&=Z_{\g,xy}^{1/2}A_j(x,\t),
\nt
\t_{B}&={Z_2\o Z_1}\t,
&
x_B=x.
}
Here, we have written $Z_1=1+\d_1$ and $Z_2=1+\d_2$, and, by taking $x=x_B$, we are renormalizing relative to the spatial scale. 
Gauge invariance constrains the photon field strength renormalization constants to be
\eq{
Z_{\g,0}^{1/2}&={Z_1\o Z_2},
&
Z_{\g,xy}^{1/2}=1,
}
and it follows that we must have $\d_{1,2}=\d_{1,2}'$. This has been explicitly verified.
The field strength renormalization of the topological currents then follows simply from the renormalization of $A_\m$ and $(x,\t)$:
\eq{
J^0_{\mathrm{top},B}&={\ptl A_y\o\ptl x}-{\ptl A_x\o\ptl y},
&
J^x_{\mathrm{top},B}&={Z_1\o Z_2}\({\ptl A_0\o\ptl y}-{\ptl A_y\o\ptl \t}\),
&
J^y_{\mathrm{top},B}&=-{Z_1\o Z_2}\({\ptl A_0\o\ptl x}-{\ptl A_x\o\ptl \t}\)\cdot
}


As discussed in the previous section, the dynamic critical exponent relates the scaling of time and space to one another:
\eq{
\m{d\o d\m}\t&=z\t \,\cdot
}
Since $\t_B$ should scale like $\m$, taking its derivative with respect to $\log \m$ gives
\eq{\label{eqn:DynExp}
z=1-\m{d\o d\m}\log\(Z_2\o Z_1\)\cdot
}
The renormalization of the disorder strengths is determined by comparing the bare action to $S_n+S_n^{\mathrm{CT}}$:
\eq{\label{eqn:BareCouplings}
g_{s}^B&=\m^{-\ep}Z_2^{-2}\(g_s+\d_s\),
&
g_{t,a}^B&=\m^{-\ep}Z_2^{-2}\(g_{t,a}+\d_{t,a}\),
\nt
g_{\A,a}^B&=\m^{-\ep}Z_2^{-2}\(g_{\A,a}+\d_{\A,a}\),
&
g_{v,a}^B&=\m^{-\ep}Z_2^{-2}\(g_{v,a}+\d_{v,a}\),
\nt
g_{\E}^B&=\m^{-\ep}\(g_\E+\d_\E\),
&
g_{\B}^B&=\m^{-\ep}Z_1^2Z_2^{-2}\(g_\B+\d_\B\)\cdot
}
The fact that the bare couplings are independent of the scale $\m$ establishes the $\beta$-functions. 
For disorder coupling to fermion bilinears, we have
\eq{
0&=-\ep\(g_\xi+\d_\xi\)-2\(g_\xi+\d_\xi\)\m{d\o d\m}\log Z_2+\m{d\o d\m}\d_\xi+\b_\xi,
&
\xi&=s,(t,a),(\A,a),(v,a),
}
where $\b_\xi=\m dg_\xi/d\m$ and $a=x,y,z$. 
Similarly, the $\b$-functions for the flux disorder are
\eq{
0&=-\ep\(g_\E+\d_\E\)+\m{d\o d\m}\d_\E+\b_\E,
\nt
0&=-\ep\(g_\B+\d_\B\)+2\(g_\B+\d_\B\)(z-1)+\m{d\o d\m}\d_\B+\b_\B.
}
In the second equation, the relation $z-1=\m d\log\(Z_1/Z_2\)/d\m$ has been used.

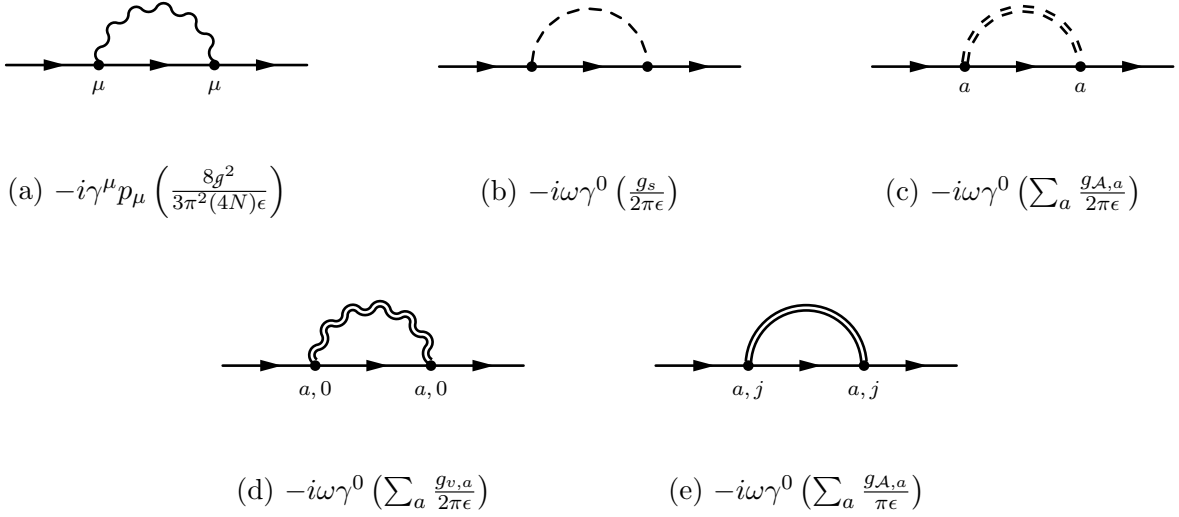
\begin{figure}
\centering
\input{fermionSelfE}
\caption{Feynman diagrams which contribute to the fermion self-energy at $\O(g_\xi,1/2N)$.}
\label{fig:fermionSelfE}
\end{figure}
The fermion self-energy diagrams which determine the counter terms $\d_1$ and $\d_2$ to leading order are shown in Fig.~\ref{fig:fermionSelfE}. 
These are evaluated in Appendix~\ref{app:fermSelfE}, and the divergent pieces are listed below the corresponding diagram in the figure.
Only the photon loop in Fig.~\ref{fig:fermSelfE-PhotonInt}  contributes to $Z_2$. 
In order to cancel this divergence, we must have
\eq{
\d_2&={8\mathpzc{g}^2\o 3\pi^2(2N)\ep}\cdot
}
The frequency counter term, on the other hand, receives contributions from all of the diagrams in Fig.~\ref{fig:fermionSelfE}:
\eq{
\d_1&={8\mathpzc{g}^2\o 3\pi^2(2N)\ep}+{1\o 2\pi \ep}\[g_s+\sum_a\(g_{t,a}+g_{v,a}+2g_{\A,a}\)\]\cdot
}
It follows from Eq.~\eqref{eqn:DynExp}, the dynamic critical exponent is
\eq{\label{eqn:DynExp2}
z=1+{1\o2\pi}\[g_s+\sum_a\(g_{t,a}+2g_{\A,a}+g_{v,a}\)\]\cdot
}
The provision that all couplings be positive implies that $z\geq1$ always.

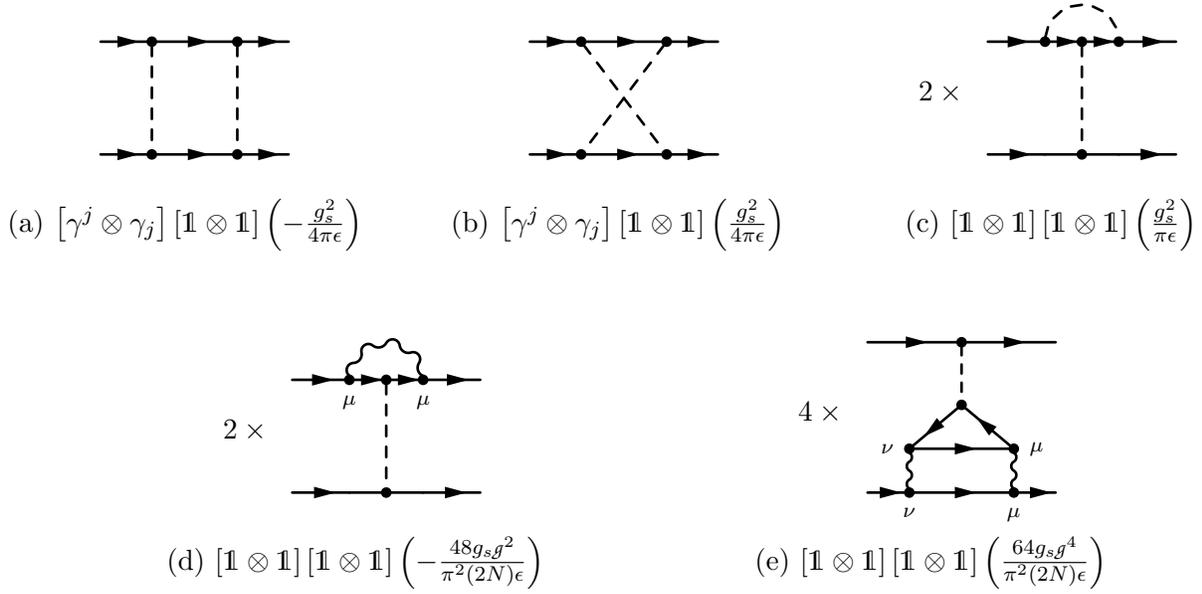
\begin{figure}
\input{4points2.tex}
\caption{
Diagrams which contribute when only SU($2N$)-preserving, bilinear disorder is considered ($g_{t,a}=g_{\A,a}=g_{v,a}=0$). Both Figs.~\ref{fig:4ptSigmaSigmaVert} and \ref{fig:4ptsSigmaPhotonArc} are accompanied by a diagram with the interaction on the other vertex. Partner diagrams to Fig.~\ref{fig:FermLoopSigmaInt1} with the fermion loop direction reversed and/or the vertex switched are also present.
These diagrams sum to $\[\id\otimes\id\]\[\id\otimes\id\]\bigg\{{g_s^2\o\pi\ep}+{64g_s\mathpzc{g}^4\o\pi^2(2N)\ep}-{48g_s\mathpzc{g}^2\o\pi^2(2N)\ep}\bigg\}$}
\label{fig:sigmasigma}
\end{figure}
The bilinear counter terms, $\d_\xi$, $\xi=s,(t,a),(\A,a),(v,a)$, are determined by adding diagrams like those in Fig.~\ref{fig:sigmasigma}. 
In particular, Fig.~\ref{fig:sigmasigma} shows all diagrams which renormalize disorder coupled to the SU($2N$)-symmetric mass when all other couplings have been tuned to zero.
The integrals are performed in Appendix~\ref{app:4ptDiagramsCalc}, and the remainder of the diagrams renormalizing the bilinear disorder are shown in Appendix~\ref{app:4ptDiagramsModel} in Tables~\ref{tab:DiagramTab1}, \ref{tab:DiagramTab2}, and \ref{tab:DiagramTab3}.
The resulting counter terms are
\eq{\label{eqn:counterTerms}
\d_s&=-{1\o\pi\ep}\[
g_s^2+{64g_s\mathpzc{g}^4\o\pi(2N)}-{48g_s\mathpzc{g}^2\o\pi(2N)}+g_s\sum_a\(g_{t,a}+g_{v,a}-2g_{\A,a}\)-2\sum_ag_{v,a}g_{\A,a}
\]
\nt
\d_{t,a}&=
-{1\o\pi\ep}\Bigg[
g_{t,a}\(2g_{t,a}-\sum_bg_{t,b}\)-2g_{t,a}\(2g_{\A,a}-\sum_bg_{\A,b}\)+g_{t,a}\(2g_{v,a}-\sum_bg_{v,b}\)
\nt&\quad
+g_{t,a}g_s+4g_{t,a}g_{\A,a}\mathpzc{g}^2-4g_{t,a}g_{v,a}\mathpzc{g}^2-2\sum_{bc}\abs{\ep^{abc}}g_{t,b}g_{\A,c}-{48g_{t,a}\mathpzc{g}^2\o\pi(2N)}
\bigg]
\nt
\d_{\mathcal{A},a}&=
-{1\o\pi\ep}\bigg[-g_sg_{v,a}-\sum_{bc}\abs{\ep^{abc}}\({g_{t,b}g_{t,c}\o2}+2g_{\A,b}g_{\A,c}+{g_{v,b}g_{v,c}\o2}\)-{16g_{\A,a}\mathpzc{g}^2\o3\pi(2N)}\bigg]
\nt
\d_{v,a}&=-{1\o\pi\ep}\bigg[
-g_{v,a}\(2g_{v,a}-\sum_bg_{v,b}\)-g_{v,a}\(2g_{t,a}-\sum_bg_{t,b}\)-2g_{v,a}\(2g_{\A,a}-\sum_bg_{\A,b}\)
\nt&\quad
-g_{v,a}g_s-2g_sg_{\A,a}
-2\sum_{bc}\abs{\ep^{abc}}g_{v,b}g_{\A,c}-{16g_{v,a}\mathpzc{g}^2\o3\pi(2N)}\bigg]\cdot
}
The graphs which renormalize the topological disorder stengths, $g_\E$ and $g_\B$, are actually three loop diagrams at leading order. 
These are calculated in Appendix~\ref{app:TopRenorm} where we find
\eq{\label{eqn:FluxCounterTerms}
\d_\E&={g_s g_\B\mathpzc{g}^4\o\pi\ep},
&
\d_\B&={g_s g_\E\mathpzc{g}^4\o\pi\ep}\cdot
}
Differentiating the bare couplings (Eq.~\eqref{eqn:BareCouplings}) with respect to $\m$, solving for the $\b$-functions to $\O(g_\xi^2, g_\xi/2N)$, and setting $\mathpzc{g}^2=1$, we obtain
\eq{\label{eqn:GeneralBetaFuns}
\pi \b_s
&=\pi\ep g_s+
g_s\[
g_s+\sum_a\(g_{t,a}+g_{v,a}-2g_{\A,a}\)+2c\]-2\sum_ag_{v,a}g_{\A,a}\
\nt
\pi\b_{t,a}&=\pi\ep g_{t,a}+
g_{t,a}\Bigg[
\(2g_{t,a}-\sum_bg_{t,b}\)+2\(2g_{\A,a}+\sum_bg_{\A,b}\)-\(6g_{v,a}+\sum_bg_{v,b}\)
\nt&\quad
+g_{t,a}g_s-2c
\bigg]-2\sum_{bc}\abs{\ep^{abc}}g_{t,b}g_{\A,c}
\nt
\pi\b_{\A,a}&=
\pi\ep g_{\A,a}-g_sg_{v,a}-\sum_{bc}\abs{\ep^{abc}}\({g_{t,b}g_{t,c}\o2}+2g_{\A,b}g_{\A,c}+{g_{v,b}g_{v,c}\o2}\)
\nt
\pi\b_{v,a}&=\pi\ep g_{v,a}-g_{v,a}\Bigg[
\(2g_{v,a}-\sum_bg_{v,b}\)+\(2g_{t,a}-\sum_bg_{t,b}\)+2g_{v,a}\(2g_{\A,a}-\sum_bg_{\A,b}\)+g_s\Bigg]
\nt&\quad
-2g_sg_{\A,a}
-2\sum_{bc}\abs{\ep^{abc}}g_{v,b}g_{\A,c},
\nt
\pi\b_\E&=\pi\ep g_\E-{3}g_sg_\B,
\nt
\pi\b_\B&=\pi\ep g_\B-{3}g_sg_\E-g_\B\[g_s+\sum_a(g_{t,a}+2g_{\A,a}+g_{v,a})\]\cdot
}
where 
\eq{
c=\frac{64}{3\pi N}. 
}
In what follows we will work in 2 spatial dimensions and set $\ep=0$.

\subsection{SU($2N$) flavour symmetry}\label{sec:SU4disorder}
Since disorder coupling to the U(1) gauge currents is irrelevant, the only finite couplings which preserve the SU($2N$) flavour symmetry of QED$_3$ are $g_s$, $g_\E$, and $g_\B$. With $g_{t,a}=g_{t,\A}=g_{v,a}=0$, the only non-trivial $\b$-functions are
\eq{
\pi\b_s&=g_s^2+2cg_s,
&
\pi\b_\E&=-{3}g_sg_\B,
&
\pi\b_\B&=-g_s\({3}g_\E+g_\B\)\cdot
}
$\b_s$ is entirely determined by the fermion self-energy diagrams in Figs.~\ref{fig:fermSelfE-PhotonInt} and \ref{fig:fermSelfE-SigmaInt} and the 4-point diagrams in Fig.~\ref{fig:sigmasigma}. 
Figs.~\ref{fig:4ptSigmaSigmaNoX} and \ref{fig:4ptSigmaSigmaX} cancel, and Fig.~\ref{fig:4ptSigmaSigmaVert} contributes the second term in $\b_s$. 
This is precisely the same term found in Ref.~\onlinecite{Ludwig1994} for free Dirac fermions. 
The second term in $\b_s$ results from interactions with the photon. 
In fact, this is simply the anomalous dimension of $N_s(r)={1\o\sqrt{2N}}\bpsi\psi(r)$ in pure QED$_3$ (Eq.~\eqref{eqn:massDim}).
Since $g_s>0$, both terms in $\b_s$ are positive, and, as the energy scale is taken to zero, $g_s$ flows to zero.

On inspecting the $\b$-functions for the topological disorder strengths, we note an apparent inconsistency with our claim that $J^\m_\mathrm{top}$ is a conserved current. 
In particular, as indicated near the beginning of Sec.~\ref{sec:disType}, the scaling dimensions of the spatial and time components of a conserved current are non-perturbatively protected to be $1+z$ and 2 respectively, and this should be reflected in their $\b$-functions.
However, this is not the case in the expression above for either $J^j_\mathrm{top}$ or $J^0_\mathrm{top}$ when $g_s\neq0$. 
Fortunately, this result makes sense in the context of the parity anomaly: when a single species of Dirac fermions is coupled to a mass, a Chern-Simons term at level 1/2 is generated $\sim{1\o2}\ep^{\m\n\r}A_\m\ptl_\n A_\r/4\pi$.
In the disordered system, this manifests itself through the induced coupling of the two topological currents.

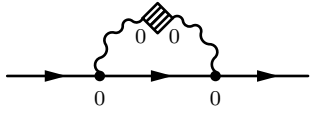
\begin{figure}
\centering
\input{gBSelfEnergy.tex}
\caption{The only disorder diagram to contribute to $\O(g_\B/2N)$ when $\B(x)$ is the only random field coupled to QED$_3$. Note that it is subleading to the self-energy diagrams we consider elsewhere in the paper (Fig.~\ref{fig:fermionSelfE}). It contributes a divergence $-ip_0\g^0\(g_\B\o 2\pi(2N)\ep\)$.}
\label{fig:gBSelfEnergy}
\end{figure}

Regardless, both of the $\b$-functions for the topological disorder are directly proportional to the SU($2N$)-symmetric mass coupling and so vanish when $g_s=0$.
However, we argue that higher order effects ultimately destabilize the clean critical point in the absence of time reversal symmetry.
To start, we observe that 
the Dirac equation has an additional discrete, anti-unitary symmetry under which both time and charge flip, leading us to refer to it as ``$\mathcal{CT}$" symmetry.
$J^{0}_\mathrm{top}$ is even under the action of $\mathcal{CT}$, while both $\bpsi\psi$ and  $J^{j}_\mathrm{top}$ are odd. 
Imposing this symmetry sets $g_s=g_\E=0$ and allows only $g_\B$ to be finite. 
The lowest order diagram which contributes is the fermion self-energy shown in Fig.~\ref{fig:gBSelfEnergy}.
Like the diagrams in Fig.~\ref{fig:fermionSelfE}, its divergence is cancelled by $Z_1$, yielding a dynamic critical exponent greater than unity: 
\eq{\label{eqn:zCTsym}
z=1+{g_\B\o2\pi(2N)}\cdot
}
Even though time-reversal is broken, the $\mathcal{CT}$ symmetry ensures that no diagrams mixing $g_\B$ and $g_\E$ are generated.
We conclude that since flux is still conserved, the only contribution to the $\b$-function of $g_\B$ arise from the corrections to the dynamic critical exponent given in Eq.~\eqref{eqn:zCTsym}.
In Sec.~\ref{sec:disType}, we showed that the dimension of disorder coupling to $J^0_\mathrm{top}$ is
 \eq{
[g_\B]=2(1+z)-2\[J_\mathrm{top}^0\]=2(z-1),
}
and, therefore, the $\b$-function is
\eq{
\pi\b_\B=-{g_\B^2\o2N}\cdot
}
It follows that this theory flows to strong coupling, albeit at a higher order  in $g_\xi$ and $1/(2N)$ than what is considered in the rest of the paper: $\O(g_\B^2/2N)\sim\O(1/(2N)^3)$ instead of $\O(1/(2N)^2)$.

This continues to be true even upon breaking $\mathcal{CT}$ and allowing finite $g_\E$ and $g_s$.
The $g_s$ disorder strength will flow to zero and need not be considered further. 
Then, the irrelevance of monopoles ensures that $g_\E$ remains marginal and that $g_\B$ flows to strong coupling (we note $g_\E$ will give an additional contribution to $z$ and, consequently, $\b_\B$).
In summary, the clean theory is unstable to SU($2N$) symmetric disorder when time reversal is broken.

Finally, when both the SU($2N$) flavour symmetry and time reversal are imposed, only disorder coupling to $J^j_\mathrm{top}$ is allowed, and the theory is exactly marginal to all orders in perturbation theory.

\subsection{SU($2$)$\times$SU($N$) flavour symmetry}\label{sec:nonAbelianFlow}
If we instead allow disorder to break the symmetry from $\mathrm{SU}(2N)\rightarrow \mathrm{SU}(2)\times \mathrm{SU}(N)$, no non-trivial fixed point is found; the system flows to strong disorder, and out of the perturbative regime.
Setting $g_{t,a}=g_t$, $g_{v,a}=g_v$, and $g_{\A,a}=g_\A$, the resulting set of $\b$-functions is
\eq{
\pi\b_s&=g_s\[g_s+3g_t+3g_v-6g_\A+2c\]-6g_v g_\A,
\nt
\pi\b_t&=g_t\[-g_t+g_s-9g_v+6g_\A-c\],
\nt
\pi\b_\A&=-4g_\A^2-g_t^2-g_v^2-g_sg_v,
\nt
\pi\b_v&=g_v\[g_v-g_s+g_t-2g_\A\]-2g_sg_\A,
\nt
\pi\b_\E&=-{3}g_sg_\B,
\nt
\pi\b_\B&=-{3}g_sg_\E-g_\B\[g_s+3(g_t+2g_\A+g_v)\]\cdot
}
The third equation indicates that if either $g_t$, $g_\A$, or $g_v$ is non-zero, $g_\A$ always flows to strong coupling. 
The four negative terms in $\b_{\A}$ can be traced to the diagrams in the first, fifth, and seventh rows of Table~\ref{tab:DiagramTab1}, and the second row of Table~\ref{tab:DiagramTab2} (shown in Appendix~\ref{app:4ptDiagramsModel}). 
In these diagrams, the anticommutation properties of the Pauli matrices ensure that the ``box" and ``crossing" diagrams do not cancel as they did for the singlet mass term (Figs.~\ref{fig:4ptSigmaSigmaNoX} and \ref{fig:4ptSigmaSigmaX}).
In fact, it is shown in Appendix~\ref{app:NonAbelianSymm} that disorder symmetric under any continuous non-abelian subgroup $\mathcal{H}$ of SU($2N$) will have this property and, consequently, flow to strong coupling.

This may appear to contradict the argument of the previous section: since $g_\A$ couples disorder to the spatial components of a conserved current, in the absence of a random mass $M_s(x)$, should it not be exactly marginal like $g_\E$?
The key difference is that because SU($2$) is non-ablelian, the SU($2$)$\times$SU($N$) flavour symmetry is only present on average.
The action for a specific realization of disorder, $\A_j^a(x)$, only has a SU($N$) flavour symmetry, and, as a result, the scaling dimension of $i\bpsi\g^j\s^a\psi$ is not protected. 

Similarly, if $g_\B$ is non-zero and any of the other four fermion bilinears couplings are non-zero, disorder coupling to $J^0_\mathrm{top}$ also becomes strong.
Again, this is because the dynamical critical exponent is greater than 1 when $g_s$,  $g_t$, $g_\A$, or $g_v$  are non-zero. 
We recall that the dimensional analysis of Sec.~\ref{sec:disType} indicated that when $z\neq1$, the critical scaling dimension is no longer 2, but instead $1+z$. Therefore, $[J^0_\mathrm{top}]=2<1+z$, making it a relevant perturbation.

\subsection{U($1$)$\times$SU($N$) symmetry}\label{sec:FixedPt}
We turn, finally, to the case of greatest interest in the present paper.
When the disorder couples to a U(1) subgroup of SU$(2N)$, we find a fixed line with both finite disorder and interactions.

We begin by considering an XY anisotropy where $g_{\cdot,z}$ is allowed to differ from $g_{\cdot,x}=g_{\cdot,y}=g_{\cdot,\perp}$.
With this restriction, the $\b$-functions in Eq.~\eqref{eqn:GeneralBetaFuns} reduce to
\eq{
\pi\b_s&=g_s\[g_s+g_{t,z}+2g_{t,\perp}-2g_{\A,z}-4g_{\A,\perp}
+g_{v,z}+2g_{v,\perp}+2c\]
-2g_{v,z}g_{\A,z}-4g_{v,\perp}g_{\A,\perp},
\nt
\pi\b_{t,z}&=g_{t,z}\[g_{t,z}-2g_{t,\perp}+6g_{\A,z}+4g_{\A,\perp}+g_s
-7g_{v,z}-2g_{v,\perp}
-c\]-4g_{t,\perp}g_{\A,\perp},
\nt
\pi\b_{t,\perp}&=g_{t,\perp}\[-g_{t,z}+g_s+8g_{\A,\perp}
-g_{v,z}-8g_{v,\perp}
-c\]-2g_{t,z}g_{\A,\perp},
\nt
\pi\b_{\A,z}&=-4g_{\A,\perp}^2-g_{t,\perp}^2-g_{v,\perp}^2-g_sg_{v,z},
\nt
\pi\b_{\A,\perp}&=-4g_{\A,\perp}g_{\A,z}-g_{t,z}g_{t,\perp}-g_{v,z}g_{v,\perp}-g_sg_{v,\perp},
\nt
\pi\b_{v,z}&=g_{v,z}\[-g_{v,z}+2g_{v,\perp}-g_{t,z}+2g_{t,\perp}-2g_{\A,z}+4g_{\A,\perp}-g_s\],
-2g_sg_{\A,z}-4g_{v,\perp}g_{\A,\perp}
\nt
\pi\b_{v,\perp}&=g_{v\perp}\[g_{v,z}+g_{t,z}-g_s\]-2g_sg_{\A,\perp}-2g_{v,z}g_{\A,\perp},
\nt
\pi\b_\E&=-{3}g_sg_\B,
\nt
\pi\b_\B&=-{3}g_sg_\E-g_\B\[g_s+g_{t,z}+2g_{\A,z}+g_{v,z}+2(g_{t,\perp}+2g_{\A,\perp}+g_{v,\perp})\]\cdot
}
These results are consistent with the RG equations obtained in Ref.~\onlinecite{Foster2008}. 
In this paper, the authors considered Dirac cones interacting through a $3d$ Coulomb term instead of a strictly 2+1 dimension gauge field; we can compare to their results by setting the Coulomb coupling in their equations to zero and $\mathpzc{g}^2=g_s=g_{v,z}=g_{v,\perp}=g_\E=g_\B=0$ in Eq.~\eqref{eqn:counterTerms}. 

\begin{figure}
\centering
\begin{subfigure}{0.48\textwidth}
\includegraphics[scale=0.4]{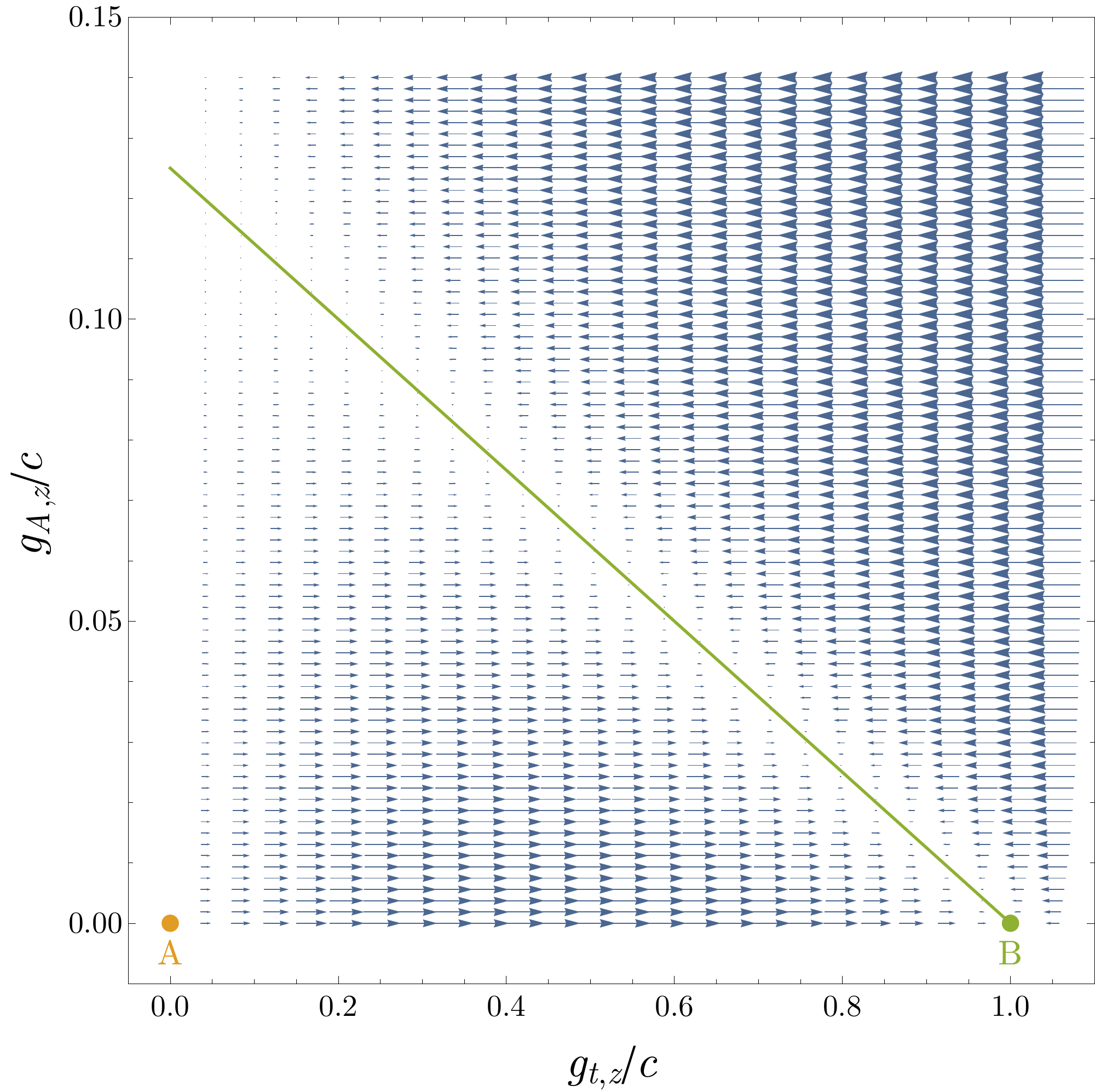}
\caption{}
\label{fig:RGflowTA}
\end{subfigure}
\begin{subfigure}{0.48\textwidth}
\includegraphics[scale=0.4]{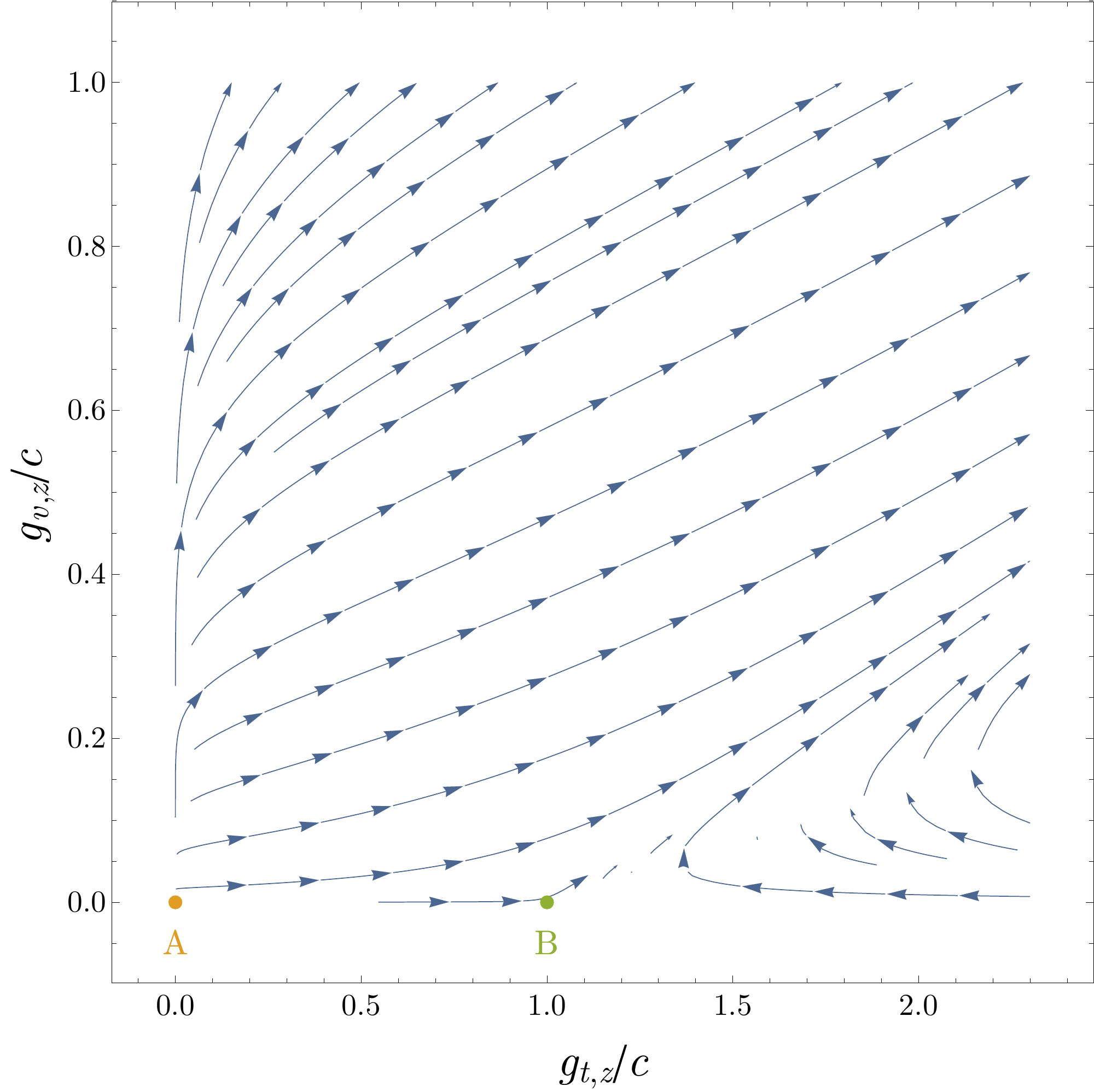}
\caption{}
\label{fig:RGflowTV}
\end{subfigure}
\\\vspace{4mm}
\begin{subfigure}{0.48\textwidth}
\includegraphics[scale=0.4]{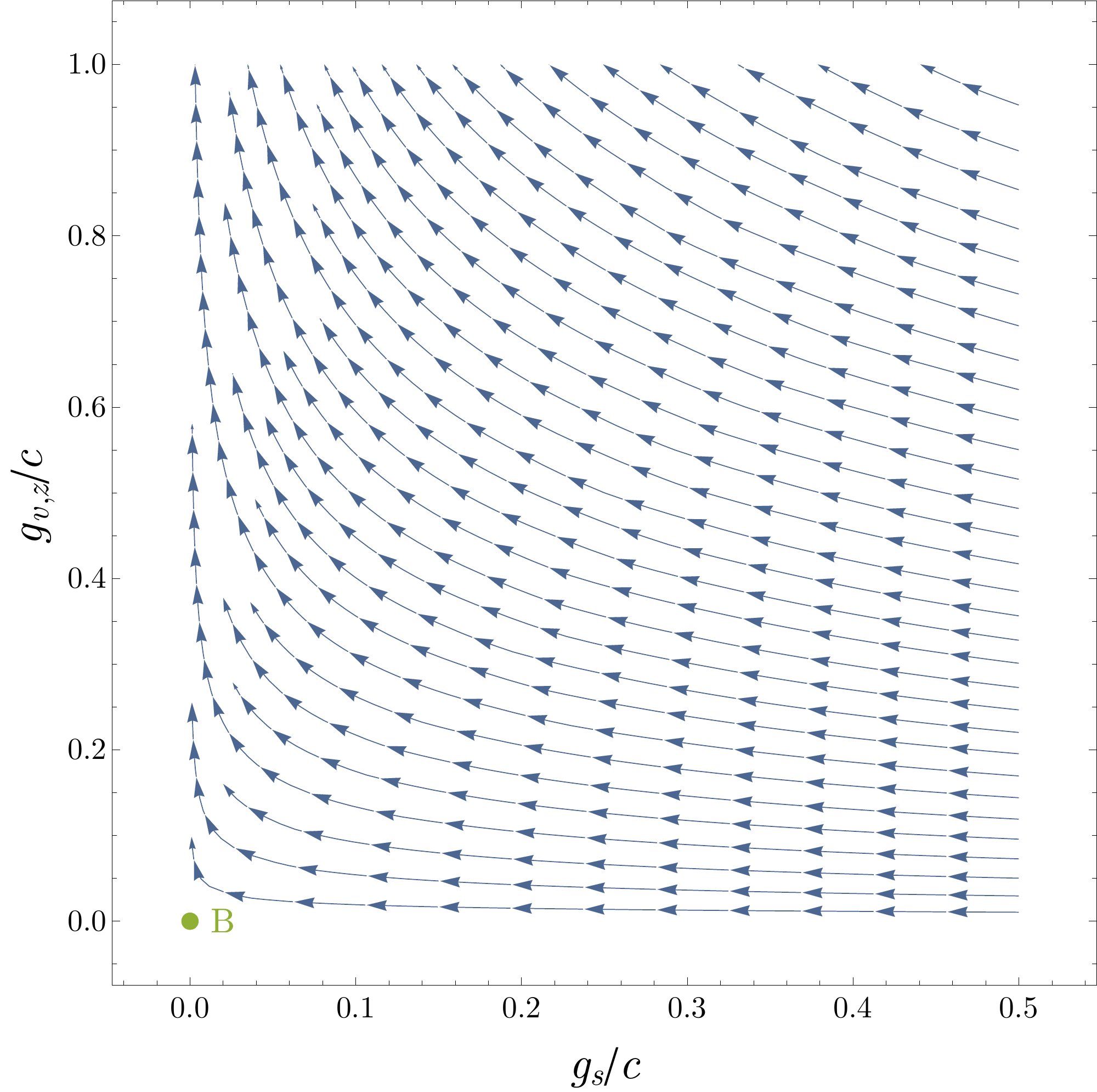}
\caption{}
\label{fig:RGflowSV}
\end{subfigure}
\begin{subfigure}{0.48\textwidth}
\includegraphics[scale=0.4]{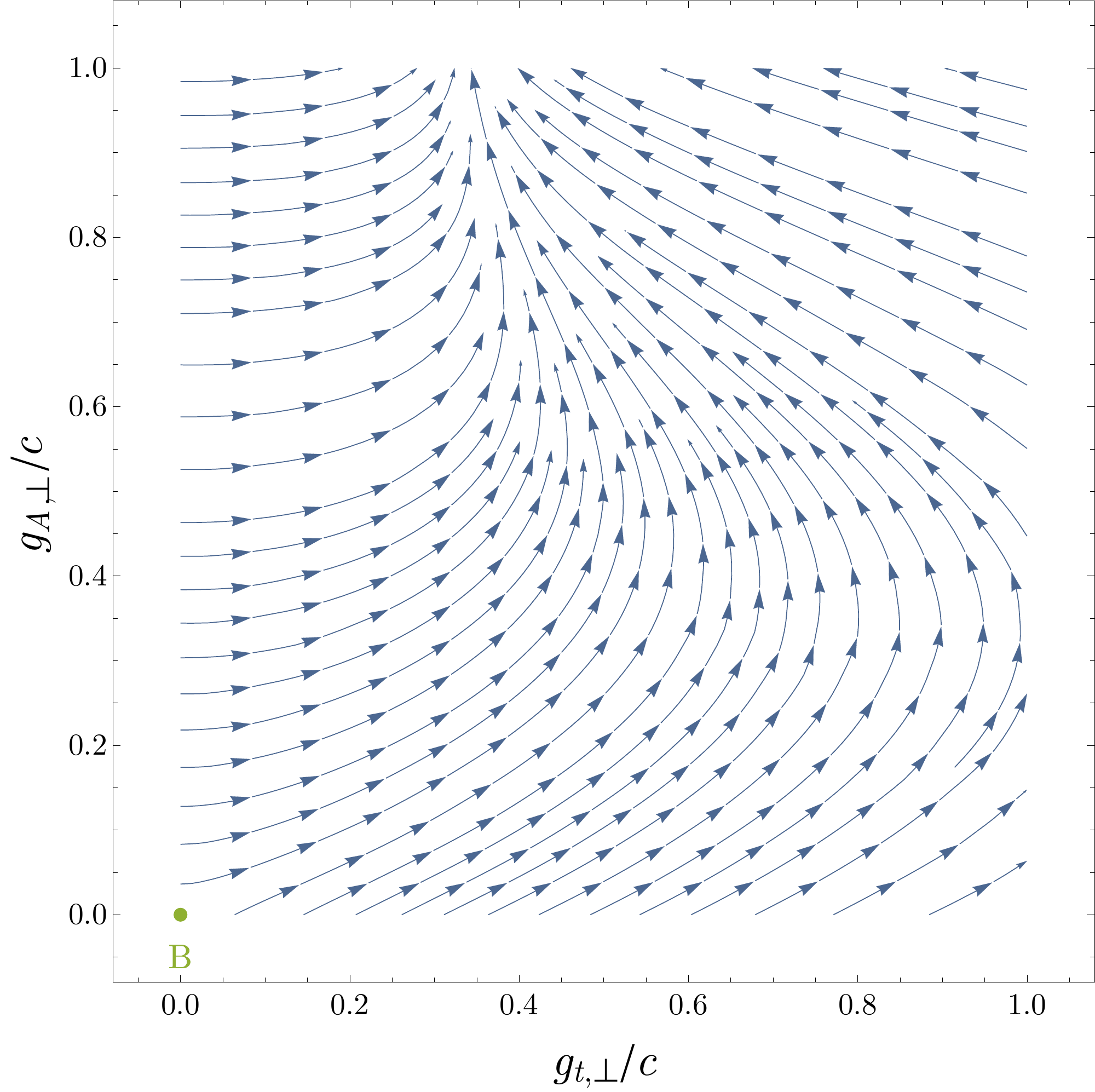}
\caption{}
\label{fig:RGflowTpAp}
\end{subfigure}
\caption{RG flow in the (a) $(g_{t,z},g_{\A,z})$ plane, (b) $(g_{t,z},g_{v,z})$ plane, and (c) $(g_{s},g_{v,z})$ plane with all other couplings set to zero. (d) shows the $(g_{t,\perp},g_{\A,\perp})$ plane with $g_{t,z}=c$ and all other couplings vanishing. The critical point with all couplings equal to zero (no disorder) is marked in orange with ``A" and the critical point with $g_{t,z}=c$ is marked in green with a ``B". In (a), the critical line is drawn in green. Here $c=128/3\pi(2N)$. }
\end{figure}
As in the previous section, the $\b$-functions for the vector potential couplings, $g_{\A,z}$ and $g_{\A,\perp}$ are all negative.
In order to ensure that they do not flow to infinity, all perpendicular couplings must vanish, $g_{\A,\perp}=g_{t,\perp}=g_{v,\perp}=0$.
This describes a situation where the U($1$)$\times$SU($N$) symmetry of the underlying theory is preserved even in the
presence of disorder. The $\b$-functions in the presence of this symmetry are
\eq{
\pi\b_s&=g_s\(g_s+g_{t,z}-2g_{\A,z}+3g_{v,z}+2c\)-2g_{\A,z}g_{v,z},
\nt
\pi\b_{t,z}&=g_{t,z}\(g_{t,z}+g_s+8g_{\A,z}-7g_{v,z}-c\),
\nt
\pi\b_{\A,z}&=-g_sg_{v,z},
\nt
\pi\b_{v,z}&=-g_{v,z}\(g_{v,z}+g_s+g_{t,z}+2g_{\A,z}\)-2g_sg_{\A,z},
\nt
\pi\b_\E&=-{3}g_sg_\B,
\nt
\pi\b_\B&=-{3}g_sg_\E-g_\B\(g_s+g_{t,z}+2g_{\A,z}+g_{v,z}\)\cdot
}
Recalling that all couplings are positive, we find a single physical solution which breaks the SU$(2N)$ flavour symmetry to U(1)$\times$SU$(N)$.
It is parametrized by the line
\eq{
g_{t,z}&=c-8g_{\A,z},
&
g_{\A,z}\leq{c\o8}\,,
}
with $g_\B$ and all other bilinear couplings equal to zero. 
Moreover, since $g_s$, $g_{v,z}$, and $g_\B$ are absent, each realization of disorder is invariant under time reversal and, consequently, $g_\E$ is exactly marginal (see Sec.~\ref{sec:SU4disorder}).
The fixed line we discuss is more correctly a fixed plane (though we will frequently refer to it only as a line). 
Referring to Eq.~\eqref{eqn:DynExp2}, the dynamical critical exponent on this surface is
\eq{
z&=1+c-6g_{\A,z}\cdot
}

In the presence of both time reversal and the U($1$)$\times$SU$(2N)$ flavor symmetry, $g_{\cdot,\perp}=0$, the critical surface has one irrelevant and two marginal directions.
It is stable to small variations in $g_{t,z}$ while perturbations in $g_\E$ and $g_{\A,z}$ are marginal. 
As we saw in the previous two sections, these couplings are associated with the spatial components of a conserved current, implying that their scaling dimensions are non-perturbatively fixed at exactly two when time reversal symmetry is present. 
The presence of these symmetries means that we do not expect the stability of the critical surface to change with the inclusion of higher order diagrams provided $N$ is sufficiently large.
However, it is possible that that it will be reduced to a single critical point.
The RG flow in the $(g_{t,z},g_{\A,z})$ plane is shown in Fig.~\ref{fig:RGflowTA}.

When time reversal only holds on average, $g_s$, $g_{v,z}$ and $g_\B$ are allowed to be finite as well. 
Disorder coupling to the SU($2N$)-symmetric mass term remains irrelevant, but the scalar potential-like disorder, $g_{v,z}$ and $g_\B$, take the theory into the strong coupling regime, as expected when the $z>1$.
The RG flows in the $(g_{t,z},g_{v,z})$ and $(g_s,g_{v,z})$ planes are shown in Figs.~\ref{fig:RGflowTV} and \ref{fig:RGflowSV}.

The fixed surface is not stable to perturbations which explicitly break the U($1$)$\times$SU($2N$) flavour symmetry of the replicated theory.
Fig.~\ref{fig:RGflowTpAp} shows the RG flow in the $\(g_{t,\perp},g_{\A,\perp}\)$ plane for $g_{t,z}=c$, $g_{\A,z}=0$ and indicates that both parameters are relevant. 
This is true along the entire critical surface.
Conversely, it can also be shown that along the critical line $g_{v,\perp}$ is irrelevant.

\section{Application to the kagome antiferromagnet}\label{sec:MicroOperators}

The large emergent symmetry of the QED$_3$ CFT implies that the currents and the fermion bilinears which we couple to disorder can be interpreted in a number of ways. 
Nonetheless, it is useful to directly relate our model to the microscopic operators of the spin-1/2 kagome Heisenberg antiferromagnet ($N=2$): $H_\mathrm{H}=J\sum_{\Braket{ij}} \v{S}_i\cdot\v{S}_j$, where $\Braket{ij}$ are nearest-neighbour sites on the kagome lattice (see Fig.~\ref{fig:bondOrientation}).  
Special attention will be given to the fixed line found in Sec.~\ref{sec:FixedPt}.
This section draws heavily from the discussion of Ref.~\cite{Hermele2008}, and more details can be found therein.

We begin by reviewing how the CFT is obtained as the low energy description of the kagome antiferrormagnet.
We start by expressing the spin operators in terms of fermions, $\v{S}_i={1\o2}f^\dag_{i\t}\v{\s}_{\t\t'}f_{i\t'}$, where $\v{\s}$ are the three Pauli matrices. 
This representation reproduces the Hilbert space of the spins provided it is accompanied by the local constraint $\sum_{\t=\uparrow,\downarrow} f^\dag_{i\t}f_{i\t}=1$.
The resulting Hamiltonian, $H_\mathrm{H}=-{J\o 4}\sum_{\Braket{ij}}f_{i\t}^\dag f_{j\t}f_{j\t}^\dag f_{i\t'}+\mathrm{const.}$, can be approximated by a mean field Hamiltonian $H_\mathrm{MF}=-\sum_{\Braket{ij}}t_{ij}f_{i\t}^\dag f_{j\t'}+H.c.$, where $t_{ij}$ is chosen so as to minimize the ground state energy while enforcing the condition $\sum_{\t=\uparrow,\downarrow}\Braket{f_{i\t}f_{i\t}}=1$ on average. 
The mean field ansatz which inserts $\pi$ and zero flux through the kagome hexagon and triangle plaquettes respectively is found to have a particularly low energy \cite{Hastings2000,Ran2007,Hermele2008}. 
In this case, the dispersion of $H_\mathrm{MF}$ has two Dirac cones per spin at a non-zero crystal momentum, $\pm\vQ$ \cite{Hastings2000,Hermele2008}.
The low energy excitations of $H_{\mathrm{MF}}$ are described by expanding about these two valleys, giving a free Dirac Lagrangian, $\mathcal{L}_\mathrm{D}=-\bpsi_\a\sd{\ptl}\psi_\a$, where $\a$ labels both spin and valley (the relation between the continuum Dirac spinors, $\psi_\a$, and the lattice fermions, $f_{i\t}$, is given in the appendix to Ref.~\onlinecite{Hermele2008}).
However, since the physical spin operators, $\v{S}_i$, are invariant under local phase rotations, $f_{i\t}\rightarrow e^{i\phi_i}f_{i\t}$, the fermions carry an emergent gauge charge, and,
consequently, the true effective theory of $H_\mathrm{H}$ must take gauge fluctuations into account.
Provided monopoles do not the confine the theory, the low energy description of the kagome antiferromagnet is QED$_3$ and not the free Dirac theory \cite{Hermele2004,Borokhov2002,Pufu14,Pufu15}.
We note that while $H_\mathrm{H}$ only had an SU(2) spin symmetry, QED$_3$ has an emergent SU(4) symmetry under which spin and valley indices are rotated into one another.

In order to calculate physical quantities, microscopic observables of the lattice theory must be associated with continuum operators of QED$_3$:
\eq{\label{eqn:MicroCFTcorr}
A_i\sim\sum_\ell c_\ell\O_\ell(\vr),
}
where $A_i$ is some function of local operators  near the lattice site $\vr$, and $\O_\ell(\vr)$ are a set of operators belonging to the CFT.
At long distances, the quantities to the left and right of Eq.~\eqref{eqn:MicroCFTcorr} must decay in the same manner. 
Given $A_i$, the set of operators $\O_\ell$ for which $c_\ell$ is non-vanishing could be determined by repeating the steps used to derive QED$_3$ from the Heisenberg model on the microscopic operators $\O_\ell$ \cite{Hermele2008}. 
However, it is easier to note that the $c_\ell$'s can be non-zero if and only if $A_i$ and $\O_\ell$ transform in the same manner under the action of the microscopic symmetries of the theory. 
In particular, the action under time reversal and space group transformations will be important. 
The symmetry operations relevant to the kagome antiferromagnet can be found in Ref.~\onlinecite{Hermele2008}.

As discussed in Sec.~\ref{sec:disType}, we only consider disorder coupling to the topological current and the fermion bilinears. 
That is, we restrict $\O_\ell$ to be either the conserved currents in Eqs.~\eqref{eqn:ConservedCurr} and~\eqref{eqn:topCurrentDef}, or the mass-like operators given in Eqs.~\eqref{eqn:MassOps} and~\eqref{eqn:MassOps2}.
By applying our large-$N$ results to the $N=2$ case, we may be neglecting important types of disorder in the form of monopole operators.

With this caveat in mind, we begin by identifying the singlet mass operator ${1\o\sqrt{2N}}\bpsi\psi$ with the chiral mass term discussed in Ref.~\onlinecite{Hastings2000}. 
Noting that ${1\o\sqrt{2N}}\bpsi\psi$ is odd under both parity and time reversal, it's not surprising that it can be associated with the scalar spin chirality, 
\eq{
{\mathcal{C}}_\mathrm{SSP}(\vx_\triangle)=\sum_{(ijk)\in\triangle}\v{S}_i\cdot\(\v{S}_j\times\v{S}_k\),
}
where $\vx_\triangle$ is the position of a triangle in the lattice, and  $(ijk)$ are ordered as indicated by the arrows in Fig.~\ref{fig:bondOrientation}.
Similarly, the flux disorder operator, $J^0_\mathrm{top}$, transforms in the same way as ${1\o\sqrt{2N}}\bpsi\psi$, indicating that it can also be associated with $\mathcal{C}_\mathrm{SSP}$.
We conclude that the random fields $M_s(x)$ and $\mathcal{B}(x)$ in Eq.~\eqref{eqn:DisorderAction} descend from disorder coupling to ${\mathcal{C}}_\mathrm{SSP}$.
The renormalization group study of Secs.~\ref{sec:SU4disorder}, \ref{sec:nonAbelianFlow}, and~\ref{sec:FixedPt} indicates that a randomly varying scalar spin chirality remains a marginal perturbation to leading order. 
However, this is not protected by any symmetry and, as discussed in Sec.~\ref{sec:SU4disorder}, higher order diagrams make it relevant.

\begin{figure}
\centering
\begin{subfigure}{0.48\textwidth}
\includegraphics[scale=0.4]{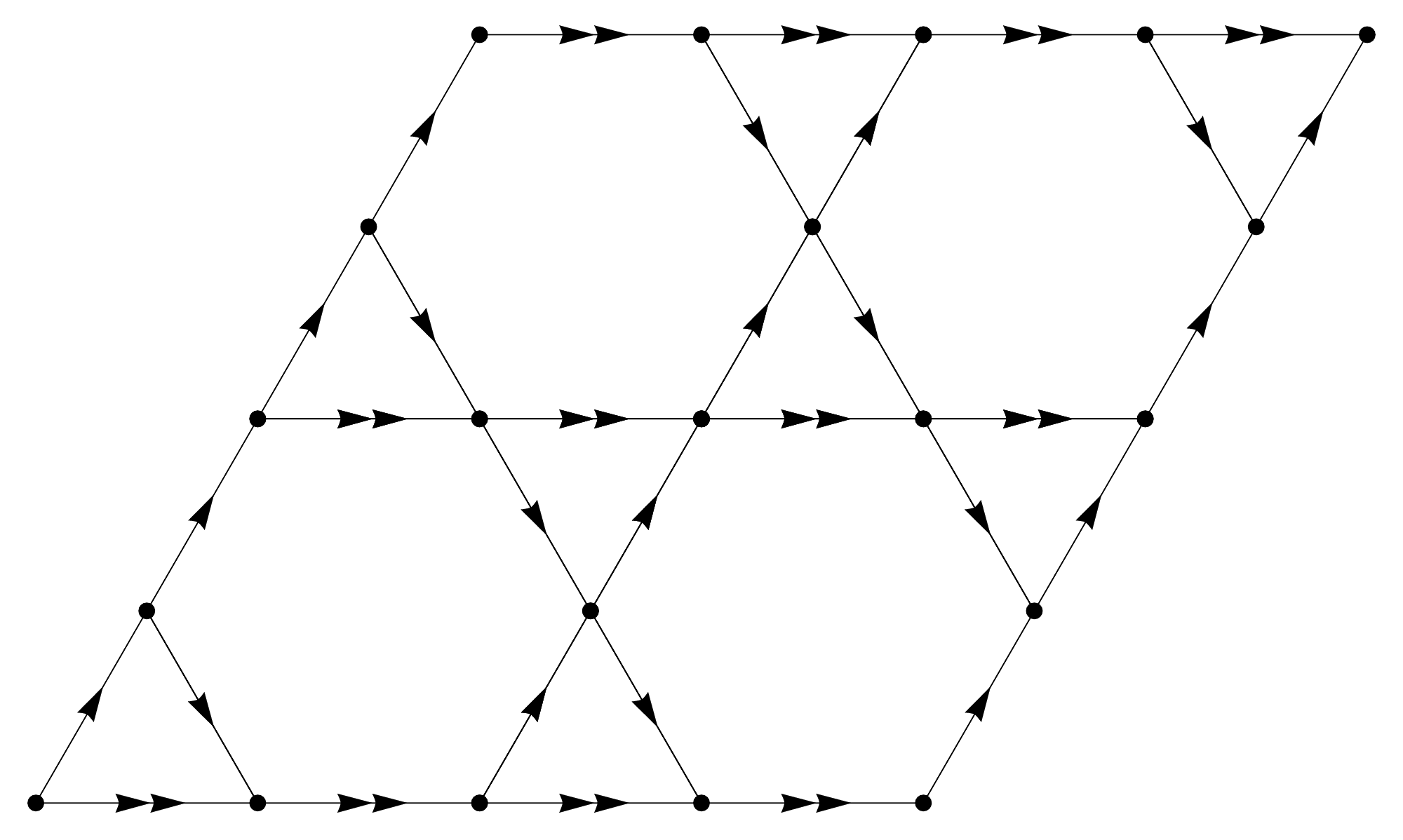}
\caption{Current in $x$-direction.}
\label{fig:CurrentJx}
\end{subfigure}
\begin{subfigure}{0.48\textwidth}
\includegraphics[scale=0.4]{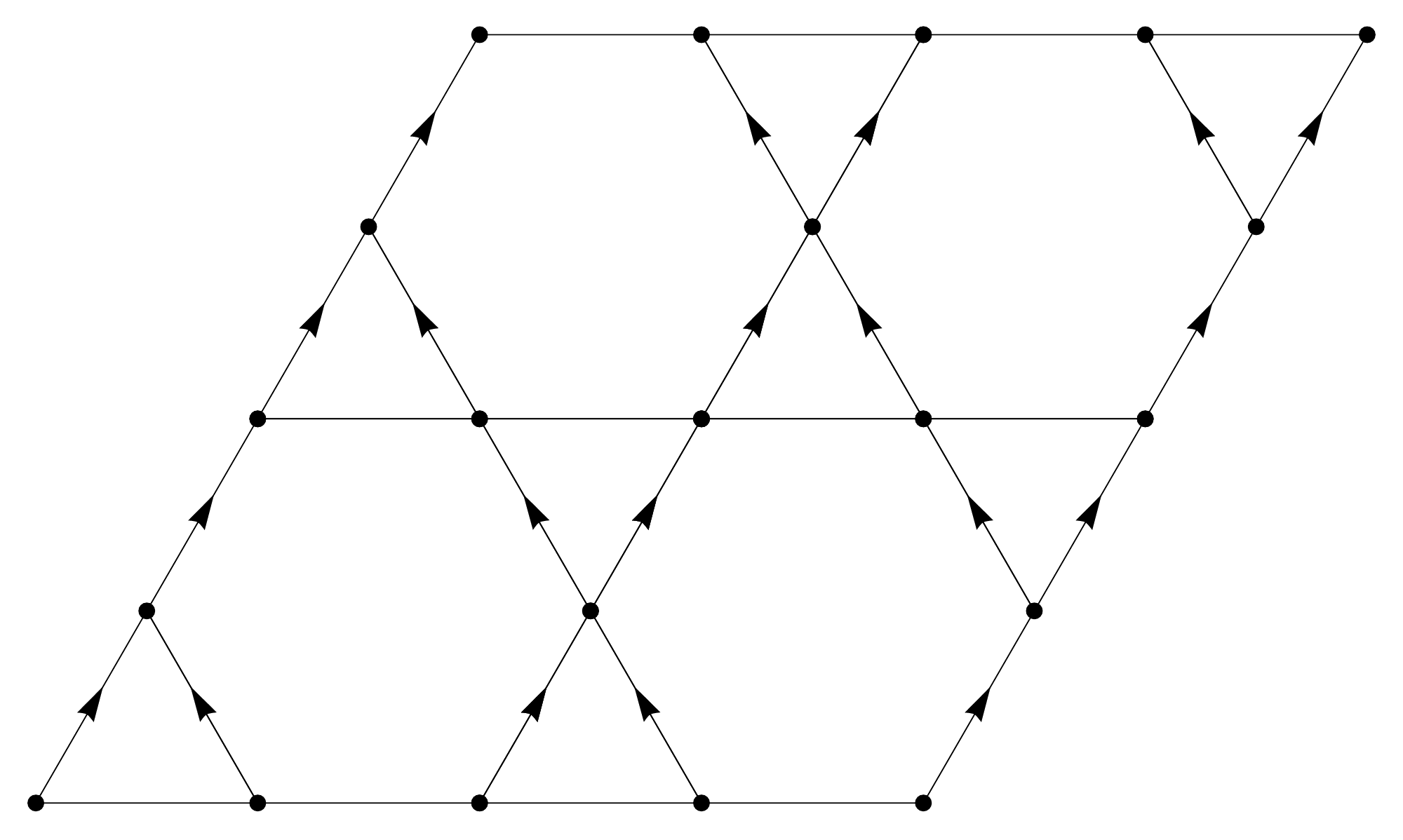}
\caption{Current in $y$-direction.}
\label{fig:CurrentJy}
\end{subfigure}
\caption{Bond ordering of bond order and vector chirality operators corresponding to the topological currents, $J^j_\mathrm{top}$, and the spin currents $J^{a,j}_B(r)$ in the $x$ and $y$ directions respectively. Our convention is that in $\v{C}_{ij}=\v{S}_i\times\v{S}_j$, the $i$th site points towards the $j$th. The double arrows in (a) identify the bonds which are weighted twice as strongly as others, while the absence of arrows on the horizonal bonds in (b) implies that they do not contribute at all. 
}
\label{fig:currentBonds}
\end{figure}

The spatial components of the topological current are time reversal invariant and transform as vectors under spatial rotations. 
The simplest operators invariant under time reversal are the bond operators, 
\eq{
{P}_{ij}&=\v{S}_i\cdot\v{S}_j,
}
where $i$ and $j$ are nearest-neighbours.
In order to find the simplest combination of ${P}_{ij}$'s which rotate in the correct fashion, we
calculate the irreducible representations governing the bond configurations within a unit cell. 
Defining
\eq{\label{eqn:bondordering}
{\mathcal{P}}_x(\vx)&=\sum_{ij\in \mathrm{hex}(\vx)}e^x_{ij}{P}_{ij},
&
{\mathcal{P}}_y(\vx)&=\sum_{ij\in \mathrm{hex}(\vx)}e^y_{ij}{P}_{ij},
\nt
\(e^x_{ij}\)^\mathrm{T}&={1\o2\sqrt{3}}\(2,1,-1,-2,-1,1\),
&
\(e^y_{ij}\)^\mathrm{T}&={1\o2}\(0,1,1,0,-1,-1\),
}
we identity ${J}^x_\mathrm{top}$ and  ${J}^y_\mathrm{top}$ with ${\mathcal{P}}_x$ and ${\mathcal{P}}_y$ respectively; these patterns are shown in Fig.~\ref{fig:currentBonds}.
This identification along with the results of Sec.~\ref{sec:SU4disorder} may then appear to indicate that random bond disorder, corresponding to a Hamiltonian of the form
\eq{
H_\mathrm{RB}&=\sum_\mathrm{ij}J_{ij}\v{S}_i\cdot\v{S}_j,
}
is an exactly marginal perturbation to the QED$_3$ fixed point when time-reversal is preserved. 
However, we will see shortly that this is not the case.

We next express the 15 generators of SU(4) as $\{ \s^a,\m^j,\s^a\m^j\}$
where $\s^a$ and $\m^j$ are commuting sets of Pauli matrices with $\s^a$ acting on spin and $\m^j$ acting on valley indices. Following the notation of Ref.~\onlinecite{Hermele2008}, 
it's useful to re-label the operators of Eqs.~\eqref{eqn:ConservedCurr} and~\eqref{eqn:MassOps} as
\eq{\label{eqn:MassOpsSU4}
J^{ia}_{A,\m}&=i\bpsi \m^i{\s^a}\g_\m\psi,
&
J^a_{B,\m}&=i\bpsi{\s^a}\g_\m\psi,
&
J_{C,\m}^i&=i\bpsi\m^i\g_\m\psi,
\nt
{N}^{ia}_A&=\bpsi \m^i{\s^a}\psi,
&
{N}^a_B&=\bpsi{\s^a}\psi,
&
N_C^i&=\bpsi\m^i\psi\;\cdot
}
Each of these operators can couple to a random field to contribute to an action of the form in Eq.~\eqref{eqn:DisorderAction}.

In Ref.~\onlinecite{Hermele2008}, the microscopic spin operators corresponding to each of the mass operators, $N_A^{ia}$, $N_B^a$, and $N_C^i$ are identified.
We will primarily be interested in $N_B^a$. This is a spin triplet and is even under time reversal. 
The simplest microscopic operator with this property is the vector chirality operator $\v{C}_{ij}=\v{S}_i\times\v{S}_{j}$, where $i$ and $j$ are nearest-neighbours. 
The linear combination of $\v{C}_{ij}$'s within a unit cell which transform in the same way as $\v{N}_B$ can be written
\eq{
\v{\mathcal{C}}_s(\vx)&=\sum_{(ij)\in \mathrm{hex}(\vx)}\v{C}_{ij},
}
where the sum is taken around the hexagon at $\vx$ following the convention in Fig.~\ref{fig:bondOrientation}. 
As we indicated in Sec.~\ref{sec:intro}, $\mathcal{\v{C}}_s$ is precisely the DM interaction term.

Similar reasoning suggests that the $B$-type currents, $\v{J}_{B,\m}(r)$, correspond to the spin operators and currents.
First, the space group symmetry acts on $\v{S}_i$ in the same way as it acts on $\v{J}_{B,0}$; in particular, both $\v{S}$ and $\v{J}_{B,0}$ are invariant under spatial rotations and odd under time reversal.
It's not surprising then that $\v{J}_{B,x}$ and $\v{J}_{B,y}$ correspond to spin currents. 
They are both even under time reversal and are spin triplets. 
As with $\v{N}_B$, this suggests a linear combination of nearest-neighbour vector chirality operators, $\v{C}_{ij}$, as their natural microscopic counterpart.
Like $J_\mathrm{top}^j$, they must transform as vectors under spatial rotations, implying that the $C_{ij}$'s should correspond to the $\v{J}_{B,j}$ in the same way the $P_{ij}$'s correspond to $J_\mathrm{top}^j$:
\eq{
\v{\mathcal{C}}_x(\vx)&=\sum_{ij\in \mathrm{hex}(\vx)}e^x_{ij}\v{C}_{ij},
&
\v{\mathcal{C}}_y(\vx)&=\sum_{ij\in \mathrm{hex}(\vx)}e^y_{ij}\v{C}_{ij},
}
where $e^x_{ij}$ and $e^y_{ij}$ are given in Eq.~\eqref{eqn:bondordering} and shown in Fig.~\ref{fig:currentBonds}.
In fact, since we assume that fermion bilinears and topological currents are the only relevant operators of the CFT, \emph{all} disorder coupling to the $\v{C}_{ij}$'s is taken into account by random fields coupling to $\v{N}_B$, $\v{J}_{B,x}$, and $\v{J}_{B,y}$.
In particular, modulo the caveats we have already discussed, the low energy theory of the kagome AF with weak disorder of the form
\eq{
H_{\mathrm{dis}}^\mathrm{DM}&=\sum_{\Braket{ij}}J^\mathrm{DM}_{ij}\v{\hat{z}}\cdot\v{S}_i\times\v{S}_j
}
where $J_{ij}^\mathrm{DM}$ are sufficiently weak random variables, should be described by fixed line of Sec.~\ref{sec:FixedPt}.
Unlike $\v{N}_B$, the remaining two mass bilinears in Eq.~\eqref{eqn:MassOpsSU4} carry valley indices.
The bilinear $\v{N}_A^i$ represents a set of three spin triplets and is odd under time reversal. 
Focusing on the $z$ component in spin space, ${N}_A^{i,z}$, three magnetic ordering patterns can be identified, each with a crystal momentum at a different $M$ point in the Brillouin zone.
Under rotations about the $z$-axis, the $N_A^{i,z}$'s transform into one another. 
Disorder resulting from magnetic defects could couple to bilinears of this form, but the fixed line resulting in Sec.~\ref{sec:FixedPt} is particularly unlikely to occur. Except in cases of extreme anisotropy, we do not expect disorder to exclusively couple to a single momentum channel.

Similarly considerations hold for $N^i_C$. 
These operators are spin singlets and, like $J^j_\mathrm{top}$, can be associated with bond ordering patterns $P_{ij}$ \cite{Hermele2008,Hastings2000}. 
In this case, two 3-dimensional irreducible representations of bonds transforming in the same way as $N^i_C$ are identified, and, again, each ordering pattern within an irreducible representation is distinguished by having a crystal momentum at one of the three $M$ points.
It follows that perturbing $H_\mathrm{H}$ by given a generic random bond Hamiltonian $H_\mathrm{RB}$ in the UV results in finite disorder strengths for $N^i_C$, $J^{i,a}_{C,\m}$, as well as $J_\mathrm{top}^i$. 
The appropriate form of disorder is not the the SU($2N$) symmetric case of Sec.~\ref{sec:SU4disorder}, but rather the situation discussed in Sec.~\ref{sec:nonAbelianFlow}. 
We therefore conclude that the kagome antiferromagnet is unstable to generic random bond disorder.

Finally, the same arguments hold for the microscopic analogues of $J^{i,a}_{A,\m}$ and $J^i_{C,\m}$. 


\section{Flavor conductivity}\label{sec:SpinConductivity}
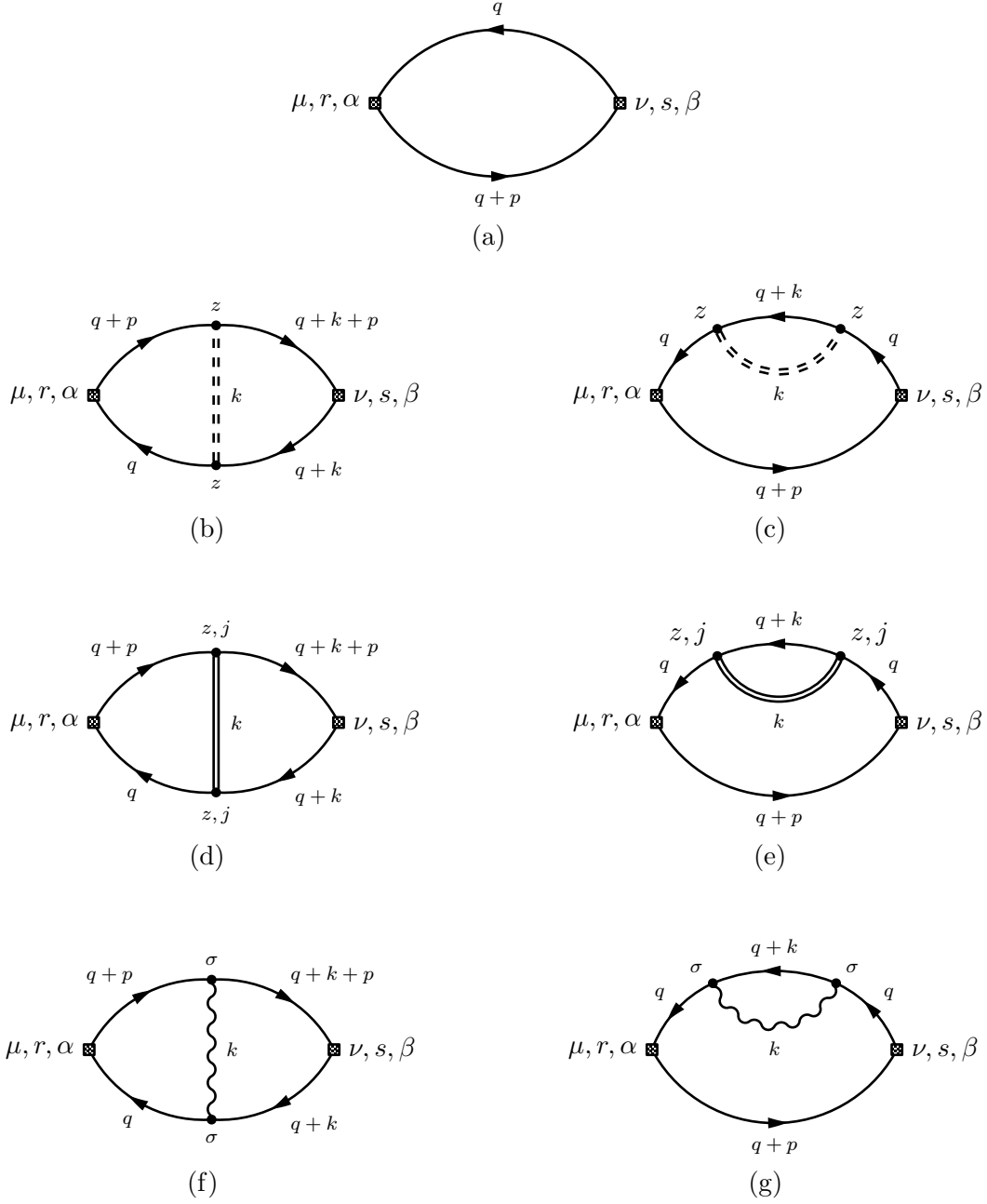
\begin{figure}
\input{currentDiagram.tex}
\caption{Diagrams which contribute to the current-current correlator.}
\label{fig:currentLoops}
\end{figure}
The flavor conductivity is a universal observable of the CFT; for the case of the kagome antiferromagnet, this conductivity 
is interpreted as a spin conductivity. 
By the usual arguments, we expect this conductivity to also be a universal observable
along the fixed line with U($1$)$\times$SU($N$) symmetry found in Sec.~\ref{sec:FixedPt}. Because of the presence of continuously
variable critical exponents along this line, we also anticipate the flavor conductivity to be continuously variable.

The flavor conductivity is determined by the two point correlators at zero external momentum of the following currents:
\eq{
J^x_{za}(p)&=i\bpsi\s^zT^a\g^x\psi(p),
&
J^x_{sa}(p)&=i\bpsi T^a\g^x\psi(p),
&
J^x_{\perp a}(p)&=i\bpsi \s^xT^a\g^x\psi(p)=i\bpsi\s^yT^a\g^x\psi(p).
} 
In particular, we calculate the \emph{optical} conductivity, valid for frequencies greater than the temperature $T$, allowing us to evaluate these correlators at zero temperature.
The diagrams which contribute to $\O(g_{t,z},g_{\A,z},1/2N)$ are shown in Fig.~\ref{fig:currentLoops}. 
To this order, a non-zero $g_\E$ will not contribute.

We recall from the discussion of Sec.~\ref{sec:disType} that the dimensions of the spatial currents $J^x_{za}(x,\t)$ and $J^x_{sa}(x,\t)$ are fixed at $1+z$ and, therefore, their correlators contain no divergences at zero external momentum. 
Moreover, an inspection of the diagrams in Appendix~\ref{app:4ptDiagramsModel} shows that the scaling dimensions of $J^x_{a\perp}$ remain unaltered to the order we are considering. 
Appendix~\ref{app:CurrentCurrentCorr} outlines how Figs.~\ref{fig:currentLoopBare} to~\ref{fig:currentLoopSelfE2} are calculated, and also verifies that counter term diagrams do not contribute.
The photon diagrams, Figs.~\ref{fig:photonVertex} and~\ref{fig:photonSelfE}, are determined in Ref.~\cite{Huh2015}.
Combining these results, we find
\eq{
\Braket{J^x_{za}(p_0)J^x_{zb}(-p_0)}
&=\Braket{J^x_{sa}(p_0)J^x_{sb}(-p_0)}
\nt
&=\d_{ab}\abs{p_0}\bigg\{
-{1\o16}-{a_\g\o2N}+ a_\mathrm{V} g_{t,z}+a_\Sigma(g_{t,z}+2g_{\A,z})\Bigg\}
\nt
&=\d_{ab}\abs{p_0}\bigg\{
-{1\o16}-{a_\g\o2N}+ c(a_\mathrm{V}+a_\Sigma)-(8a_\mathrm{V}+6a_\Sigma)g_{\A,z}\bigg\}
}
and
\eq{
\Braket{J^x_{\perp a}(p_0)J^x_{\perp b}(-p_0)}&=
\d_{ab}\abs{p_0}\bigg\{
-{1\o16}-{a_\g\o2N}- a_\mathrm{V} g_{t,z}+a_\Sigma(g_{t,z}+2g_{\A,z})\bigg\}
\nt
&=\d_{ab}\abs{p_0}\bigg\{
-{1\o16}-{a_\g\o2N}+c(-a_\mathrm{V}+a_\Sigma)+(8a_\mathrm{V}-6a_\Sigma)g_{\A,z}\bigg\}
}
where
$a_\mathrm{V}$, $a_\Sigma$, and $a_\g$ are derived from Figs.~\ref{fig:currentLoopVert} and \ref{fig:currentLoopSelfE}, Figs.~\ref{fig:currentLoopVert2} and \ref{fig:currentLoopSelfE2}, and Figs.~\ref{fig:photonVertex} and \ref{fig:photonSelfE} respectively.
The two disorder contributions are equal,
\eq{
a_\mathrm{dis}&=a_V=a_\Sigma={1\o96\pi},
}
and the photon contribution is \cite{Huh2015}
 \eq{
a_\g&=\(0.0370767-{5\o18\pi^2}\).
}
From the Kubo formula, it follows that the conductivities are
\eq{
\s_z(0)=\s_s(0)&={1\o16}+{a_\g\o2N}- 2a_\mathrm{dis}(c-7g_{\A,z}),
\nt
\s_\perp(0)&={1\o16}+{a_\g\o2N}-2a_\mathrm{dis}g_{\A,z}\cdot
}
In both flavor channels, disorder suppresses the conductivity and, except when $g_{\A,z}=c/8$, and $g_{t,z}=0$, the singlet and spin-$z$ channels are affected more strongly.
This is physically reasonable since we naturally expect transport in channels coupling directly to disorder to decrease the most.

\section{Conclusion}\label{sec:Conclusion}

This paper examined the influence of quenched disorder on the 2+1 dimensional CFT of $2N$ massless two-component Dirac
fermions coupled to a U(1) field. The existence of this CFT can be established for sufficiently large $N$ by the $1/N$ expansion,
and we combined the $1/N$ expansion with a weak disorder expansion. 

For generic disorder, our renormalization group analysis shows a flow to strong coupling, and so we were unable to determine
the fate of the theory. However, if we restrict the disorder to obey certain global symmetries, then we were able to obtain
controlled results. 

For disorder respecting time reversal and the full SU($2N$) flavor symmetry of the CFT, we found in Sec.~\ref{sec:SU4disorder} 
that all allowed disorder
perturbations were marginal to the order we considered.
Such a result does $\emph{not}$ apply to the CFT of $2N$ free Dirac fermions: in that case, disorder coupling to a randomly varying chemical potential leads to a flow to strong coupling \cite{PhysRevB.50.7526}. 
However, once disorder is allowed to break time reversal, we again find a runaway flow towards strong disorder, albeit at a higher order in perturbation theory. 

Our main results, in Sec.~\ref{sec:FixedPt}, 
concerned the case in which disorder respects time-reversal and $\text{U}(1)\times\text{SU}(N)$ symmetry.
In this case, to leading order in $1/N$, we found a non-trivial fixed line with both interactions and disorder.
This fixed line had continuously varying exponents, in particular a dynamic critical exponent $z>1$.
It also had a continuously
varying, but cutoff independent, flavor conductivity. 

We also discussed the possible relevance of our results to the spin-1/2 kagome lattice antiferromagnet.
In this case, the $\text{U}(1)\times\text{SU}(N)$ symmetric disorder corresponds to a randomly varying Dzyaloshinkii-Moriya field,
as we described in Secs.~\ref{sec:intro} and~\ref{sec:MicroOperators}.

\section*{Acknowledgements}

This research was supported by the NSF under Grant DMR-1360789 and the MURI grant W911NF-14-1-0003 from ARO. Research at Perimeter Institute is supported by the Government of Canada through Industry Canada and by the Province of Ontario through the Ministry of Research and Innovation. 
SS also acknowledges support from Cenovus Energy at Perimeter Institute. AT is supported by NSERC.

As our paper was being completed, we learnt of related work by Goswami, Goldman, and Raghu \cite{raghu2017}: we thank the authors useful discussions and
for sharing their draft with us.

In addition, after completing our paper, an additional study of QED$_3$ with disorder by Zhao, Wang, and Liu \cite{Zhao1} was brought to our attention.
\appendix
%
%
%

\section{General non-abelian subgroup of SU($2N$)}\label{app:NonAbelianSymm}
In this appendix, we briefly discuss the RG flow which results upon breaking the flavor symmetry from SU($2N$)$\rightarrow\mathcal{G}\times$SU$(2N)/\mathcal{G}$, where $\mathcal{G}$ is a continuous non-abelian subgroup of SU($2N$). 
The most general form the disorder could take is
\eq{\label{eqn:SubGpDisorderAction}
S^\mathcal{G}_\mathrm{dis}[\psi,\bpsi]&=\int d^dx\,d\t\bigg[
M^\mathcal{G}_a(x)\bpsi\T^a\psi(x,\t)
+i\mathcal{A}^\mathcal{G}_{j a}(x)\T^a\g^j\psi(x,\t)
+V^\mathcal{G}_a(x)\T^a\g^0\psi(x,\t)
\bigg]
}
where $\T^a$ are the generators of $\mathcal{G}$. Averaging over disorder, we assume
\eq{
\overline{M^\mathcal{G}_a(x)M_b^\mathcal{G}(x')}&={\m^{-\ep}\lam_{t}\o2}\d_{ab}\d^d(x-x'),
&
\overline{M^\mathcal{G}_a(x)\A_{jb}^\mathcal{G}(x')}&=0,
\nt
\overline{\A^\mathcal{G}_{ia}(x)\A_{jb}^\mathcal{G}(x')}&={\m^{-\ep}\lam_{\A}\o2}\d_{ab}\d_{ij}\d^d(x-x'),
&
\overline{M^\mathcal{G}_a(x)V_{b}^\mathcal{G}(x')}&=0,
\nt
\overline{V^\mathcal{G}_a(x)V_b^\mathcal{G}(x')}&={\m^{-\ep}\lam_{v}\o2}\d_{ab}\d^d(x-x'),
&
\overline{\A^\mathcal{G}_{ja}(x)V_{b}^\mathcal{G}(x')}&=0.
}
We can study this theory in the same way we did in Secs.~\ref{sec:disType} and \ref{sec:RenormAction}.
The Feynman rules will be analogous to those shown in Fig.~\ref{fig:FeynRulesAll}.

From the calculations in Appendix~\ref{app:4ptDiagramsCalc}, we see that only the diagrams in Figs.~\ref{fig:4ptNoX} and \ref{fig:4ptX}, and Figs.~\ref{fig:4ptGamGamNoX} and \ref{fig:4ptGamGamX}  contribute to the renormalization of $\lam_\A$.
In particular, letting $\Gamma_\A$ be the vertex function whose spinor indices are proportional to $i\g^j\otimes i\g_j$, we find
\eq{
\Gamma_a\,i\g^j\otimes i\g_j&=-{1\o 4\pi\ep}\(\lam_t^2+4\lam_\A^2+\lam_v^2\)\g^j\otimes\g_j\sum_{ab}\[\T^a\T^b\otimes\T^a\T^b-\T^a\T^b\otimes\T^b\T^a\]
\nt
&=+{1\o 8\pi\ep}\(\lam_t^2+4\lam_\A^2+\lam_v^2\)i\g^j\otimes i\g_j\sum_{ab}\[\T^a,\T^b\]\otimes\[\T^a,\T^b\]
\nt
&=-{1\o 8\pi\ep}\(\lam_t^2+4\lam_\A^2+\lam_v^2\)i\g^j\otimes i\g_j\sum_{a}\T^a\otimes \T^a
}
where we've used the fact that
\eq{
\sum_{ab}\[\T^a,\T^b\]\otimes\[\T^a,\T^b\]&=\sum_{abcd}if^{abc}\,if^{abd}\T^c\otimes\T^d
=-\sum_{cd}\d_{cd}\T^c\otimes\T^d=-\sum_a\T^a\otimes\T^a,
}
where $f^{abc}$ are the structure constants of the algebra. 
It follows that 
\eq{
\pi\b_\A&=-\(\lam_t^2+4\lam_\A^2+\lam_v^2\).
}

\section{Fermion self-energy}\label{app:fermSelfE}
In this section, we calculate the fermion self-energy diagrams given in Fig.~\ref{fig:fermionSelfE}.

\subsubsection*{Self-energy contribution from photon: Fig.~\ref{fig:fermSelfE-PhotonInt}}
\eq{
\text{Fig.~\ref{fig:fermSelfE-PhotonInt}}&=
{16\o 2N}\m^{-\ep}\mathpzc{g}^2\int {d^Dq\o(2\pi)^D} \,i\g^\m{i(p+q)_\a\g^\a\o(p+q)^2}i\g^\n{\d_{\m\n}\o\abs{q}}\cdot
}
Using the identity 
\eq{
{1\o AB^n}&=\int_0^1dx\,{n(1-x)^{n-1}\o\[A+x(B-A)\]^{n+1}}
}
and the fact that  $\g^\m\g^\a\g_\m=\(2\d^{\a\m}-\g^\a\g^\m\)\g_\m=-(D-2)\g^\a$, we write
\eq{
\text{Fig.~\ref{fig:fermSelfE-PhotonInt}}&={i(D-2)16\m^{-\ep}\mathpzc{g}^2\o 2N}\g^\m\int{d^Dq\o(2\pi)^D}
\int_0^1dx\,{1\o 2\sqrt{1-x}}{q_\m+(1-x)p_\m\o\[q^2+x(1-x)p^2\]^{3/2}}
\nt
&=-i\g^\m p_\m\({8\mathpzc{g}^2\o3\pi^2(2N)\ep}\)+\mathrm{finite}.
\label{eqn:QEDfermSelfE}
}

\subsubsection*{Self-energy contribution from singlet mass disorder: Fig.~\ref{fig:fermSelfE-SigmaInt}}
\eq{
\text{Fig.~\ref{fig:fermSelfE-SigmaInt}}&=g_{s}\int {d^Dq\o(2\pi)^D}\,2\pi\d(q_0){i(q+p)_\m\g^\m\o (q+p)^2}
\nt
&=ig_{s}\int {d^dq\o(2\pi)^d}\,{i\[(q+p)_i\g^i+p_0\g^0\]\o (q+p)^2+p_0^2}
=-ip_0\g^0\({g_{s}\o2\pi\ep}\)+\mathrm{finite}.
}
\subsubsection*{Self-energy contribution from SU(2) mass disorder: Fig.~\ref{fig:fermSelfE-PhiInt}}
The contribution from the SU(2) mass disorder is the same, since the Pauli matrices square to the identity:
\eq{
\text{Fig.~\ref{fig:fermSelfE-PhiInt}}&=g_{t,a}\s^a\s^a\int {d^Dq\o(2\pi)^D}\,2\pi\d(q_0){i(q+p)_\m \g^\m\o (q+p)^2}
=-ip_0\g^0\({g_{t,a}\o2\pi\ep}\)+\mathrm{finite}.
}

\subsubsection*{Self-energy contribution from scalar potential disorder: Fig.~\ref{fig:fermSelfE-ScalarPotInt}}
\eq{
\text{Fig.~\ref{fig:fermSelfE-ScalarPotInt}}
&=-g_{v,a}\int{d^Dq\o(2\pi)^D}2\pi\d(q_0)i\g^0{i(q+p)_\a\g^\a\o(q+p)^2}i\g^0
=-ip_0\g^0\({g_{v,a}\o2\pi\ep}\)+\mathrm{finite}.
}

\subsubsection*{Self-energy contribution from vector potential disorder: Fig.~\ref{fig:fermSelfE-VectorPotInt}}
\eq{
\text{Fig.~\ref{fig:fermSelfE-VectorPotInt}}
&=g_{\A,a}\int{d^Dq\o(2\pi)^D}2\pi\d(q_0)i\g^j{i(q+p)_\a\g^\a\o(q+p)^2}i\g_j
=-ip_0\g^0\(g_{\A,a}\o\pi\ep\)+\mathrm{finite}.
}
\section{Diagrams without flavor indices}\label{app:4ptDiagramsCalc}
\begin{figure}
\input{feynRulesNoFlavour}
\caption{Feynman rules for diagrams without flavor indices. $a,b,c,d$ on the graphs label the spinor indices, and $\ell$ and $m$ label the replica indices.
The vertex on the left describes mass-like disorders, such as $M_s(r)$ and $M_{t,a}(r)$, and the diagram on the right corresponds to the SU(2) scalar and vector potential disorder, $V_a(r)$, and $\A_{j,a}(r)$. }
\label{fig:feynRulesNoFlavour}
\end{figure}
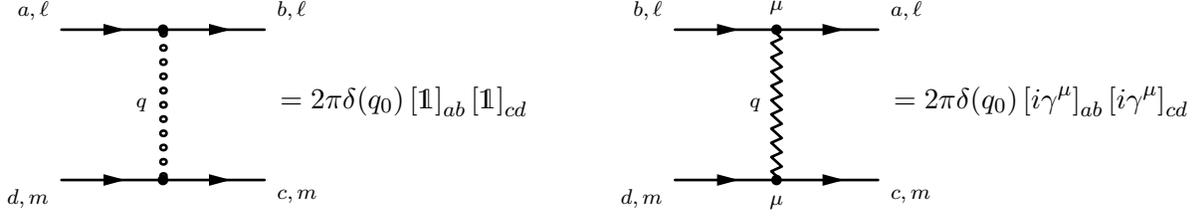 
Since the spinor and flavor structure of the interactions factor, it's convenient to first calculate the diagrams which correct the four-point interaction without reference to the fermion flavor indices. 
We denote these generalized vertices with the Feynman graphs shown in Fig.~\ref{fig:feynRulesNoFlavour}. The set of diagrams with only internal mass-like disorder  and photon lines is shown in Fig.~\ref{fig:massDisorder4pt}, while diagrams with only gauge-like disorder and photon lines are shown in Fig.~\ref{fig:gaugeDisorder4pt}. Finally, Fig.~\ref{fig:mixDisorder4pt} lists those diagrams which have contributions from both mass and gauge-like disorder. 
While there are many repetitions, all integrals have been included for completeness.
\begin{figure}
\input{integralDiagrams}
\caption{4-point diagrams with photon and mass-like disorder internal lines. Below each diagram, the divergent piece, if present, is given. (The factor of $2\pi\d(q_0)$ has been been suppressed for simplicity.)}
\label{fig:massDisorder4pt}
\end{figure}
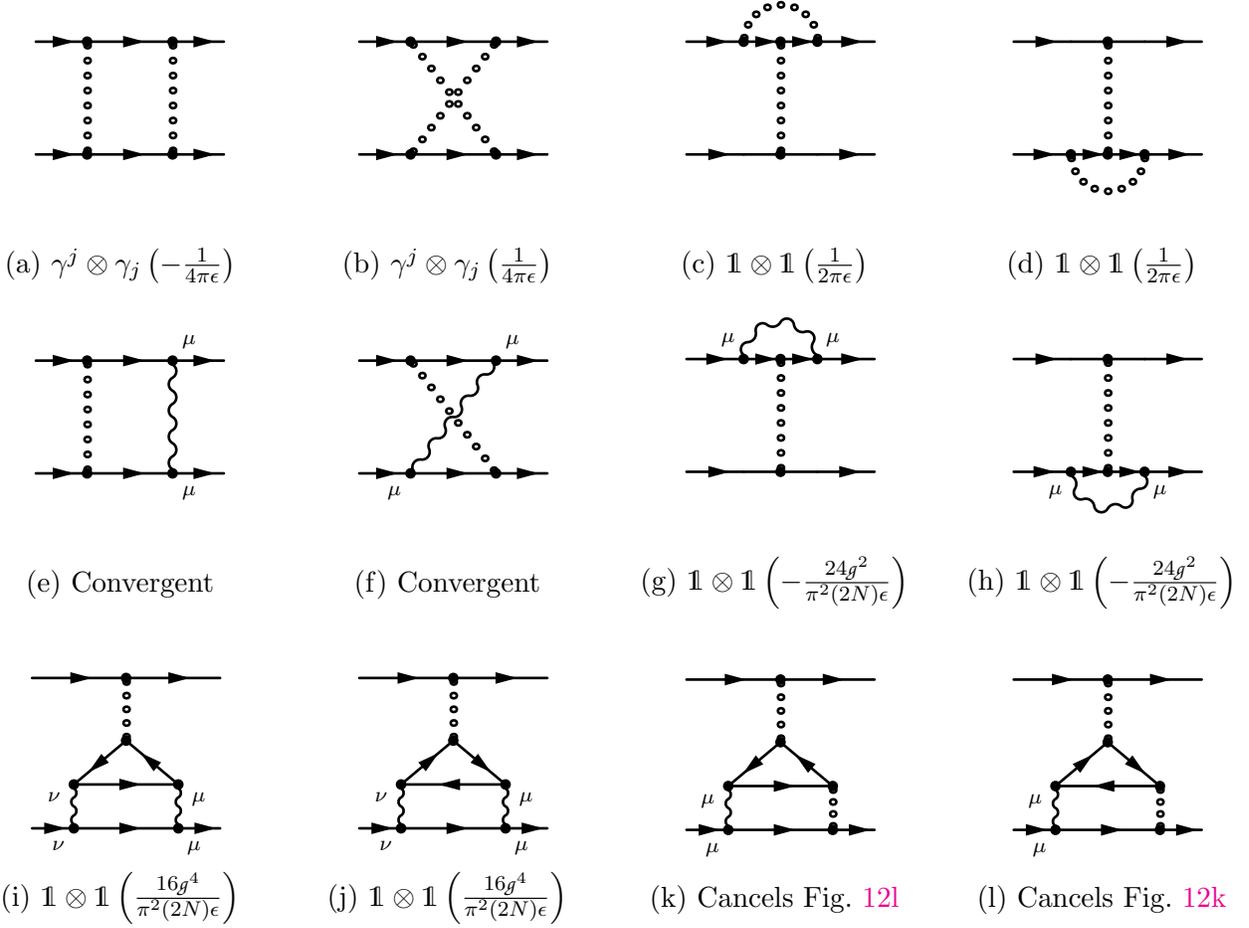
\begin{figure}
\input{integralDiagrams2}
\caption{4-point diagrams with photon and gauge-like disorder internal lines. Below each diagram, the divergent piece, if present, is given. (The factor of $2\pi\d(q_0)$ has been been suppressed for simplicity.)}
\label{fig:gaugeDisorder4pt}
\end{figure}
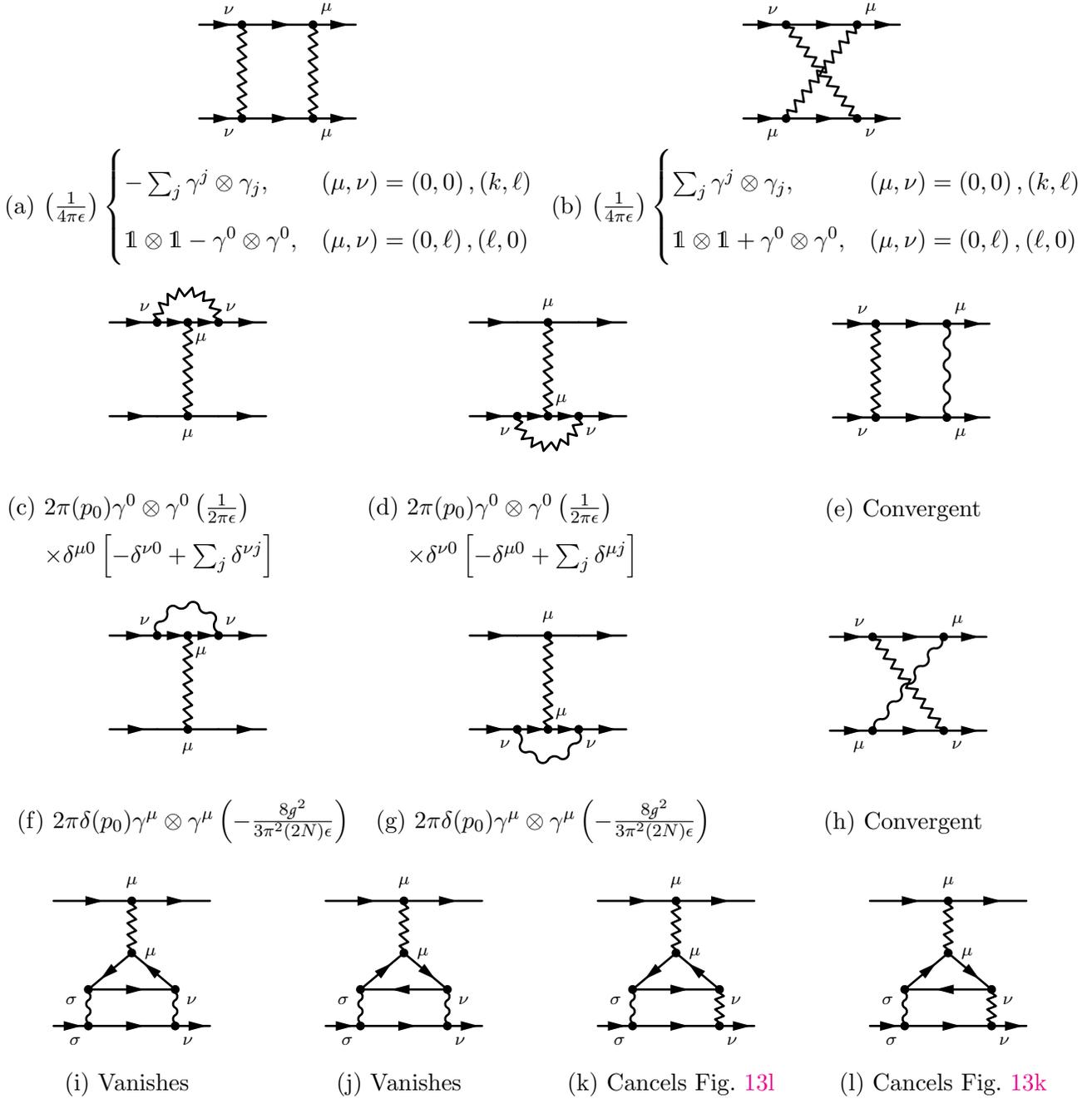 
\begin{figure}
\input{IntegralDiagramsMix.tex}
\caption{4-point diagrams with both mass-like and gauge-like disorder internal lines.
Below each diagram, the divergent piece, if present, is given. (The factor of $2\pi\d(q_0)$ has been been suppressed for simplicity.)
The $\tr\[\O_{fl}\]$ term in Figs.~\ref{fig:FermLoopMassVerPotentialInt1} and \ref{fig:FermLoopMassVerPotentialInt2} indicates that once the action on the flavour indices has been specified, a trace over this operator should be taken.
}
\label{fig:mixDisorder4pt} 
\end{figure}
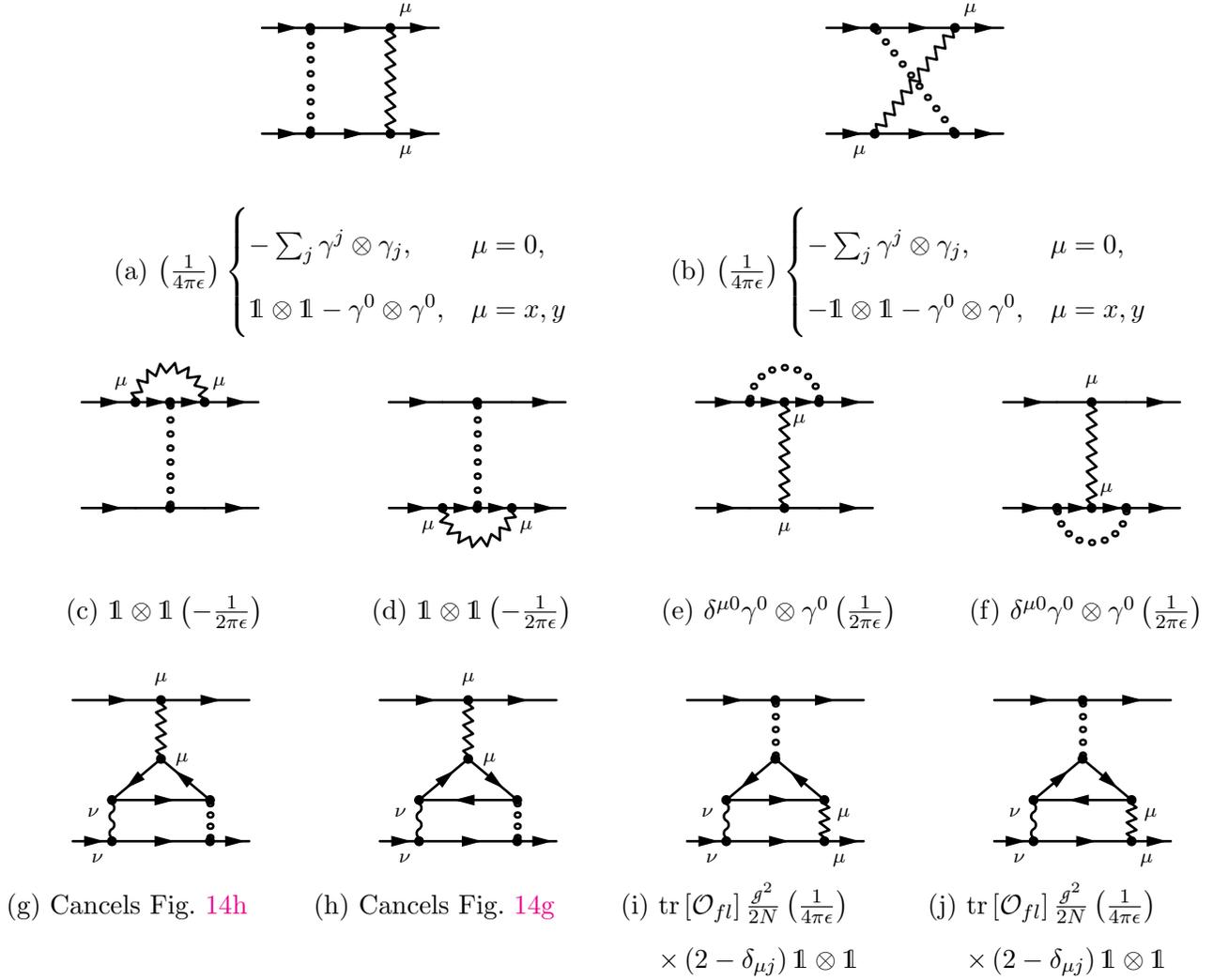
\subsection{Diagrams with mass-type disorder and photon lines:  Fig:~\ref{fig:massDisorder4pt}}
In this section, we evaluate the diagrams  with only internal mass disorder and photon lines. These are listed in Fig:~\ref{fig:massDisorder4pt}.
\subsubsection*{Two internal mass lines, no crossing: Fig.~\ref{fig:4ptNoX}}
\eq{
\text{Fig.~\ref{fig:4ptNoX}}&=\m^{-\ep}\int {d^Dq\o(2\pi)^D}2\pi\d(q_0)2\pi\d(q_0+p_0){i\[-q-p\)_\a\g^\a\o\(q+p\)^2}\otimes{i(q+p)_\b\g^\b\o(q+p)^2}
\nt
&=2\pi(p_0)\m^{-\ep}\int{d^dq\o(2\pi)^d}\int_0^1dx\,\g^j\otimes\g_j{q^2/2\o\[q^2+x(1-x)p^2\]^2}
\nt
&=2\pi(p_0){ \g^j\otimes\g_j}\(-{1\o4\pi\ep}\)+\mathrm{finite}.
}

\subsubsection*{Two internal mass lines, with crossing: Fig.~\ref{fig:4ptX}}
\eq{
\text{Fig.~\ref{fig:4ptX}}&=\m^{-\ep}\int {d^Dq\o(2\pi)^D}2\pi\d(q_0)2\pi\d(q_0+p_0){iq_\a\g^\a\o q^2}\otimes{i(q+p)_\b\g^\b\o(q+p)^2}
\nt
&=2\pi(p_0){\g^i\otimes\g_i}\({1 \o4\pi\ep}\)+\mathrm{finite}.
}

\subsubsection*{Vertex correction from disorder: Figs.~\ref{fig:4ptVert} and \ref{fig:4ptVert2}}
\eq{
\text{Fig.~\ref{fig:4ptVert}}&=\m^{-\ep}\int {d^Dq\o(2\pi)^D}2\pi\d(q_0)2\pi\d(p_0){iq_\a\g^\a\o q^2}{i\(q-p\)_\b\g^\b\o(q-p)^2}\otimes\id
\nt
&=2\pi(p_0)\,\m^{-\ep}\[-{\g^j\otimes\g_j\o2}\int{d^dq\o(2\pi)^d}\int_0^1dx\,{q^2\o\[q^2+x(1-x)p^2\]^2}\]\otimes\id+\mathrm{finite}
\nt
&=2\pi(p_0)\,{\id\otimes\id}\({1\o2\pi\ep}\)+\mathrm{finite}.
}
The other vertex gives the same correction:
\eq{
\text{Fig.~\ref{fig:4ptVert2}}&=2\pi(p_0)\,{\id\otimes\id}\({1\o2\pi\ep}\)+\mathrm{finite}.
}

\subsubsection*{One internal gauge-like disorder line and one photon line: Figs.~\ref{fig:4ptSigmaPhotonNoX} and \ref{fig:4ptSigmaPhotonX}}

The diagrams are both convergent.
We see this by writing
\eq{\label{eqn:DisPhotConverge}
\text{Fig.~\ref{fig:4ptSigmaPhotonNoX}}&=
{\m^{-\ep}\mathpzc{g}^2\o 2N}\int {d^Dk\o(2\pi)^D}\,2\pi\d(k_0)\,{1\o \abs{k}}i\g^\m{\(-ik_\a\g^\a\)\o k^2}\otimes i\g^\m{i(k+q)_\b\g^\b\o(k+q)^2}
\nt
&={\m^{-\ep}\mathpzc{g}^2\o 2N}\int {d^dk\o(2\pi)^d}\g^\m\g^\a\otimes \g_\m\g_\a\cdot{1\o \abs{k}(q_0^2+k^2)}
\nt
&=\mathrm{finite},
}
where we have assumed that $q=(q_0,0)$. The same reasoning shows that Fig.~\ref{fig:4ptSigmaPhotonX} is convergent as well.

\subsubsection*{Vertex correction from photon: Figs.~\ref{fig:4ptsPhotonArc} and~\ref{fig:4ptsPhotonArc2}}

\eq{
\text{Fig.~\ref{fig:4ptsPhotonArc}}&=
2\pi\d(p_0){16\m^{-\ep}\mathpzc{g}^2\o2N}\int {d^Dq\o(2\pi)^D}i\g^\m{iq_\a\g^\a\o q^2}{i(q+p)_\b\g^\b\o(q+p)^2}i\g_\m{1\o\abs{q}}\otimes\id
\nt
&=2\pi\d(p_0){16\m^{-\ep}\mathpzc{g}^2\o2N}\g^\m\g^\a\g^\b\g_\m\otimes\id
\int {d^Dq\o(2\pi)^D}\int_0^1dx\,{3\o2}\sqrt{1-x}\,{q_\a q_\b-x(1-x)p_\a p_\b\o\[q^2+x(1-x)p^2\]^{5/2}}
\nt
&=2\pi\d(p_0)\id\otimes\id\(-{24\mathpzc{g}^2\o\pi^2(2N)\ep}\)+\mathrm{finite}.
}
Similarly,
\eq{
\text{Fig.~\ref{fig:4ptsPhotonArc2}}&=2\pi\d(p_0)\id\otimes\id\(-{24\mathpzc{g}^2\o\pi^2(2N)\ep}\)+\mathrm{finite}.
}

\subsubsection*{Internal fermion loop with two photon legs: Figs.~\ref{fig:FermLoopPhotonPhoton1} and \ref{fig:FermLoopPhotonPhoton2}}
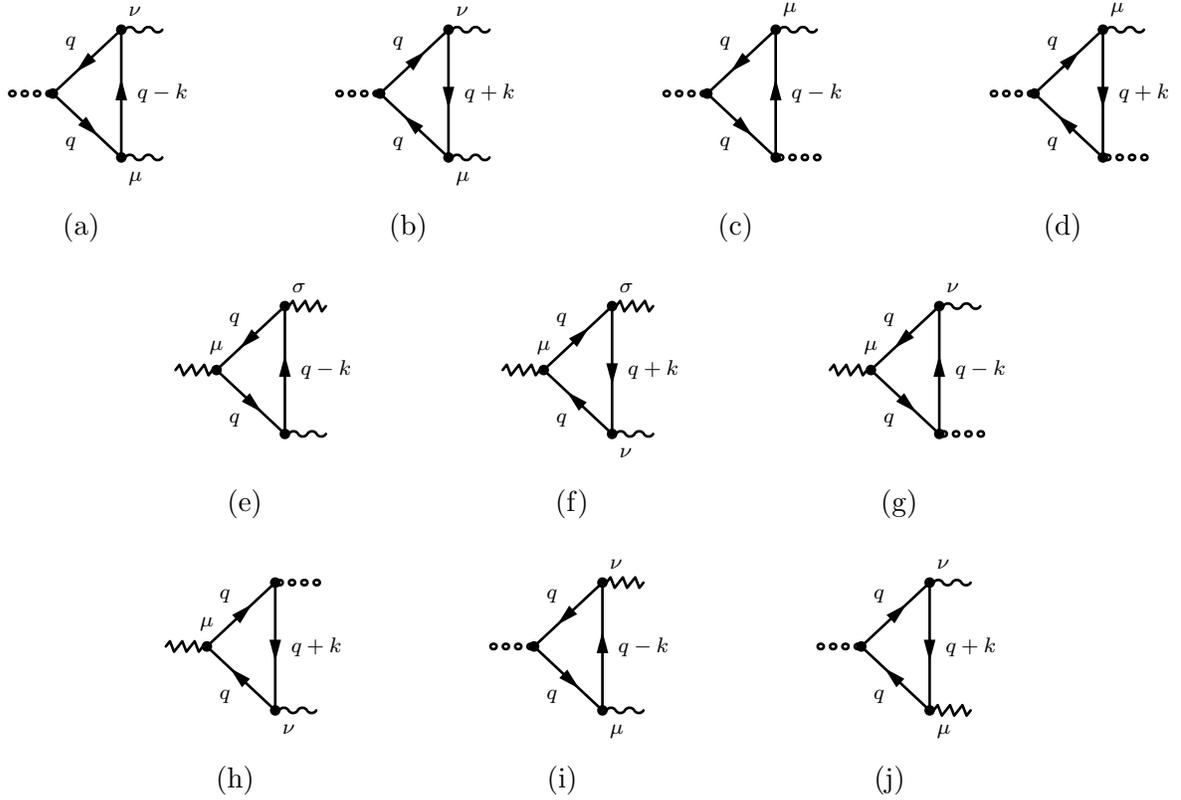
\begin{figure}
\centering
\input{InternalFermionLoops.tex}
\caption{Fermion loop subdiagrams which appear in the $\O(g_\xi^2,g_\xi/2N)$ bilinear counter terms.}
\label{fig:IntFermionLoop3Gauge}
\end{figure}

Because of the sum over $N$ in the internal fermion loop, several two-loop diagrams contribute to the order in perturbation theory we are considering. 
Since the frequency $\delta$-function which renormalizes disorder must come entirely from the single disorder leg in Figs.~\ref{fig:FermLoopPhotonPhoton1} and \ref{fig:FermLoopPhotonPhoton2}, we can determine the divergence by sending zero (spatial) momentum through this diagram. Therefore, it becomes easier to first calculate the vertices shown in Figs.~\ref{fig:FermLoopPhotonPhoton1} and \ref{fig:FermLoopPhotonPhoton2}. 

We have
\eq{\label{eqn:IntFermionLoopMass2Phot1}
\text{Fig.~\ref{fig:IntFermionLoopMass2Phot1}}
&=-{\m^{-\ep}\mathpzc{g}^2\o2N}2N\int {d^Dq\o(2\pi)^D}\tr\[{iq_\a\g^\a\o q^2}{iq_\b\g^\b\o q^2}i\g^\n{i(q-k)_\g\g^\r\o(q-k)^2}\g^\m\]
\nt&
=-i\m^{-\ep}\mathpzc{g}^2\int {d^Dq\o(2\pi)^D}\tr\[\g^\n\g^\r\g^\m\]{q_\a q_\b(q-k)_\r\o(q^2)^2(q-k)^2}
\nt
&=2\m^{-\ep}\mathpzc{g}^2\ep^{\m\n\r}\int {d^Dq\o(2\pi)^D}{(q-k)_\r\o q^2(q-k)^2}
=-\m^{-\ep}\mathpzc{g}^2\int {d^Dq\o(2\pi)^D}\int_0^1dx\,{2\ep^{\m\n\r}(1-x)k_\r\o\[q^2+x(1-x)k^2\]^2}
}
We note that since the photons are diagonal in flavour space, the mass disorder in the loop must also be diagonal. 
It follows that this diagram will only contribute to disorder coupling to the singlet mass operator, $\bpsi\psi$, and, for this reason, we have taken the flavour trace to be $2N$.
The full diagram is then
\eq{
\text{Fig.~\ref{fig:FermLoopPhotonPhoton1}}&=
-2\pi\d(p_0)\id\otimes \m^{-2\ep}{\mathpzc{g}^4\o 2N} \int{d^Dk\o(2\pi)^D}i\g^\n{i(k+p)_\s\g^\s\o(k+p)^2}i\g^\m{(16)^2\o\abs{k}^2}\cdot \int {d^Dq\o(2\pi)^D}\int_0^1dx\,{2\ep^{\m\n\r}(1-x)k_\r\o\[q^2+x(1-x)k^2\]^2}
}
We set $p=0$ and use an IR cutoff. Then, we can take $k_\s k_\r\rightarrow \d_{\s\r}k^2/d$ and
\eq{
-{i\o d}\g^\n\g^\s\g^\m \ep_{\m\n\s}={1\o d}\g^\n\g_\lam \ep^{\s\m\lam}\ep_{\s\m\n}=\id.
}
Inserting this into the expression above, we find
\eq{
\text{Fig.~\ref{fig:FermLoopPhotonPhoton1}}&=
-2\pi\d(p_0)\id\otimes\id\cdot 2(16)^2\m^{-2\ep}{\mathpzc{g}^4\o 2N}\int_0^1dx\int{d^Dk\o(2\pi)^D}{d^Dq\o(2\pi)^D}\,{1-x\o k^2\[q^2+x(1-x)k^2\]^2}
\nt
&=2\pi\d(p_0)\id\otimes\id\(16\mathpzc{g}^4\o\pi^2(2N)\ep\)+\mathrm{finite}\cdot
}
For the second diagram, we calculate the vertex in Fig.~\ref{fig:IntFermionLoopMass2Phot2}.
\eq{
\text{Fig.~\ref{fig:IntFermionLoopMass2Phot2}}
&=-{\m^{-\ep}\mathpzc{g}^2\o2N}2N\int {d^Dq\o(2\pi)^D}\tr\[{iq_\a\g^\a\o q^2}{iq_\b\g^\b\o q^2}i\g^\m{i(q+k)_\g\g^\r\o(q+k)^2}i\g^\n\]
}
This is identical to Eq.~\eqref{eqn:IntFermionLoopMass2Phot1} except with $k\rightarrow-k$ and $\m\leftrightarrow \n$:
\eq{\label{eqn:IntFermionLoopMass2Phot2}
\text{Fig.~\ref{fig:IntFermionLoopMass2Phot2}}
&=\m^{-\ep}\mathpzc{g}^2\int {d^Dq\o(2\pi)^D}\int_0^1dx\,{2\ep^{\n\m\r}(1-x)k_\r\o\[q^2+x(1-x)k^2\]^2}
=\text{Fig.~\ref{fig:IntFermionLoopMass2Phot1}}.
}
It follows that
\eq{
\text{Fig.~\ref{fig:FermLoopPhotonPhoton2}}&=\text{Fig.~\ref{fig:FermLoopPhotonPhoton1}}
=2\pi\d(p_0)\id\otimes\id\(16\mathpzc{g}^4\o\pi^2(2N)\ep\)+\mathrm{finite}\cdot
}

\subsubsection*{Internal fermion loop with one photon and one disorder line: Figs.~\ref{fig:FermLoopPhotonDisorder1} and \ref{fig:FermLoopPhotonDisorder2}}
As above, we approach the two-loop diagrams by first calculating the relevant fermion loop vertices, shown in Figs.~\ref{fig:IntFermionLoopMass1Mass1Phot1} and~\ref{fig:IntFermionLoopMass1Mass1Phot2}. 
We have
\eq{
\text{Fig.~\ref{fig:IntFermionLoopMass1Mass1Phot1}}&=-{\m^{-\ep/2}\mathpzc{g}^2\o\sqrt{2N}}\tr\[\O_{fl}\]\int{d^Dq\o(2\pi)^D}\tr\[{iq_\a\g^\a\o q^2}{iq_\b\g^\b\o q^2}i\g^\m {i(q-k)_\s\g^\s\o (q-k)^2}\]
}
Here, we leave the flavour index behaviour of the vertices arbitrary by letting $\O_{fl}$ be a general $2N\times 2N$ Hermitian matrix.
Similarly
\eq{
\text{Fig.~\ref{fig:IntFermionLoopMass1Mass1Phot2}}&=-{\m^{-\ep/2}\mathpzc{g}^2\o\sqrt{2N}}\tr\[\O_{fl}\]\int{d^Dq\o(2\pi)^D}\tr\[ {i(q+k)_\s\g^\s\o (q+k)^2}i\g^\m{iq_\b\g^\b\o q^2}{iq_\a\g^\a\o q^2}\]
}
Taking $q\rightarrow-q$ and noting that $\tr[\g^\s\g^\m\g^\b\g^\a]=\tr[\g^\a\g^\b\g^\m\g^\s]$, this becomes
\eq{
\text{Fig.~\ref{fig:IntFermionLoopMass1Mass1Phot2}}&=
{\m^{-\ep/2}\mathpzc{g}^2\o\sqrt{2N}}\tr\[\O_{fl}\]\int{d^Dq\o(2\pi)^D}\tr\[{iq_\a\g^\a\o q^2}{iq_\b\g^\b\o q^2}i\g^\m {i(q-k)_\s\g^\s\o (q-k)^2}\]
=-\text{Fig.~\ref{fig:IntFermionLoopMass1Mass1Phot1}}.
}
It follows that the divergences in Figs.~\ref{fig:FermLoopPhotonDisorder1} and \ref{fig:FermLoopPhotonDisorder2} cancel.

\subsection{Diagrams with gauge-like disorder and photon lines: Fig.~\ref{fig:gaugeDisorder4pt}}
\subsubsection*{Two internal gauge-like disorder lines, no crossing: Fig.~\ref{fig:4ptGamGamNoX}}
\eq{
\text{Fig.~\ref{fig:4ptGamGamNoX}}
&=2\pi(p_0){1\o2}\sum_j{ \g^\m\g^j\g^\n\otimes\g^\m\g_j\g^\n}\(-{1\o2\pi\ep}\)+\mathrm{finite}
\nt
&=2\pi(p_0)\({1\o4\pi\ep}\)
\begin{cases}
-\sum_j\g^j\otimes\g_j, & \(\m,\n\)=\(0,0\) \\
\id\otimes\id-\g^0\otimes\g^0, & \(\m,\n\)=\(0,\ell\),\(\ell,0\) \\
-\sum_j\g^j\otimes\g_j, & \(\m,\n\)=\(\ell,k\)
\end{cases}
+\mathrm{finite}
\label{eqn:4ptGamGamNoX}
}

\subsubsection*{Two internal gauge-like disorder lines, with crossing: Fig.~\ref{fig:4ptGamGamX}}
\eq{
\text{Fig.~\ref{fig:4ptGamGamX}}
&=2\pi(p_0){1\o2}\sum_j{ \g^\m\g^j\g^\n\otimes\g^\n\g_j\g^\m}\({1\o2\pi\ep}\)+\mathrm{finite}
\nt
&=2\pi(p_0)\({1\o4\pi\ep}\)
\begin{cases}
\sum_j\g^j\otimes\g_j, & \(\m,\n\)=\(0,0\) \\
\id\otimes\id+\g^0\otimes\g^0, & \(\m,\n\)=\(0,\ell\),\(\ell,0\) \\
\sum_j\g^j\otimes\g_j, & \(\m,\n\)=\(k,\ell\)
\end{cases}
+\mathrm{finite}
\label{eqn:4ptGamGamX}
}

\subsubsection*{Vertex correction from gauge-like disorder: Figs.~\ref{fig:4ptGamGamVert} and \ref{fig:4ptGamGamVert2}}
\eq{
\text{Fig.~\ref{fig:4ptGamGamVert}}&=\int {d^Dq\o(2\pi)^D}2\pi\d(q_0)2\pi\d(p_0)i\g^\n{iq_\a\g^\a\o q^2}i\g^\m{i\[q-p\]_\b\g^\b\o(q-p)^2}i\g_\n\otimes i\g_\m
\nt
&=2\pi(p_0)\,\[-{1\o2}\int{d^dq\o(2\pi)^d}\int_0^1dx\,{q^2\o\[q^2+x(1-x)p^2\]^2}\]\g^\n\g^j\g^\m\g_j\g_\n\otimes\g^\n+\mathrm{finite}
\nt
&=2\pi(p_0)\({1\o2\pi\ep}\)\times\begin{cases}
0 & \(\m,\n\)=\(\ell,0\),\(\ell,k\) \\
-\g^0\otimes\g^0 & \(\m,\n\)=\(0,0\)\\
\g^0\otimes\g^0 & \(\m,\n\)=\(0,\ell\)
\end{cases}+\mathrm{finite}
\nt
&=2\pi(p_0)\g^0\otimes\g^0\({1\o2\pi\ep}\)\d^{\m0}\[-\d^{\n0}+\sum_j\d^{\n j}\]
}
The other vertex gives the same correction:
\eq{
\text{Fig.~\ref{fig:4ptGamGamVert2}}&=2\pi(p_0)\g^0\otimes\g^0\({1\o2\pi\ep}\)\d^{\n0}\[-\d^{\m0}+\sum_j\d^{\m j}\]+\mathrm{finite}
}

\subsubsection*{One internal gauge-like disorder line and one photon line: Figs.~\ref{fig:4ptGaugePhotonNoX} and \ref{fig:4ptGaugePhotonX}}

This situation is identical to the one in Eq.~\eqref{eqn:DisPhotConverge} except for some $\g$ matrices: both Fig.~\ref{fig:4ptGaugePhotonNoX} and Fig.~\ref{fig:4ptGaugePhotonX} are finite.

\subsubsection*{Vertex correction from photon: Figs.~\ref{fig:4ptsGammaPhotonArc} and Fig.~\ref{fig:4ptsGammaPhotonArc2}}

\eq{
\text{Fig.~\ref{fig:4ptsGammaPhotonArc}}&=
2\pi\d(p_0){16\m^{-\ep}\mathpzc{g}^2\o2N}\sum_\n\int {d^Dq\o(2\pi)^D}i\g^\n{iq_\a\g^\a\o q^2}i\g^\m{i(q+p)_\b\g^\b\o(q+p)^2}i\g_\n{1\o\abs{q}}\otimes\g^\m
\nt
&=2\pi\d(p_0){16\m^{-\ep}\mathpzc{g}^2\o2N}i\sum_\n\g^\n\g^\a\g^\m\g^\b\g_\n\otimes\g^\m
\int {d^Dq\o(2\pi)^D}\int_0^1dx\,{3\o2}\sqrt{1-x}\,{q_\a q_\b-x(1-x)p_\a p_\b\o\[q^2+x(1-x)p^2\]^{5/2}}
\nt
&=2\pi\d(p_0)\g^\m\otimes\g^\m\(-{8\mathpzc{g}^2\o3\pi^2(2N)\ep}\)+\mathrm{finite}.
}
Similarly,
\eq{
\text{Fig.~\ref{fig:4ptsGammaPhotonArc2}}&=2\pi\d(p_0)\g^\m\otimes\g^\m\(-{8\mathpzc{g}^2\o3\pi^2(2N)\ep}\)+\mathrm{finite}.
}

\subsubsection*{Internal fermion loop with one disorder and two  photon legs: Figs.~\ref{fig:FermLoopGammaPhotonPhoton1} and~\ref{fig:FermLoopGammaPhotonPhoton2}}

None of the gauge-like disorder terms are diagonal in the flavour indices. As we remarked above, this is because the global U(1) current has scaling dimension 3, making it extremely irrelevant. 
Therefore, the gauge-like disorder in Figs.~\ref{fig:FermLoopGammaPhotonPhoton1} and~\ref{fig:FermLoopGammaPhotonPhoton2} inserts an $2N\times 2N$ traceless Hermitian matrix into the fermion loop. Upon taking the trace, both vanish.

\subsubsection*{Internal fermion loops with two disorder and one photon leg: Figs.~\ref{fig:FermLoopGammaPhotonDisorder1} and~\ref{fig:FermLoopGammaPhotonDisorder2}}

As we did for the two loop diagrams with mass-like disorder above, we first calculate the fermion loop vertices. 
The vertices relevant to our diagrams are shown in Figs.~\ref{fig:IntFermionLoop3Gauge1} and ~\ref{fig:IntFermionLoop3Gauge2}. We have
\eq{
\text{Fig.~\ref{fig:IntFermionLoop3Gauge1}}&=
-{\m^{-\ep/2}\mathpzc{g}\o\sqrt{2N}}\tr\[\O_{fl}\]\int {d^Dq\o(2\pi)^D}\tr\[{iq_\b\g^\b\o q^2}i\g^\m {iq_\a\g^\a\o q^2}i\g^\n{i(q-k)_\r\g^\r\o(q-k)^2}i\g^\s\]
\nt
&={\m^{-\ep/2}\mathpzc{g}\o\sqrt{2N}}\tr\[\O_{fl}\]\int {d^Dq\o(2\pi)^D}\tr\[\g^\n\g^\r\g^\s\g^\b\g^\m\g^\a\]{q_\a q_\b(q-k)_\s\o (q^2)^2(q-k)^2},
}
where $\O_{fl}$ is the matrix in flavour space coming from disorder vertices.
Similarly, reversing the direction of the fermion loop, we have
\eq{
\text{Fig.~\ref{fig:IntFermionLoop3Gauge2}}&=
-{\m^{-\ep/2}\mathpzc{g}\o\sqrt{2N}}\tr\[\O_{fl}\]\int {d^Dq\o(2\pi)^D}\tr\[{iq_\a\g^\a\o q^2}i\g^\m {iq_\b\g^\b\o q^2}i\g^\s{i(q+k)_\r\g^\r\o(q+k)^2}i\g^\n\]
\nt
&={\m^{-\ep/2}\mathpzc{g}\o\sqrt{2N}}\tr\[\O_{fl}\]\int {d^Dq\o(2\pi)^D}\tr\[\g^\a\g^\m\g^\b\g^\s\g^\r\g^\n\]{q_\a q_\b(q+k)_\s\o (q^2)^2(q+k)^2}\cdot
}
Noting that
\eq{
\tr\[\g^{\m_1}\g^{\m_2}\cdots\g^{\m_n}\]=(-1)^n\tr\[\g^{\m_n}\g^{\m_{n-1}}\cdots\g^{\m_1}\]
}
and taking $q\rightarrow-q$, we have
\eq{
\text{Fig.~\ref{fig:IntFermionLoop3Gauge2}}&=
-{\m^{-\ep/2}\mathpzc{g}\o\sqrt{2N}}\tr\[\O_{fl}\]\int {d^Dq\o(2\pi)^D}\tr\[\g^\n\g^\r\g^\s\g^\b\g^\m\g^\a\]{q_\a q_\b(q-k)_\s\o (q^2)^2(q-k)^2}
=-\text{Fig.~\ref{fig:IntFermionLoop3Gauge1}}.
}
We conclude that Figs.~\ref{fig:FermLoopGammaPhotonDisorder1} and~\ref{fig:FermLoopGammaPhotonDisorder2} cancel one another.

\subsection{Both potential and mass disorder diagrams}

\subsubsection*{One internal mass-like and gauge-like disorder lines, no crossing: Fig.~\ref{fig:4ptMixNoX}}

\eq{\label{eqn:4ptMixNoX}
\text{Fig.~\ref{fig:4ptMixNoX}}
&=\int{d^Dq\o(2\pi)^D}2\pi\d(q_0)2\pi\d\(q+p\)\,i\g^\m\,{i\[-q\]_\a\g^a\o q^2}\otimes\,i\g^\m{i(q+p)_\b\g^\b\o(q+p)^2}
\nt
&=-2\pi\d(p_0)\g^\m\g^i\otimes\g^\m\g^j\cdot{\d_{ij}\o2}\int{d^dq\o(2\pi)^2}\int_0^1dx\,{q^2\o\[q+x(1-x)p^2\]^2}+\mathrm{finite}
\nt
&=2\pi\d(p_0)\sum_j\g^\m\g^j\otimes\g^\m\g_j\(1\o4\pi\ep\)+\mathrm{finite}
\nt
&=2\pi\d(p_0)\({1\o4\pi\ep}\)\begin{cases}
-\sum_j\g^j\otimes\g_j, & \m=0, \\
\id\otimes\id-\g^0\otimes\g^0, &\m=\ell
\end{cases}
}

\subsubsection*{One internal mass-like and gauge-like disorder lines, with crossing: Fig.~\ref{fig:4ptMixX}}

\eq{\label{eqn:4ptMixX}
\text{Fig.~\ref{fig:4ptMixX}}
&=-2\pi\d(p_0)\sum_j{ \g^\m\g^j\otimes\g^j\g_\m}\({1\o4\pi\ep}\)+\mathrm{finite}
\nt
&=2\pi\d(p_0)\({1\o4\pi\ep}\)\begin{cases}
-\sum_j\g^j\otimes\g_j, & \m=0, \\
-\id\otimes\id-\g^0\otimes\g^0, &\m=\ell
\end{cases}
}

\subsubsection*{Mass disorder vertex correction from potential disorder: Figs.~\ref{fig:4ptMixVert1a} and~\ref{fig:4ptMixVert2a}}

\eq{
\text{Fig.~\ref{fig:4ptMixVert1a}}
&=2\pi(p_0)\,{\id\otimes\id}\(-{1\o2\pi\ep}\)+\mathrm{finite}
}
and
\eq{
\text{Fig.~\ref{fig:4ptMixVert2a}}
&=2\pi(p_0)\,{\id\otimes\id}\(-{1\o2\pi\ep}\)+\mathrm{finite}
}

\subsubsection*{Potential disorder vertex correction from mass disorder: Figs.~\ref{fig:4ptMixVert1b} and~\ref{fig:4ptMixVert2b}}
\eq{
\text{Fig.~\ref{fig:4ptMixVert1b}}&=\int {d^Dq\o(2\pi)^D}2\pi\d(q_0)2\pi\d(p_0){iq_\a\g^\a\o q^2}i\g^\m{i\[q-p\]_\b\g^\b\o(q-p)^2}\otimes i\g_\m
\nt
&=2\pi(p_0)\,\[{1\o2}\int{d^dq\o(2\pi)^d}\int_0^1dx\,{q^2\o\[q^2+x(1-x)p^2\]^2}\]\g^j\g^\m\g_j\otimes\g^\m+\mathrm{finite}
\nt
&=2\pi(p_0)\d^{\m0}\g^0\otimes\g^0\({1\o2\pi\ep}\)+\mathrm{finite}
}
Similarly,
\eq{
\text{Fig.~\ref{fig:4ptMixVert2b}}&=2\pi\d(p_0)\d^{\m0}\g^0\otimes\g^0\({1\o2\pi\ep}\)+\mathrm{finite}
}

\subsubsection*{Internal fermion loop with internal gauge and photon legs: Figs.~\ref{fig:FermLoopPotentialVertMassInt1} and \ref{fig:FermLoopPotentialVertMassInt2}}

In order to calculate Figs.~\ref{fig:FermLoopPotentialVertMassInt1} and \ref{fig:FermLoopPotentialVertMassInt2}, we being by determining the subdiagrams in Figs.~\ref{fig:IntFermLoopGaugeMassPhot1} and~\ref{fig:IntFermLoopGaugeMassPhot2}:
\eq{
\text{Fig.~\ref{fig:IntFermLoopGaugeMassPhot1}}&=
-{\m^{-\ep/2}\mathpzc{g}\o\sqrt{2N}}\tr\[\O_{fl}\]\int {d^Dq\o(2\pi)^D}\tr\[{iq_\a \g^\a \o q^2}i\g^\m{iq_\b \g^\b \o q^2}i\g^\n{i(q-k)_\s\g^\s\o(q-k)^2}\]
\nt&
}
where $\O_{fl}$ is the matrix in flavour space resulting from disorder vertices.
Similarly, the other diagram gives
\eq{
\text{Fig.~\ref{fig:IntFermLoopGaugeMassPhot2}}&=
-{\m^{-\ep/2}\mathpzc{g}\o\sqrt{2N}}\tr\[\O_{fl}\]\int {d^Dq\o(2\pi)^D}\tr\[{iq_\a \g^\a \o q^2}i\g^\m{iq_\b \g^\b \o q^2}i\g^\n{i(q+k)_\s\g^\s\o(q+k)^2}\]
\nt&
={\m^{-\ep/2}\mathpzc{g}\o\sqrt{2N}}\tr\[\O_{fl}\]\int {d^Dq\o(2\pi)^D}\tr\[{iq_\a \g^\a \o q^2}i\g^\m{iq_\b \g^\b \o q^2}i\g^\n{i(q-k)_\s\g^\s\o(q-k)^2}\]
=-\text{Fig.~\ref{fig:IntFermLoopGaugeMassPhot2}}
}
where in the last line we took $q\rightarrow-q$. It follows that these diagrams cancel with each other.

\subsubsection*{Internal fermion loop with internal mass and gauge disorder and photon lines: Figs.~\ref{fig:FermLoopMassVerPotentialInt1} and~\ref{fig:FermLoopMassVerPotentialInt2}}

We start by evaluating the fermion loop vertices in Figs.~\ref{fig:IntFermLoopMassGaugePho1} and~\ref{fig:IntFermLoopMassGaugePho2}.
Actually, it's not difficult to see that up to the photon vertex coupling, $\m^{-\ep/2}\mathpzc{g}/\sqrt{2N}$, these diagrams are identical to the vertices in Figs.~\ref{fig:IntFermionLoopMass2Phot1} and~\ref{fig:IntFermionLoopMass2Phot1}, determined in Eqs.~\eqref{eqn:IntFermionLoopMass2Phot1} and ~\eqref{eqn:IntFermionLoopMass2Phot2}:
\eq{
\text{Fig.~\ref{fig:IntFermLoopMassGaugePho1}}&=\text{Fig.~\ref{fig:IntFermLoopMassGaugePho2}}
\nt
&=-{\m^{-\ep/2}\mathpzc{g}\o\sqrt{2N}}\tr\[\O_{fl}\]\int {d^Dq\o(2\pi)^D}\int_0^1dx\,{2\ep^{\m\n\r}(1-x)k_\r\o\[q^2+x(1-x)k^2\]^2}\cdot
}
Proceeding as we did for this case, we have
\eq{
\text{Fig.~\ref{fig:FermLoopMassVerPotentialInt1}}&=\text{Fig.~\ref{fig:FermLoopMassVerPotentialInt2}}
\nt
&=-2\pi\d(p_0)\id\otimes {\m^{-\ep}\mathpzc{g}^2\o2N}\tr\[\O_{fl}\]\int{d^Dk\o(2\pi)^D}2\pi\d(k_0)i\g^\n{ik_\s\g^\s\o k^2}i\g^\m{16\o\abs{k}}\cdot \int {d^Dq\o(2\pi)^D}\int_0^1dx\,{2\ep^{\m\n\r}(1-x)k_\r\o\[q^2+x(1-x)k^2\]^2}
\nt
&=-2\pi\d(p_0)\id\otimes {32\m^{-\ep}\mathpzc{g}^2\o2N}\tr\[\O_{fl}\]
\nt
&\quad\times\int{d^dk\o(2\pi)^d}\int {d^Dq\o(2\pi)^D}\int_0^1dx\,(-i)\g^\n\g^\s\g^\m\ep_{\m\n\r}  {\d_\s^{\;j}\d_j^\r\o  d }{1\o\abs{k}} {1-x\o\[q^2+x(1-x)k^2\]^2}
}
where $\tr\[\O_{fl}\]$ indicates that, in order to allow disorder vertices which are off-diagonal in the flavour indices, we have not yet explicitly taken the trace over the flavours. Moreover, we sum over $\n$, $\s$, and $\r$ but \emph{not} $\m$. With this in mind, we note
\eq{
-{i\o d}\sum_{\s\r\n j}\g^\n\g^\s\g^\m\ep_{\m\n\r}\d_\s^j\d_j^\r={1\o d}\sum_{\n\lam j}\ep^{\n j\lam}\ep_{\m\n j}\g_\lam \g^\m={1\o d}\(d-\d^j_\m\)\id.
}
Performing the $q$, $k$, and $x$ integrals, we obtain,
\eq{
\text{Fig.~\ref{fig:FermLoopMassVerPotentialInt1}}&=\text{Fig.~\ref{fig:FermLoopMassVerPotentialInt2}}
\nt
&=2\pi\d(p_0)\id\otimes\id \tr\[\O_{fl}\]{\mathpzc{g}^2\o2N}\({1\o4\pi\ep}\)(2-\d_{\m j})+\mathrm{finite}
\nt&=
\begin{cases}
2\pi\d(p_0)\tr\[\O_{fl}\]{\mathpzc{g}^2\o2N}\({1\o2\pi\ep}\)\id\otimes\id, & \mu=0,\\
2\pi\d(p_0)\tr\[\O_{fl}\]{\mathpzc{g}^2\o2N}\({1\o4\pi\ep}\)\id\otimes\id, & \mu=x,y
\end{cases}
}

\section{4-point diagrams contributing to fermion bilinear counter terms}\label{app:4ptDiagramsModel}

The diagrams which contribute to the $\b$-functions at $\O(g^2_\xi,g_\xi/N)$ are shown in Fig.~\ref{fig:sigmasigma} and in Tables.~\ref{tab:DiagramTab1} through \ref{tab:DiagramTab3}. The divergences are based on the integrals determined in Sec.~\ref{app:4ptDiagramsCalc} and only diagrams which do not vanish are shown. 
The label ``$n_d$" indicates the degeneracy of the diagram or else the existence of a diagram with a nearly identical form.

Some of the diagrams result in divergences proportional to $\[\g^\m\otimes\g^\m\]\[\id\otimes\id\]$ and would appear to imply that disorder coupling to the U(1) gauge current $J^\m$ is generated. 
While counter terms are technically required to render the theory finite, we emphasize that it is not necessary to consider them since $J^\m$ already has a large scaling dimension at the QED$_3$ fixed point.
\input{DiagramTable.tex}

\section{Diagrams renormalizing flux disorder, $g_\E$ and $g_\B$}\label{app:TopRenorm}
\input{FluxDisorder2.tex}
\input{appendixConductivity.tex}

\bibliographystyle{apsrev4-1_custom}
\bibliography{qedDisorderRef}

\end{document}

%% file: headersDraft2.tex
\usepackage[utf8]{inputenc}
\usepackage{amssymb,bm,xspace,soul} 
\usepackage{color} 
\usepackage[papersize={8.5in,11in}]{geometry}
\usepackage{natbib}
\usepackage{graphicx}
\usepackage{amsmath}
\usepackage[table]{xcolor}
\usepackage{verbatim}
\usepackage{braket}
\usepackage{enumitem}
\usepackage{feynmp}
\usepackage{math tools} 
\usepackage{dsfont} 
\usepackage{mathrsfs} 
\usepackage{multirow}
\usepackage{slashed}

\newcommand{\N}{\mathcal{N}}

\usepackage{subcaption}
\usepackage[justification=raggedright]{caption}
\usepackage{pifont}
\newcommand{\A}{\mathcal{A}}

\newcommand{\B}{\mathcal{B}}

\usepackage[colorlinks=true]{hyperref} 
\hypersetup{
    bookmarks=true,         
    unicode=false,          
    pdftoolbar=true,        
    pdfmenubar=true,        
    pdffitwindow=false,     
    pdfstartview={FitH},    
    pdftitle={My title},    
    pdfauthor={Author},     
    pdfsubject={Subject},   
    pdfcreator={Creator},   
    pdfproducer={Producer}, 
    pdfkeywords={keyword1} {key2} {key3}, 
    pdfnewwindow=true,      
    colorlinks=true,       
    linkcolor=magenta, 
    citecolor=blue,        
    filecolor=magenta,      
    urlcolor=cyan           
} 

\makeatletter
    \def\CT@@do@color{%
      \global\let\CT@do@color\relax
            \@tempdima\wd\z@
            \advance\@tempdima\@tempdimb
            \advance\@tempdima\@tempdimc
    \advance\@tempdimb\tabcolsep
    \advance\@tempdimc\tabcolsep
    \advance\@tempdima2\tabcolsep
            \kern-\@tempdimb
            \leaders\vrule
                    \hskip\@tempdima\@plus  1fill
            \kern-\@tempdimc
            \hskip-\wd\z@ \@plus -1fill }
    \makeatother

\renewcommand{\O}{\mathcal{O}}

\newcommand{\T}{\mathcal{T}}

\newcommand{\E}{\mathcal{E}}

\newcommand{\vk}{{\boldsymbol{k}}}
\renewcommand{\vr}{{\boldsymbol{r}}}
\newcommand{\vQ}{{\boldsymbol{Q}}}

\newcommand{\vq}{{\boldsymbol{q}}}
\newcommand{\vp}{{\boldsymbol{p}}}
\newcommand{\vx}{{\boldsymbol{x}}}

\newcommand{\bpsi}{\bar{\psi}}

\newcommand{\tr}{\text{tr}}
\newcommand{\sd}[1]{\slashed{#1}}

\renewcommand{\o}{\over}
\newcommand{\eq}[1]{\begin{align}#1\end{align}}

\renewcommand{\(}{\left(}
\renewcommand{\)}{\right)}
\renewcommand{\[}{\left[}
\renewcommand{\]}{\right]}
\newcommand{\abs}[1]{\left| #1 \right|}

\newcommand{\nt}{\notag\\}

\let\v\boldsymbol

\newcommand{\ph}{\phantom}

\newcommand{\ep}{\epsilon}

\renewcommand{\a}{\alpha}
\renewcommand{\b}{\beta}
\renewcommand{\d}{\delta}
\newcommand{\g}{\gamma}
\newcommand{\n}{\nu}
\newcommand{\m}{\mu}
\renewcommand{\t}{\tau}
\newcommand{\s}{\sigma}
\newcommand{\w}{\omega}

\newcommand{\lam}{\lambda}
\renewcommand{\r}{\rho}

\let\ptl\partial
\renewcommand{\dag}{\dagger}

\newcommand{\id}{\mathds{1}}

\newcommand{\mf}[1]{\raisebox{-0.4\height}{\hbox{ {#1}}}}
\newcommand{\mff}[1]{\raisebox{-0.48\height}{\hbox{ {#1}}}}
\newcommand{\mfff}[1]{\raisebox{-0.44\height}{\hbox{ {#1}}}}

\usepackage{tocloft} 
\addtocontents{toc}{\cftpagenumbersoff{chapter}}
\addtocontents{toc}{\cftpagenumbersoff{section}}

\setlist[enumerate]{leftmargin=*}

%% file: effPhotonProp.tex
\centering
\begin{fmffile}{effPhotonProp}
\fmfset{dot_size}{1.5thick}
\fmfset{arrow_len}{3mm}
\scriptsize
\eq{
\mfff{
\begin{fmfgraph*}(20,20)
\fmfpen{thin}
\fmfstraight
\fmfleft{i1}
\fmfright{o1}
\fmfv{label=$\m$,l.a=90}{i1}
\fmfv{label=$\n$,l.a=90}{o1}
\fmf{photon}{i1,o1}
\end{fmfgraph*}
}\quad
=\quad
\mff{
\begin{fmfgraph*}(20,20)
\fmfpen{thin}
\fmfstraight
\fmfleft{i1}
\fmfright{o1}
\fmfv{label=$\m$,l.a=90}{i1}
\fmfv{label=$\n$,l.a=90}{o1}
\fmf{dots}{i1,o1}
\end{fmfgraph*}
}\quad
+\quad
\mff{
\begin{fmfgraph*}(25,20)
\fmfpen{thin}
\fmfstraight
\fmfleft{i1}
\fmfright{o1}
\fmfv{label=$\m$,l.a=90}{i1}
\fmfv{label=$\n$,l.a=90}{o1}
\fmf{dots}{i1,v1}
\fmf{phantom,tension=.922}{v1,v2}
\fmf{dots}{v2,o1}
\fmfdot{v1}
\fmfdot{v2}
\fmffreeze
\fmf{fermion,right}{v1,v2}
\fmf{fermion,right}{v2,v1}
\end{fmfgraph*}
}\quad+\quad
\mff{
\begin{fmfgraph*}(40,20)
\fmfpen{thin}
\fmfstraight
\fmfleft{i1}
\fmfright{o1}
\fmfv{label=$\m$,l.a=90}{i1}
\fmfv{label=$\n$,l.a=90}{o1}
\fmf{dots}{i1,v1}
\fmf{phantom,tension=0.85}{v1,v2}
\fmf{dots}{v2,v3}
\fmf{phantom,tension=0.85}{v3,v4}
\fmf{dots}{v4,o1}
\fmfdot{v1}
\fmfdot{v2}
\fmfdot{v3}
\fmfdot{v4}
\fmffreeze
\fmf{fermion,right}{v1,v2}
\fmf{fermion,right}{v2,v1}
\fmf{fermion,right}{v3,v4}
\fmf{fermion,right}{v4,v3}
\end{fmfgraph*}
}\quad+\quad\cdots
\notag}
\end{fmffile}

%% file: feynRules.tex
\centering
\begin{fmffile}{feynRules}
\fmfset{dot_size}{1.5thick}
\fmfset{arrow_len}{3mm}
\begin{subfigure}[b]{0.31\textwidth}
\eq{
\mfff{
\scriptsize
\begin{fmfgraph*}(20,20)
\fmfpen{thin}
\fmfstraight
\fmfleft{i1}
\fmfright{o1}
\fmf{fermion,label=$p$}{i1,o1}
\end{fmfgraph*}
}
={ip_\m\g^\m\o p^2}
\notag}
\end{subfigure}
\hspace{3mm}
\begin{subfigure}[b]{0.31\textwidth}
\eq{
\mfff{
\scriptsize
\begin{fmfgraph*}(20,20)
\fmfpen{thin}
\fmfstraight
\fmfleft{i1}
\fmfright{o1}
\fmfv{label=$\m$}{i1}
\fmfv{label=$\n$}{o1}
\fmf{photon,label=$p$}{i1,o1}
\end{fmfgraph*}
}
\quad
={16\o\abs{p}}\,\d^{\m\n}
\notag}
\end{subfigure}
\hspace{3mm}
\begin{subfigure}[b]{0.31\textwidth}
\centering
\eq{
\mfff{
\scriptsize
\begin{fmfgraph*}(20,20)
\fmfpen{thin}
\fmfstraight
\fmfleft{i1}
\fmfright{o2,o1}
\fmf{photon,label=$p$}{i1,v1}
\fmf{fermion,label=$q$,label.side=left}{v1,o1}
\fmf{fermion,label=$p+q$,label.side=left}{o2,v1}
\fmfdot{v1}
\fmfv{label=$\mu$}{v1}
\end{fmfgraph*}
}
={\m^{-\ep/2}\mathpzc{g}\o\sqrt{2N}}\,i\g^\m
\notag}
\end{subfigure}
\\
\begin{tabular}{ m{.12\textwidth} m{.35\textwidth} m{.12\textwidth} m{.35\textwidth} }
\vspace{2mm}
\scriptsize
\begin{fmfgraph*}(20,17)
\fmfpen{thin}
\fmfstraight
\fmfleft{i1}
\fmfright{o1}
\fmf{photon}{i1,v1}
\fmf{phantom,tension=1.5}{v1,v2}
\fmf{photon}{v2,o1}
\fmffreeze
\fmfpoly{shade}{v1,vb,v2,vt}
\fmfv{label=$0$,l.a=135}{v1}
\fmfv{label=$0$,l.a=45}{v2}
\end{fmfgraph*}
&
$=-2\pi\d(q_0)\m^{-\ep}g_\B\vq^2$
&
\vspace{2mm}
\scriptsize
\begin{fmfgraph*}(20,17)
\fmfpen{thin}
\fmfstraight
\fmfleft{i1}
\fmfright{o1}
\fmf{photon}{i1,v1}
\fmf{phantom,tension=1.3}{v1,v2}
\fmf{photon}{v2,o1}
\fmffreeze
\fmfpoly{shade}{v1,vb,v2,vt}
\fmfv{label=$i$,l.a=135}{v1}
\fmfv{label=$j$,l.a=45}{v2}
\end{fmfgraph*}
&
$=2\pi\d(q_0)\m^{-\ep}g_\E\vq^2\(\d^{ij}-{q^iq^j\o q^2}\)$
\\
\vspace{5mm}
\scriptsize
\begin{fmfgraph*}(20,17)
\fmfpen{thin}
\fmfstraight
\fmfleft{i2,i1}
\fmfright{o2,o1}
\fmfv{label=$\b,,\ell$}{i1}
\fmfv{label=$\a,,\ell$}{o1}
\fmfv{label=$\r ,,m$}{i2}
\fmfv{label=$\s,, m$}{o2}
\fmf{phantom}{i1,v1,o1}
\fmf{phantom}{i2,v2,o2}
\fmffreeze
\fmf{fermion}{i1,v1}
\fmf{fermion}{v1,o1}
\fmf{fermion}{i2,v2}
\fmf{fermion}{v2,o2}
\fmf{dashes,label=$q$}{v1,v2}
\fmfdot{v1}
\fmfdot{v2}
\end{fmfgraph*}
\vspace{4mm}
&
$=2\pi\d(q_0)\m^{-\ep}g_s\[\id\]_{\a\b}\[\id\]_{\s\r}$
&
\scriptsize
\vspace{5mm}
\begin{fmfgraph*}(20,17)
\fmfpen{thin}
\fmfstraight
\fmfleft{i2,i1}
\fmfright{o2,o1}
\fmfv{label=$\b,,\ell$}{i1}
\fmfv{label=$\a,,\ell$}{o1}
\fmfv{label=$\r ,,m$}{i2}
\fmfv{label=$\s,, m$}{o2}
\fmf{phantom}{i1,v1,o1}
\fmf{phantom}{i2,v2,o2}
\fmffreeze
\fmf{fermion}{i1,v1}
\fmf{fermion}{v1,o1}
\fmf{fermion}{i2,v2}
\fmf{fermion}{v2,o2}
\fmf{dbl_wiggly,label=$q$}{v1,v2}
\fmfdot{v1}
\fmfdot{v2}
\fmfv{l=$a,,0$}{v1}
\fmfv{l=$a,,0$}{v2}
\end{fmfgraph*}
\vspace{4mm}
&
$=-2\pi\d(q_0)\m^{-\ep}g_{v,a} \[i\g^0\s^a\]_{\a\b}\[i\g^0\s^a\]_{\s\r}$
\\
\vspace{8mm}
\scriptsize
\begin{fmfgraph*}(20,17)
\fmfpen{thin}
\fmfstraight
\fmfleft{i2,i1}
\fmfright{o2,o1}
\fmfv{label=$\b,,\ell$}{i1}
\fmfv{label=$\a,,\ell$}{o1}
\fmfv{label=$\r ,,\,,m$}{i2}
\fmfv{label=$\s,,\,, m$}{o2}
\fmf{phantom}{i1,v1,o1}
\fmf{phantom}{i2,v2,o2}
\fmffreeze
\fmf{fermion}{i1,v1}
\fmf{fermion}{v1,o1}
\fmf{fermion}{i2,v2}
\fmf{fermion}{v2,o2}
\fmf{dbl_dashes,label=$q$}{v1,v2}
\fmfdot{v1}
\fmfdot{v2}
\fmfv{l=$a$}{v1}
\fmfv{l=$a$}{v2}
\end{fmfgraph*}
\vspace{4mm}
&
$=2\pi\d(q_0)\m^{-\ep}g_{t,a}\[\s^a\]_{\a\b}\[\s^a\]_{\s\r}$
&
\vspace{8mm}
\scriptsize
\begin{fmfgraph*}(20,17)
\fmfpen{thin}
\fmfstraight
\fmfleft{i2,i1}
\fmfright{o2,o1}
\fmfv{label=$\b,,\ell$}{i1}
\fmfv{label=$\a,,\ell$}{o1}
\fmfv{label=$\r ,,m$}{i2}
\fmfv{label=$\s,, m$}{o2}
\fmf{phantom}{i1,v1,o1}
\fmf{phantom}{i2,v2,o2}
\fmffreeze
\fmf{fermion}{i1,v1}
\fmf{fermion}{v1,o1}
\fmf{fermion}{i2,v2}
\fmf{fermion}{v2,o2}
\fmf{dbl_plain,label=$q$}{v1,v2}
\fmfdot{v1}
\fmfdot{v2}
\fmfv{l=$a,,j$}{v1}
\fmfv{l=$a,,j$}{v2}
\end{fmfgraph*}
\vspace{4mm}
%
&
$=2\pi\d(q_0)\m^{-\ep}g_{\A,a} \[i\g^j\s^a\]_{\a\b}\[i\g^j\s^a\]_{\s\r}$
\end{tabular}
\end{fmffile}

%% file: vanishingBubble.tex
\centering
\begin{fmffile}{vanishingBubble}
\fmfset{dot_size}{1.5thick}
\fmfset{arrow_len}{3mm}
\fmfset{zigzag_width}{.7mm}
\centering
\eq{
\mff{
\scriptsize
\begin{fmfgraph*}(27,20)
\fmfpen{thin}
\fmfstraight
\fmfleft{i2,i1}
\fmfright{o2,o1}
\fmf{phantom}{i1,v1,o1}
\fmf{phantom}{i2,v2,o2}
\fmffreeze
\fmf{phantom}{v1,c1}
\fmf{phantom}{c2,v2}
\fmf{phantom,tension=0.7}{c1,c2}
\fmffreeze
\fmf{fermion}{i1,v1}
\fmf{fermion}{v1,o1}
\fmf{fermion}{i2,v2}
\fmf{fermion}{v2,o2}
\fmf{dashes}{v1,c1}
\fmf{fermion,right,tension=0}{c1,c2}
\fmf{fermion,right,tension=0}{c2,c1}
\fmf{dashes}{c2,v2}
\fmfdot{c1}
\fmfdot{c2}
\fmfdot{v1}
\fmfdot{v2}
\end{fmfgraph*}
}
\notag}
\end{fmffile}

%% file: fermionSelfE.tex
\centering
\begin{fmffile}{FermionSelfEnergy}
\fmfset{dot_size}{1.5thick}
\fmfset{arrow_len}{3mm}
\begin{subfigure}[b]{0.32\textwidth}
\eq{
\mff{
\scriptsize
\begin{fmfgraph*}(40,15)
\fmfpen{thin}
\fmfstraight
\fmfleft{i1}
\fmfright{o1}
\fmf{fermion}{i1,v1a}
\fmf{fermion,tension=0.8}{v1a,v1b}
\fmf{fermion}{v1b,o1}
\fmf{photon,left,tension=0}{v1a,v1b}
\fmfdot{v1a}
\fmfdot{v1b}
\fmfv{l=$\m$,l.a=-90}{v1a}
\fmfv{l=$\m$,l.a=-90}{v1b}
\end{fmfgraph*}}
\notag}
\caption{$-i\g^\m p_\m\({8 \mathpzc{g}^2\o3\pi^2(4N)\ep}\)$}
\label{fig:fermSelfE-PhotonInt}
\end{subfigure}
\begin{subfigure}[b]{0.32\textwidth}
\eq{
\mff{
\scriptsize
\begin{fmfgraph*}(40,15)
\fmfpen{thin}
\fmfstraight
\fmfleft{i1}
\fmfright{o1}
\fmf{fermion}{i1,v1a}
\fmf{fermion,tension=0.8}{v1a,v1b}
\fmf{fermion}{v1b,o1}
\fmf{dashes,left,tension=0}{v1a,v1b}
\fmfdot{v1a}
\fmfdot{v1b}
\fmfv{l=x}{v1a}
\fmfv{l=x}{v1b}
\fmfv{l=}{v1a}
\fmfv{l=}{v1b}
\end{fmfgraph*}}
\notag}
\caption{$-i\w\g^0\( g_s\o2\pi\ep\)$}
\label{fig:fermSelfE-SigmaInt}
\end{subfigure}
\begin{subfigure}[b]{0.32\textwidth}
\eq{
\mff{
\scriptsize
\begin{fmfgraph*}(40,15)
\fmfpen{thin}
\fmfstraight
\fmfleft{i1}
\fmfright{o1}
\fmf{fermion}{i1,v1a}
\fmf{fermion,tension=0.8}{v1a,v1b}
\fmf{fermion}{v1b,o1}
\fmf{dbl_dashes,left,tension=0}{v1a,v1b}
\fmfdot{v1a}
\fmfdot{v1b}
\fmfv{l=$a$,l.a=-90}{v1a}
\fmfv{l=$a$,l.a=-90}{v1b}
\end{fmfgraph*}}
\notag}
\caption{$-i\w\g^0\( \sum_a{g_{\A,a}\o2\pi\ep}\)$}
\label{fig:fermSelfE-PhiInt}
\end{subfigure}
\\
\vspace{8mm}
\begin{subfigure}[b]{0.32\textwidth}
\eq{
\mff{
\scriptsize
\begin{fmfgraph*}(40,15)
\fmfpen{thin}
\fmfstraight
\fmfleft{i1}
\fmfright{o1}
\fmf{fermion}{i1,v1a}
\fmf{fermion,tension=0.8}{v1a,v1b}
\fmf{fermion}{v1b,o1}
\fmf{dbl_wiggly,left,tension=0}{v1a,v1b}
\fmfdot{v1a}
\fmfdot{v1b}
\fmfv{l=$a,,0$,l.a=-90}{v1a}
\fmfv{l=$a,,0$,l.a=-90}{v1b}
\end{fmfgraph*}}
\notag}
\caption{$-i\w\g^0\( \sum_a{g_{v,a}\o2\pi\ep}\)$}
\label{fig:fermSelfE-ScalarPotInt}
\end{subfigure}
\begin{subfigure}[b]{0.32\textwidth}
\eq{
\mff{
\scriptsize
\begin{fmfgraph*}(40,15)
\fmfpen{thin}
\fmfstraight
\fmfleft{i1}
\fmfright{o1}
\fmf{fermion}{i1,v1a}
\fmf{fermion,tension=0.8}{v1a,v1b}
\fmf{fermion}{v1b,o1}
\fmf{dbl_plain,left,tension=0}{v1a,v1b}
\fmfdot{v1a}
\fmfdot{v1b}
\fmfv{l=$a,,j$,l.a=-90}{v1a}
\fmfv{l=$a,,j$,l.a=-90}{v1b}
\end{fmfgraph*}}
\notag}
\caption{$-i\w\g^0\( \sum_a{g_{\A,a}\o\pi\ep}\)$}
\label{fig:fermSelfE-VectorPotInt}
\end{subfigure}

\end{fmffile}
\vspace{5mm}

%% file: 4points2.tex
\centering
\begin{fmffile}{4pts}
\fmfset{dot_size}{1.5thick}
\fmfset{arrow_len}{3mm}
\begin{subfigure}[b]{0.32\textwidth}
\eq{
\mff{
\scriptsize
\begin{fmfgraph*}(25,15)
\fmfpen{thin}
\fmfstraight
\fmfleft{i2,i1}
\fmfright{o2,o1}
\fmf{phantom}{i1,v1a}
\fmf{phantom}{i2,v2a}
\fmf{phantom,tension=.6}{v1a,v1b}
\fmf{phantom,tension=.6}{v2a,v2b}
\fmf{phantom}{v1b,o1}
\fmf{phantom}{v2b,o2}
\fmffreeze
\fmf{fermion}{i1,v1a}
\fmf{fermion}{v1a,v1b}
\fmf{fermion}{v1b,o1}
\fmf{fermion}{i2,v2a}
\fmf{fermion}{v2a,v2b}
\fmf{fermion}{v2b,o2}
\fmf{dashes}{v1a,v2a}
\fmf{dashes}{v1b,v2b}
\fmfdot{v1a}
\fmfdot{v2a}
\fmfdot{v1b}
\fmfdot{v2b}
\end{fmfgraph*}}
\notag}
\caption{$\[\g^j\otimes\g_j\]\[\id\otimes\id\]\(-{g_s^2\o4\pi\ep}\)$}
\label{fig:4ptSigmaSigmaNoX}
\end{subfigure}
\begin{subfigure}[b]{0.32\textwidth}
\eq{
\mff{
\scriptsize
\begin{fmfgraph*}(25,15)
\fmfpen{thin}
\fmfstraight
\fmfleft{i2,i1}
\fmfright{o2,o1}
\fmf{phantom}{i1,v1a}
\fmf{phantom}{i2,v2a}
\fmf{phantom,tension=.6}{v1a,v1b}
\fmf{phantom,tension=.6}{v2a,v2b}
\fmf{phantom}{v1b,o1}
\fmf{phantom}{v2b,o2}
\fmffreeze
\fmf{fermion}{i1,v1a}
\fmf{fermion}{v1a,v1b}
\fmf{fermion}{v1b,o1}
\fmf{fermion}{i2,v2a}
\fmf{fermion}{v2a,v2b}
\fmf{fermion}{v2b,o2}
\fmf{dashes}{v1a,v2b}
\fmf{dashes}{v1b,v2a}
\fmfdot{v1a}
\fmfdot{v2a}
\fmfdot{v1b}
\fmfdot{v2b}
\end{fmfgraph*}
}
\notag}
\caption{$\[\g^j\otimes\g_j\]\[\id\otimes\id\]\({ g_s^2\o4\pi\ep}\)$}
\label{fig:4ptSigmaSigmaX}
\end{subfigure}
\begin{subfigure}[b]{0.32\textwidth}
\eq{2\times
\mff{
\scriptsize
\begin{fmfgraph*}(25,15)
\fmfpen{thin}
\fmfstraight
\fmfleft{i2,i1}
\fmfright{o2,o1}
\fmf{phantom}{i1,v1a}
\fmf{phantom}{i2,v2a}
\fmf{phantom,tension=1.55}{v1a,v1c,v1b}
\fmf{phantom,tension=1.55}{v2a,v2c,v2b}
\fmf{phantom}{v1b,o1}
\fmf{phantom}{v2b,o2}
\fmffreeze
\fmf{fermion}{i1,v1a}
\fmf{fermion}{v1a,v1c}
\fmf{fermion}{v1c,v1b}
\fmf{fermion}{v1b,o1}
\fmf{fermion}{i2,v2a}
\fmf{plain}{v2a,v2c}
\fmf{plain}{v2c,v2b}
\fmf{fermion}{v2b,o2}
\fmf{dashes}{v1c,v2c}
\fmf{dashes,left,tension=0}{v1a,v1b}
\fmfdot{v1a}
\fmfdot{v1b}
\fmfdot{v1c}
\fmfdot{v2c}
\end{fmfgraph*}
}
\notag}
\caption{$\[\id\otimes\id\]\[\id\otimes\id\]\({ g_s^2\o\pi\ep}\)$}
\label{fig:4ptSigmaSigmaVert}
\end{subfigure}
\\\vspace{8mm}
\begin{subfigure}[b]{0.45\textwidth}
\eq{2\times
\mff{
\scriptsize
\begin{fmfgraph*}(25,15)
\fmfpen{thin}
\fmfstraight
\fmfleft{i2,i1}
\fmfright{o2,o1}
\fmf{phantom}{i1,v1a}
\fmf{phantom}{i2,v2a}
\fmf{phantom,tension=1.55}{v1a,v1c,v1b}
\fmf{phantom,tension=1.55}{v2a,v2c,v2b}
\fmf{phantom}{v1b,o1}
\fmf{phantom}{v2b,o2}
\fmffreeze
\fmf{fermion}{i1,v1a}
\fmf{fermion}{v1a,v1c}
\fmf{fermion}{v1c,v1b}
\fmf{fermion}{v1b,o1}
\fmf{fermion}{i2,v2a}
\fmf{plain}{v2a,v2c}
\fmf{plain}{v2c,v2b}
\fmf{fermion}{v2b,o2}
\fmf{dashes}{v1c,v2c}
\fmf{photon,left,tension=0}{v1a,v1b}
\fmfv{l=$\m$,l.a=-90}{v1b}
\fmfv{l=$\m$,l.a=-90}{v1a}
\fmfdot{v1a}
\fmfdot{v1b}
\fmfdot{v1c}
\fmfdot{v2c}
\end{fmfgraph*}
}
\notag}
\caption{$\[\id\otimes\id\]\[\id\otimes\id\]\(-{48 g_s\mathpzc{g}^2\o\pi^2(2N)\ep}\)$}
\label{fig:4ptsSigmaPhotonArc}
\end{subfigure}
\hspace{4mm}
\begin{subfigure}[b]{0.345\textwidth}
\eq{4\times
\mff{
\scriptsize
\begin{fmfgraph*}(25,20)
\fmfpen{thin}
\fmfstraight
\fmfleft{i2,i1}
\fmfright{o2,o1}
\fmf{phantom}{i1,v1a}
\fmf{phantom}{i2,v2a}
\fmf{phantom,tension=.8}{v1a,v1c,v1b}
\fmf{phantom,tension=.8}{v2a,v2c,v2b}
\fmf{phantom}{v1b,o1}
\fmf{phantom}{v2b,o2}
\fmffreeze
\fmf{phantom,tension=0.7}{v1a,l1}
\fmf{phantom,tension=0.7}{v1b,r1}
\fmf{phantom,tension=0.7}{v1c,c1}
\fmf{phantom}{l1,l2,v2a}
\fmf{phantom}{r1,r2,v2b}
\fmf{phantom}{c1,c2,v2c}
\fmffreeze
\fmf{fermion}{i1,v1c}
\fmf{fermion}{v1c,o1}
\fmf{fermion}{i2,v2a}
\fmf{fermion}{v2a,v2b}
\fmf{fermion}{v2b,o2}
\fmf{dashes}{v1c,c1}
\fmf{fermion}{c1,l2}
\fmf{fermion}{l2,r2}
\fmf{fermion}{r2,c1}
\fmf{photon}{l2,v2a}
\fmf{photon}{r2,v2b}
\fmfdot{v1c}
\fmfdot{v2a}
\fmfdot{v2b}
\fmfdot{c1}
\fmfdot{r2}
\fmfdot{l2}
\fmfv{l=$\m$,l.a=0}{r2}
\fmfv{l=$\m$,l.a=-90}{v2b}
\fmfv{l=$\n$,l.a=180}{l2}
\fmfv{l=$\n$,l.a=-90}{v2a}
\end{fmfgraph*}
}
\notag}
\caption{$\[\id\otimes\id\]\[\id\otimes\id\]\(64 g_s\mathpzc{g}^4\o\pi^2(2N)\ep\)$}
\label{fig:FermLoopSigmaInt1}
\end{subfigure}
\end{fmffile}
\vspace{5mm}

%% file: gBSelfEnergy.tex
\centering
\begin{fmffile}{gBSelfEnergy}
\fmfset{dot_size}{1.5thick}
\fmfset{arrow_len}{3mm}
\eq{
\mff{
\scriptsize
\begin{fmfgraph*}(40,10)
\fmfpen{thin}
\fmfstraight
\fmfleft{i1}
\fmfright{o1}
\fmf{fermion}{i1,v1a}
\fmf{fermion,tension=0.8}{v1a,v1b}
\fmf{fermion}{v1b,o1}
\fmffreeze
\fmfpoly{phantom,smooth,pull=?,tension=2.7}{v1b,v2,v1a,v3}
\fmffreeze
\fmf{photon,left=.5}{v1a,v2}
\fmf{photon,left=.5}{v2,v1b}
\fmfdot{v1a}
\fmfdot{v1b}
\fmfv{l=$0$,l.a=-90}{v1a}
\fmfv{l=$0$,l.a=-90}{v1b}
\fmfv{decor.shape=diamond,decor.filled=shaded,decor.size=4mm}{v2}
\fmf{phantom}{v2,v2X,v2}
\fmfv{l=$0$,l.a=-135}{v2}
\fmfv{l=$0$,l.a=-45}{v2X}
\end{fmfgraph*}}
\notag}
\end{fmffile}

%% file: currentDiagram.tex
\begin{fmffile}{currentDiagrams}
\fmfset{dot_size}{1.5thick}
\fmfset{arrow_len}{3mm}
\begin{subfigure}[b]{0.45\textwidth}
\eq{
\mff{
\small
\begin{fmfgraph*}(35,20)
\fmfcurved
\fmfleft{i3}
\fmfright{o3}
\fmflabel{$\mu,r,\a$}{i3}
\fmflabel{$\nu,s,\b$}{o3}
\fmftop{o2,y2,x2,z2,i2}
\fmfbottom{o1,y1,x1,z1,i1}
\fmf{phantom}{i1,y1,x1,z1,o1}
\fmf{phantom}{o2,x2,i2}
\fmfv{decor.size=1.5mm,decor.shape=square,decor.filled=gray50}{i3}
\fmfv{decor.size=1.5mm,decor.shape=square,decor.filled=gray50}{o3}
\fmffreeze
\fmf{fermion,right=0.6,label={\scriptsize$q+p$}}{i3,o3}
\fmf{fermion,right=0.6,label={\scriptsize$q$}}{o3,i3}
\end{fmfgraph*}}
\notag\vspace{5mm}}
\caption{}
\label{fig:currentLoopBare}
\end{subfigure}
\vspace{5mm}
\\
\begin{subfigure}[b]{0.45\textwidth}
\eq{
\mff{
\small
\begin{fmfgraph*}(35,20)
\fmfcurved
\fmfleft{i1,i3,i2}
\fmfright{o1,o3,o2}
\fmflabel{$\mu,r,\a$}{i3}
\fmflabel{$\nu,s,\b$}{o3}
\fmf{phantom}{i1,x1,o1}
\fmf{phantom}{o2,x2,i2}
\fmfv{decor.size=1.5mm,decor.shape=square,decor.filled=gray50}{i3}
\fmfv{decor.size=1.5mm,decor.shape=square,decor.filled=gray50}{o3}
\fmffreeze
\fmf{fermion,left=0.3,tension=0.3,label={\scriptsize$q$}}{x1,i3}
\fmf{fermion,left=0.3,tension=0.3,label={\scriptsize$q+k$}}{o3,x1}
\fmf{fermion,left=0.3,tension=0.3,label={\scriptsize$q+k+p$}}{x2,o3}
\fmf{fermion,left=0.3,tension=0.3,label={\scriptsize$q+p$}}{i3,x2}
\fmf{dbl_dashes,label={\scriptsize$k$}}{x1,x2}
\fmfdot{x1}
\fmfdot{x2}
\fmfv{l={\scriptsize$z$}}{x1}
\fmfv{l={\scriptsize$z$}}{x2}
\end{fmfgraph*}}
\notag\vspace{5mm}}
\caption{}
\label{fig:currentLoopVert}
\end{subfigure}
\vspace{5mm}
\begin{subfigure}[b]{0.45\textwidth}
\eq{
\mff{
\small
\begin{fmfgraph*}(35,20)
\fmfcurved
\fmfleft{i3}
\fmfright{o3}
\fmftop{o2,y2,x2,z2,i2}
\fmfbottom{o1,y1,x1,z1,i1}
\fmf{phantom}{i1,y1,x1,z1,o1}
\fmf{phantom}{o2,x2,i2}
\fmfv{decor.size=1.5mm,decor.shape=square,decor.filled=gray50}{i3}
\fmfv{decor.size=1.5mm,decor.shape=square,decor.filled=gray50}{o3}
\fmffreeze
\fmf{fermion,right=0.6,label={\scriptsize$q+p$}}{i3,o3}
\fmf{fermion,right=0.2,tension=0.3,label={\scriptsize$q$}}{y2,i3}
\fmf{fermion,right=0.2,label={\scriptsize$q+k$}}{z2,y2}
\fmf{fermion,right=0.2,label={\scriptsize$q$}}{o3,z2}
\fmf{dbl_dashes,right=0.7,tension=0.2,label={\scriptsize$k$}}{y2,z2}
\fmfv{label=$\mu,,r,,\a$}{i3}
\fmfv{label=$\nu,,s,,\b$}{o3}
\fmfv{label=${\scriptsize z}$}{y2}
\fmfv{label=${\scriptsize z}$}{z2}
\fmfdot{z2}
\fmfdot{y2}
\end{fmfgraph*}}
\notag\vspace{5mm}}
\caption{}
\label{fig:currentLoopSelfE}
\end{subfigure}
\vspace{5mm}
\\
\begin{subfigure}[b]{0.45\textwidth}
\eq{
\mff{
\small
\begin{fmfgraph*}(35,20)
\fmfcurved
\fmfleft{i1,i3,i2}
\fmfright{o1,o3,o2}
\fmflabel{$\mu,r,\a$}{i3}
\fmflabel{$\nu,s,\b$}{o3}
\fmf{phantom}{i1,x1,o1}
\fmf{phantom}{o2,x2,i2}
\fmfv{decor.size=1.5mm,decor.shape=square,decor.filled=gray50}{i3}
\fmfv{decor.size=1.5mm,decor.shape=square,decor.filled=gray50}{o3}
\fmffreeze
\fmf{fermion,left=0.3,tension=0.3,label={\scriptsize$q$}}{x1,i3}
\fmf{fermion,left=0.3,tension=0.3,label={\scriptsize$q+k$}}{o3,x1}
\fmf{fermion,left=0.3,tension=0.3,label={\scriptsize$q+k+p$}}{x2,o3}
\fmf{fermion,left=0.3,tension=0.3,label={\scriptsize$q+p$}}{i3,x2}
\fmf{dbl_plain,label={\scriptsize$k$}}{x1,x2}
\fmfdot{x1}
\fmfdot{x2}
\fmfv{l={\scriptsize$z,,j$}}{x1}
\fmfv{l={\scriptsize$z,,j$}}{x2}
\end{fmfgraph*}}
\notag\vspace{5mm}}
\caption{}
\label{fig:currentLoopVert2}
\end{subfigure}
\vspace{5mm}
\begin{subfigure}[b]{0.45\textwidth}
\eq{
\mff{
\small
\begin{fmfgraph*}(35,20)
\fmfcurved
\fmfleft{i3}
\fmfright{o3}
\fmftop{o2,y2,x2,z2,i2}
\fmfbottom{o1,y1,x1,z1,i1}
\fmf{phantom}{i1,y1,x1,z1,o1}
\fmf{phantom}{o2,x2,i2}
\fmfv{decor.size=1.5mm,decor.shape=square,decor.filled=gray50}{i3}
\fmfv{decor.size=1.5mm,decor.shape=square,decor.filled=gray50}{o3}
\fmffreeze
\fmf{fermion,right=0.6,label={\scriptsize$q+p$}}{i3,o3}
\fmf{fermion,right=0.2,tension=0.3,label={\scriptsize$q$}}{y2,i3}
\fmf{fermion,right=0.2,label={\scriptsize$q+k$}}{z2,y2}
\fmf{fermion,right=0.2,label={\scriptsize$q$}}{o3,z2}
\fmf{dbl_plain,right=0.7,tension=0.2,label={\scriptsize$k$}}{y2,z2}
\fmfv{label=$\mu,,r,,\a$}{i3}
\fmfv{label=$\nu,,s,,\b$}{o3}
\fmfv{label=${\scriptsize z,,j}$}{y2}
\fmfv{label=${\scriptsize z,,j}$}{z2}
\fmfdot{z2}
\fmfdot{y2}
\end{fmfgraph*}}
\notag\vspace{5mm}}
\caption{}
\label{fig:currentLoopSelfE2}
\end{subfigure}
\vspace{5mm}
\\
\begin{subfigure}[b]{0.45\textwidth}
\eq{
\mff{
\small
\begin{fmfgraph*}(35,20)
\fmfcurved
\fmfleft{i1,i3,i2}
\fmfright{o1,o3,o2}
\fmflabel{$\mu,r,\a$}{i3}
\fmfv{decor.size=1.5mm,decor.shape=square,decor.filled=gray50}{i3}
\fmfv{decor.size=1.5mm,decor.shape=square,decor.filled=gray50}{o3}
\fmflabel{$\nu,s,\b$}{o3}
\fmf{phantom}{i1,x1,o1}
\fmf{phantom}{o2,x2,i2}
\fmffreeze
\fmf{fermion,left=0.3,tension=0.3,label={\scriptsize$q$}}{x1,i3}
\fmf{fermion,left=0.3,tension=0.3,label={\scriptsize$q+k$}}{o3,x1}
\fmf{fermion,left=0.3,tension=0.3,label={\scriptsize$q+k+p$}}{x2,o3}
\fmf{fermion,left=0.3,tension=0.3,label={\scriptsize$q+p$}}{i3,x2}
\fmf{photon,label={\scriptsize$k$}}{x1,x2}
\fmfdot{x1}
\fmfdot{x2}
\fmfv{l={\scriptsize$\s$}}{x1}
\fmfv{l={\scriptsize$\s$}}{x2}
\end{fmfgraph*}}
\notag
\vspace{5mm}}
\caption{}
\label{fig:photonVertex}
\end{subfigure}
\begin{subfigure}[b]{0.45\textwidth}
\eq{
\mff{
\small
\begin{fmfgraph*}(35,20)
\fmfcurved
\fmfleft{i3}
\fmfright{o3}
\fmftop{o2,y2,x2,z2,i2}
\fmfbottom{o1,y1,x1,z1,i1}
\fmf{phantom}{i1,y1,x1,z1,o1}
\fmf{phantom}{o2,x2,i2}
\fmfv{decor.size=1.5mm,decor.shape=square,decor.filled=gray50}{i3}
\fmfv{decor.size=1.5mm,decor.shape=square,decor.filled=gray50}{o3}
\fmffreeze
\fmf{fermion,right=0.6,label={\scriptsize$q+p$}}{i3,o3}
\fmf{fermion,right=0.2,tension=0.3,label={\scriptsize$q$}}{y2,i3}
\fmf{fermion,right=0.2,label={\scriptsize$q+k$}}{z2,y2}
\fmf{fermion,right=0.2,label={\scriptsize$q$}}{o3,z2}
\fmf{photon,right=0.7,tension=0.2,label={\scriptsize$k$}}{y2,z2}
\fmfv{label=$\mu,,r,,\a$}{i3}
\fmfv{label=$\nu,,s,,\b$}{o3}
\fmfv{label={\scriptsize$\s$}}{y2}
\fmfv{label={\scriptsize$\s$}}{z2}
\fmfdot{z2}
\fmfdot{y2}
\end{fmfgraph*}}
\notag\vspace{5mm}}
\caption{}
\label{fig:photonSelfE}
\end{subfigure}
\end{fmffile}
\vspace{5mm}

%% file: feynRulesNoFlavour.tex
\centering
\begin{fmffile}{feynRulesNoFlavour}
\fmfset{zigzag_width}{.7mm}
\fmfset{dot_size}{1.5thick}
\fmfset{arrow_len}{3mm}
\begin{subfigure}[b]{0.45\textwidth}
\centering
\eq{
\mff{
\scriptsize
\begin{fmfgraph*}(27,20)
\fmfpen{thin}
\fmfstraight
\fmfleft{i2,i1}
\fmfright{o2,o1}
\fmfv{label=$a,,\ell$}{i1}
\fmfv{label=$b,,\ell$}{o1}
\fmfv{label=$d ,,m$}{i2}
\fmfv{label=$c,, m$}{o2}
\fmf{phantom}{i1,v1,o1}
\fmf{phantom}{i2,v2,o2}
\fmffreeze
\fmf{fermion}{i1,v1}
\fmf{fermion}{v1,o1}
\fmf{fermion}{i2,v2}
\fmf{fermion}{v2,o2}
\fmf{dbl_dots,label=$q$}{v1,v2}
\fmfdot{v1}
\fmfdot{v2}
\end{fmfgraph*}
}
=2\pi\d(q_0)\[\id\]_{ab}\[\id\]_{cd}
\notag}
\vspace{8mm}
\end{subfigure}
\hspace{3mm}
\begin{subfigure}[b]{0.45\textwidth}
\centering
\eq{
\mff{
\scriptsize
\begin{fmfgraph*}(27,20)
\fmfpen{thin}
\fmfstraight
\fmfleft{i2,i1}
\fmfright{o2,o1}
\fmfv{label=$b,,\ell$}{i1}
\fmfv{label=$a,,\ell$}{o1}
\fmfv{label=$d ,,m$}{i2}
\fmfv{label=$c,, m$}{o2}
\fmf{phantom}{i1,v1,o1}
\fmf{phantom}{i2,v2,o2}
\fmffreeze
\fmf{fermion}{i1,v1}
\fmf{fermion}{v1,o1}
\fmf{fermion}{i2,v2}
\fmf{fermion}{v2,o2}
\fmf{zigzag,label=$q$}{v1,v2}
\fmfdot{v1}
\fmfdot{v2}
\fmfv{l=$\m$}{v1}
\fmfv{l=$\m$}{v2}
\end{fmfgraph*}
}
=2\pi\d(q_0)\[i\g^\m\]_{ab}\[i\g^\m\]_{cd}
\notag}
\vspace{8mm}
\end{subfigure}
\end{fmffile}

%% file: integralDiagrams.tex
\centering
\begin{fmffile}{intdiagrams}
\fmfset{dot_size}{1.5thick}
\fmfset{arrow_len}{3mm}
\begin{subfigure}[b]{0.24\textwidth}
\eq{
\mff{
\scriptsize
\begin{fmfgraph*}(25,15)
\fmfpen{thin}
\fmfstraight
\fmfleft{i2,i1}
\fmfright{o2,o1}
\fmf{phantom}{v1a,i1}
\fmf{phantom}{i2,v2a}
\fmf{phantom,tension=.6}{v1a,v1b}
\fmf{phantom,tension=.6}{v2a,v2b}
\fmf{phantom}{v1b,o1}
\fmf{phantom}{v2b,o2}
\fmffreeze
\fmf{fermion}{i1,v1a}
\fmf{fermion}{v1a,v1b}
\fmf{fermion}{v1b,o1}
\fmf{fermion}{i2,v2a}
\fmf{fermion}{v2a,v2b}
\fmf{fermion}{v2b,o2}
\fmf{dbl_dots}{v1a,v2a}
\fmf{dbl_dots}{v2b,v1b}
\fmfdot{v1a}
\fmfdot{v1b}
\fmfdot{v2a}
\fmfdot{v2b}
\end{fmfgraph*}}
\notag}\vspace{5mm}
\caption{${ \g^j\otimes\g_j}\(-{1\o4\pi\ep}\)$}
\label{fig:4ptNoX}
\end{subfigure}
\begin{subfigure}[b]{0.24\textwidth}
\eq{
\mff{
\scriptsize
\begin{fmfgraph*}(25,15)
\fmfpen{thin}
\fmfstraight
\fmfleft{i2,i1}
\fmfright{o2,o1}
\fmf{phantom}{v1a,i1}
\fmf{phantom}{i2,v2a}
\fmf{phantom,tension=.6}{v1a,v1b}
\fmf{phantom,tension=.6}{v2a,v2b}
\fmf{phantom}{v1b,o1}
\fmf{phantom}{v2b,o2}
\fmffreeze
\fmf{fermion}{i1,v1a}
\fmf{fermion}{v1a,v1b}
\fmf{fermion}{v1b,o1}
\fmf{fermion}{i2,v2a}
\fmf{fermion}{v2a,v2b}
\fmf{fermion}{v2b,o2}
\fmf{dbl_dots}{v1a,v2b}
\fmf{dbl_dots}{v1b,v2a}
\fmfdot{v1a}
\fmfdot{v1b}
\fmfdot{v2a}
\fmfdot{v2b}
\end{fmfgraph*}
}
\notag}\vspace{5mm}
\caption{${\g^j\otimes\g_j}\({1\o4\pi\ep}\)$}
\label{fig:4ptX}
\end{subfigure}
\begin{subfigure}[b]{0.24\textwidth}
\eq{
\mff{
\scriptsize
\begin{fmfgraph*}(25,15)
\fmfpen{thin}
\fmfstraight
\fmfleft{i2,i1}
\fmfright{o2,o1}
\fmf{phantom}{i1,v1a}
\fmf{phantom}{i2,v2a}
\fmf{phantom,tension=1.55}{v1a,v1c,v1b}
\fmf{phantom,tension=1.55}{v2a,v2c,v2b}
\fmf{phantom}{v1b,o1}
\fmf{phantom}{v2b,o2}
\fmffreeze
\fmf{fermion}{i1,v1a}
\fmf{fermion}{v1a,v1c}
\fmf{fermion}{v1c,v1b}
\fmf{fermion}{v1b,o1}
\fmf{fermion}{i2,v2a}
\fmf{plain}{v2a,v2c}
\fmf{plain}{v2c,v2b}
\fmf{fermion}{v2b,o2}
\fmf{dbl_dots}{v1c,v2c}
\fmf{dbl_dots,left,tension=0}{v1a,v1b}
\fmfdot{v1a}
\fmfdot{v1b}
\fmfdot{v1c}
\fmfdot{v2c}
\end{fmfgraph*}
}
\notag}
\vspace{5mm}
\caption{${\id\otimes\id}\({1\o2\pi\ep}\)$}
\label{fig:4ptVert}
\end{subfigure}
\begin{subfigure}[b]{0.24\textwidth}
\eq{
\mff{
\scriptsize
\begin{fmfgraph*}(25,15)
\fmfpen{thin}
\fmfstraight
\fmfleft{i2,i1}
\fmfright{o2,o1}
\fmf{phantom}{i1,v1a}
\fmf{phantom}{i2,v2a}
\fmf{phantom,tension=1.55}{v1a,v1c,v1b}
\fmf{phantom,tension=1.55}{v2a,v2c,v2b}
\fmf{phantom}{v1b,o1}
\fmf{phantom}{v2b,o2}
\fmffreeze
\fmf{fermion}{i1,v1a}
\fmf{plain}{v1a,v1c}
\fmf{plain}{v1c,v1b}
\fmf{fermion}{v1b,o1}
\fmf{fermion}{i2,v2a}
\fmf{fermion}{v2a,v2c}
\fmf{fermion}{v2c,v2b}
\fmf{fermion}{v2b,o2}
\fmf{dbl_dots}{v1c,v2c}
\fmf{dbl_dots,right,tension=0}{v2a,v2b}
\fmfdot{v2a}
\fmfdot{v2b}
\fmfdot{v1c}
\fmfdot{v2c}
\end{fmfgraph*}
}
\notag}
\vspace{5mm}
\caption{${\id\otimes\id}\({1\o2\pi\ep}\)$}
\label{fig:4ptVert2}
\end{subfigure}
\\\vspace{5mm}
\begin{subfigure}[b]{0.24\textwidth}
\eq{
\mff{
\scriptsize
\begin{fmfgraph*}(25,15)
\fmfpen{thin}
\fmfstraight
\fmfleft{i2,i1}
\fmfright{o2,o1}
\fmf{phantom}{i1,v1a}
\fmf{phantom}{i2,v2a}
\fmf{phantom,tension=.6}{v1a,v1b}
\fmf{phantom,tension=.6}{v2a,v2b}
\fmf{phantom}{v1b,o1}
\fmf{phantom}{v2b,o2}
\fmffreeze
\fmf{fermion}{i1,v1a}
\fmf{fermion}{v1a,v1b}
\fmf{fermion}{v1b,o1}
\fmf{fermion}{i2,v2a}
\fmf{fermion}{v2a,v2b}
\fmf{fermion}{v2b,o2}
\fmf{dbl_dots}{v1a,v2a}
\fmf{photon}{v1b,v2b}
\fmfv{l=$\m$}{v1b}	
\fmfv{l=$\m$}{v2b}
\fmfdot{v1a}
\fmfdot{v1b}
\fmfdot{v2a}
\fmfdot{v2b}
\end{fmfgraph*}}
\notag}\vspace{5mm}
\caption{Convergent}
\label{fig:4ptSigmaPhotonNoX}
\end{subfigure}
\begin{subfigure}[b]{0.24\textwidth}
\eq{
\mff{
\scriptsize
\begin{fmfgraph*}(25,15)
\fmfpen{thin}
\fmfstraight
\fmfleft{i2,i1}
\fmfright{o2,o1}
\fmf{phantom}{i1,v1a}
\fmf{phantom}{i2,v2a}
\fmf{phantom,tension=.6}{v1a,v1b}
\fmf{phantom,tension=.6}{v2a,v2b}
\fmf{phantom}{v1b,o1}
\fmf{phantom}{v2b,o2}
\fmffreeze
\fmf{fermion}{i1,v1a}
\fmf{fermion}{v1a,v1b}
\fmf{fermion}{v1b,o1}
\fmf{fermion}{i2,v2a}
\fmf{fermion}{v2a,v2b}
\fmf{fermion}{v2b,o2}
\fmf{dbl_dots}{v1a,v2b}
\fmf{photon}{v1b,v2a}
\fmfv{l=$\m$}{v1b}	
\fmfv{l=$\m$}{v2a}
\fmfdot{v1a}
\fmfdot{v1b}
\fmfdot{v2a}
\fmfdot{v2b}
\end{fmfgraph*}
}
\notag}\vspace{5mm}
\caption{Convergent}
\label{fig:4ptSigmaPhotonX}
\end{subfigure}
\begin{subfigure}[b]{0.24\textwidth}
\eq{
\mff{
\scriptsize
\begin{fmfgraph*}(25,15)
\fmfpen{thin}
\fmfstraight
\fmfleft{i2,i1}
\fmfright{o2,o1}
\fmf{phantom}{i1,v1a}
\fmf{phantom}{i2,v2a}
\fmf{phantom,tension=1.55}{v1a,v1c,v1b}
\fmf{phantom,tension=1.55}{v2a,v2c,v2b}
\fmf{phantom}{v1b,o1}
\fmf{phantom}{v2b,o2}
\fmffreeze
\fmf{fermion}{i1,v1a}
\fmf{fermion}{v1a,v1c}
\fmf{fermion}{v1c,v1b}
\fmf{fermion}{v1b,o1}
\fmf{fermion}{i2,v2a}
\fmf{plain}{v2a,v2c}
\fmf{plain}{v2c,v2b}
\fmf{fermion}{v2b,o2}
\fmf{dbl_dots}{v1c,v2c}
\fmf{photon,left,tension=0}{v1a,v1b}
\fmfv{l=$\m$}{v1b}
\fmfv{l=$\m$}{v1a}
\fmfdot{v1a}
\fmfdot{v1b}
\fmfdot{v1c}
\fmfdot{v2c}
\end{fmfgraph*}
}
\notag}\vspace{5mm}
\caption{$\id\otimes\id\(-{24\mathpzc{g}^2\o\pi^2(2N)\ep}\)$}
\label{fig:4ptsPhotonArc}
\end{subfigure}
\begin{subfigure}[b]{0.24\textwidth}
\eq{
\mff{
\scriptsize
\begin{fmfgraph*}(25,15)
\fmfpen{thin}
\fmfstraight
\fmfleft{i2,i1}
\fmfright{o2,o1}
\fmf{phantom}{i1,v1a}
\fmf{phantom}{i2,v2a}
\fmf{phantom,tension=1.55}{v1a,v1c,v1b}
\fmf{phantom,tension=1.55}{v2a,v2c,v2b}
\fmf{phantom}{v1b,o1}
\fmf{phantom}{v2b,o2}
\fmffreeze
\fmf{fermion}{i1,v1a}
\fmf{fermion}{v2a,v2c}
\fmf{fermion}{v2c,v2b}
\fmf{fermion}{v1b,o1}
\fmf{fermion}{i2,v2a}
\fmf{plain}{v1a,v1c}
\fmf{plain}{v1c,v1b}
\fmf{fermion}{v2b,o2}
\fmf{dbl_dots}{v1c,v2c}
\fmf{photon,right,tension=0}{v2a,v2b}
\fmfv{l=$\m$}{v2b}
\fmfv{l=$\m$}{v2a}
\fmfdot{v2a}
\fmfdot{v2b}
\fmfdot{v1c}
\fmfdot{v2c}
\end{fmfgraph*}
}
\notag}\vspace{5mm}
\caption{$\id\otimes\id\(-{24\mathpzc{g}^2\o\pi^2(2N)\ep}\)$}
\label{fig:4ptsPhotonArc2}
\end{subfigure}
\\\vspace{5mm}
\begin{subfigure}[b]{0.24\textwidth}
\eq{
\mff{
\scriptsize
\begin{fmfgraph*}(25,20)
\fmfpen{thin}
\fmfstraight
\fmfleft{i2,i1}
\fmfright{o2,o1}
\fmf{phantom}{i1,v1a}
\fmf{phantom}{i2,v2a}
\fmf{phantom,tension=.8}{v1a,v1c,v1b}
\fmf{phantom,tension=.8}{v2a,v2c,v2b}
\fmf{phantom}{v1b,o1}
\fmf{phantom}{v2b,o2}
\fmffreeze
\fmf{phantom,tension=0.7}{v1a,l1}
\fmf{phantom,tension=0.7}{v1b,r1}
\fmf{phantom,tension=0.7}{v1c,c1}
\fmf{phantom}{l1,l2,v2a}
\fmf{phantom}{r1,r2,v2b}
\fmf{phantom}{c1,c2,v2c}
\fmffreeze
\fmf{fermion}{i1,v1c}
\fmf{fermion}{v1c,o1}
\fmf{fermion}{i2,v2a}
\fmf{fermion}{v2a,v2b}
\fmf{fermion}{v2b,o2}
\fmf{dbl_dots}{v1c,c1}
\fmf{fermion}{c1,l2}
\fmf{fermion}{l2,r2}
\fmf{fermion}{r2,c1}
\fmf{photon}{l2,v2a}
\fmf{photon}{r2,v2b}
\fmfv{l=$\m$}{v2b}
\fmfv{l=$\m$}{r2}
\fmfv{l=$\n$}{l2}
\fmfv{l=$\n$}{v2a}
\fmfdot{v1c}
\fmfdot{c1}
\fmfdot{l2}
\fmfdot{r2}
\fmfdot{v2a}
\fmfdot{v2b}
\end{fmfgraph*}
}
\notag}
\caption{$\id\otimes\id\({16\mathpzc{g}^4\o\pi^2(2N)\ep}\)$}
\label{fig:FermLoopPhotonPhoton1}
\end{subfigure}
\begin{subfigure}[b]{0.24\textwidth}
\eq{
\mff{
\scriptsize
\begin{fmfgraph*}(25,20)
\fmfpen{thin}
\fmfstraight
\fmfleft{i2,i1}
\fmfright{o2,o1}
\fmf{phantom}{i1,v1a}
\fmf{phantom}{i2,v2a}
\fmf{phantom,tension=.8}{v1a,v1c,v1b}
\fmf{phantom,tension=.8}{v2a,v2c,v2b}
\fmf{phantom}{v1b,o1}
\fmf{phantom}{v2b,o2}
\fmffreeze
\fmf{phantom,tension=0.7}{v1a,l1}
\fmf{phantom,tension=0.7}{v1b,r1}
\fmf{phantom,tension=0.7}{v1c,c1}
\fmf{phantom}{l1,l2,v2a}
\fmf{phantom}{r1,r2,v2b}
\fmf{phantom}{c1,c2,v2c}
\fmffreeze
\fmf{fermion}{i1,v1c}
\fmf{fermion}{v1c,o1}
\fmf{fermion}{i2,v2a}
\fmf{fermion}{v2a,v2b}
\fmf{fermion}{v2b,o2}
\fmf{dbl_dots}{v1c,c1}
\fmf{fermion}{c1,r2}
\fmf{fermion}{r2,l2}
\fmf{fermion}{l2,c1}
\fmf{photon}{l2,v2a}
\fmf{photon}{r2,v2b}
\fmfv{l=$\m$}{v2b}
\fmfv{l=$\m$}{r2}
\fmfv{l=$\n$}{l2}
\fmfv{l=$\n$}{v2a}
\fmfdot{v1c}
\fmfdot{c1}
\fmfdot{l2}
\fmfdot{r2}
\fmfdot{v2a}
\fmfdot{v2b}
\end{fmfgraph*}
}
\notag}
\caption{$\id\otimes\id\({16\mathpzc{g}^4\o\pi^2(2N)\ep}\)$}
\label{fig:FermLoopPhotonPhoton2}
\end{subfigure}
\begin{subfigure}[b]{0.24\textwidth}
\eq{
\mff{
\scriptsize
\begin{fmfgraph*}(25,20)
\fmfpen{thin}
\fmfstraight
\fmfleft{i2,i1}
\fmfright{o2,o1}
\fmf{phantom}{i1,v1a}
\fmf{phantom}{i2,v2a}
\fmf{phantom,tension=.8}{v1a,v1c,v1b}
\fmf{phantom,tension=.8}{v2a,v2c,v2b}
\fmf{phantom}{v1b,o1}
\fmf{phantom}{v2b,o2}
\fmffreeze
\fmf{phantom,tension=0.7}{v1a,l1}
\fmf{phantom,tension=0.7}{v1b,r1}
\fmf{phantom,tension=0.7}{v1c,c1}
\fmf{phantom}{l1,l2,v2a}
\fmf{phantom}{r1,r2,v2b}
\fmf{phantom}{c1,c2,v2c}
\fmffreeze
\fmf{fermion}{i1,v1c}
\fmf{fermion}{v1c,o1}
\fmf{fermion}{i2,v2a}
\fmf{fermion}{v2a,v2b}
\fmf{fermion}{v2b,o2}
\fmf{dbl_dots}{v1c,c1}
\fmf{fermion}{c1,l2}
\fmf{fermion}{l2,r2}
\fmf{fermion}{r2,c1}
\fmf{photon}{l2,v2a}
\fmf{dbl_dots}{r2,v2b}
\fmfv{l=$\m$}{v2a}
\fmfv{l=$\m$}{l2}
\fmfdot{v1c}
\fmfdot{c1}
\fmfdot{l2}
\fmfdot{r2}
\fmfdot{v2a}
\fmfdot{v2b}
\end{fmfgraph*}
}
\notag}
\caption{Cancels Fig.~\ref{fig:FermLoopPhotonDisorder2}}
\label{fig:FermLoopPhotonDisorder1}
\end{subfigure}
\begin{subfigure}[b]{0.24\textwidth}
\eq{
\mff{
\scriptsize
\begin{fmfgraph*}(25,20)
\fmfpen{thin}
\fmfstraight
\fmfleft{i2,i1}
\fmfright{o2,o1}
\fmf{phantom}{i1,v1a}
\fmf{phantom}{i2,v2a}
\fmf{phantom,tension=.8}{v1a,v1c,v1b}
\fmf{phantom,tension=.8}{v2a,v2c,v2b}
\fmf{phantom}{v1b,o1}
\fmf{phantom}{v2b,o2}
\fmffreeze
\fmf{phantom,tension=0.7}{v1a,l1}
\fmf{phantom,tension=0.7}{v1b,r1}
\fmf{phantom,tension=0.7}{v1c,c1}
\fmf{phantom}{l1,l2,v2a}
\fmf{phantom}{r1,r2,v2b}
\fmf{phantom}{c1,c2,v2c}
\fmffreeze
\fmf{fermion}{i1,v1c}
\fmf{fermion}{v1c,o1}
\fmf{fermion}{i2,v2a}
\fmf{fermion}{v2a,v2b}
\fmf{fermion}{v2b,o2}
\fmf{dbl_dots}{v1c,c1}
\fmf{fermion}{c1,r2}
\fmf{fermion}{r2,l2}
\fmf{fermion}{l2,c1}
\fmf{photon}{l2,v2a}
\fmf{dbl_dots}{r2,v2b}
\fmfv{l=$\m$}{l2}
\fmfv{l=$\m$}{v2a}
\fmfdot{v1c}
\fmfdot{c1}
\fmfdot{l2}
\fmfdot{r2}
\fmfdot{v2a}
\fmfdot{v2b}
\end{fmfgraph*}
}
\notag}
\caption{Cancels Fig.~\ref{fig:FermLoopPhotonDisorder1}}
\label{fig:FermLoopPhotonDisorder2}
\end{subfigure}

\end{fmffile}

%% file: integralDiagrams2.tex
\centering
\begin{fmffile}{intdiagrams2}
\fmfset{zigzag_width}{.7mm}
\fmfset{dot_size}{1.5thick}
\fmfset{arrow_len}{3mm}
\begin{subfigure}[b]{0.49\textwidth}
\eq{
\mff{
\scriptsize
\begin{fmfgraph*}(25,15)
\fmfpen{thin}
\fmfstraight
\fmfleft{i2,i1}
\fmfright{o2,o1}
\fmf{phantom}{i1,v1a}
\fmf{phantom}{i2,v2a}
\fmf{phantom,tension=.6}{v1a,v1b}
\fmf{phantom,tension=.6}{v2a,v2b}
\fmf{phantom}{v1b,o1}
\fmf{phantom}{v2b,o2}
\fmffreeze
\fmf{fermion}{i1,v1a}
\fmf{fermion}{v1a,v1b}
\fmf{fermion}{v1b,o1}
\fmf{fermion}{i2,v2a}
\fmf{fermion}{v2a,v2b}
\fmf{fermion}{v2b,o2}
\fmf{zigzag}{v1a,v2a}
\fmf{zigzag}{v1b,v2b}
\fmfdot{v1a}
\fmfdot{v1b}
\fmfdot{v2a}
\fmfdot{v2b}
\fmfv{l=$\mu$}{v1b}
\fmfv{l=$\m$}{v2b}
\fmfv{l=$\nu$}{v1a}
\fmfv{l=$\n$}{v2a}
\end{fmfgraph*}}
\notag}
\caption{$\({1\o4\pi\ep}\)
\begin{cases}
-\sum_j\g^j\otimes\g_j, & \(\m,\n\)=\(0,0\),\(k,\ell\) \\
\id\otimes\id-\g^0\otimes\g^0, & \(\m,\n\)=\(0,\ell\),\(\ell,0\) \\
\end{cases}$
}
\label{fig:4ptGamGamNoX}
\end{subfigure}
\begin{subfigure}[b]{0.49\textwidth}
\eq{
\mff{
\scriptsize
\begin{fmfgraph*}(25,15)
\fmfpen{thin}
\fmfstraight
\fmfleft{i2,i1}
\fmfright{o2,o1}
\fmf{phantom}{i1,v1a}
\fmf{phantom}{i2,v2a}
\fmf{phantom,tension=.6}{v1a,v1b}
\fmf{phantom,tension=.6}{v2a,v2b}
\fmf{phantom}{v1b,o1}
\fmf{phantom}{v2b,o2}
\fmffreeze
\fmf{fermion}{i1,v1a}
\fmf{fermion}{v1a,v1b}
\fmf{fermion}{v1b,o1}
\fmf{fermion}{i2,v2a}
\fmf{fermion}{v2a,v2b}
\fmf{fermion}{v2b,o2}
\fmf{zigzag}{v1a,v2b}
\fmf{zigzag}{v1b,v2a}
\fmfdot{v1a}
\fmfdot{v1b}
\fmfdot{v2a}
\fmfdot{v2b}
\fmfv{l=$\mu$}{v1b}
\fmfv{l=$\m$}{v2a}
\fmfv{l=$\nu$}{v1a}
\fmfv{l=$\n$}{v2b}
\end{fmfgraph*}
}
\notag}
\caption{$
\({1\o4\pi\ep}\)
\begin{cases}
\sum_j\g^j\otimes\g_j, & \(\m,\n\)=\(0,0\),\(k,\ell\) \\
\id\otimes\id+\g^0\otimes\g^0, & \(\m,\n\)=\(0,\ell\),\(\ell,0\) \\
\end{cases}$}
\label{fig:4ptGamGamX}
\end{subfigure}
\\\vspace{5mm}
\begin{subfigure}[b]{0.32\textwidth}
\eq{
\mff{
\scriptsize
\begin{fmfgraph*}(25,15)
\fmfpen{thin}
\fmfstraight
\fmfleft{i2,i1}
\fmfright{o2,o1}
\fmf{phantom}{i1,v1a}
\fmf{phantom}{i2,v2a}
\fmf{phantom,tension=1.55}{v1a,v1c,v1b}
\fmf{phantom,tension=1.55}{v2a,v2c,v2b}
\fmf{phantom}{v1b,o1}
\fmf{phantom}{v2b,o2}
\fmffreeze
\fmf{fermion}{i1,v1a}
\fmf{fermion}{v1a,v1c}
\fmf{fermion}{v1c,v1b}
\fmf{fermion}{v1b,o1}
\fmf{fermion}{i2,v2a}
\fmf{plain}{v2a,v2c}
\fmf{plain}{v2c,v2b}
\fmf{fermion}{v2b,o2}
\fmf{zigzag}{v1c,v2c}
\fmf{zigzag,left,tension=0}{v1a,v1b}
\fmfdot{v1a}
\fmfdot{v1b}
\fmfdot{v1c}
\fmfdot{v2c}
\fmfv{l=$\mu$,l.a=-55}{v1c}
\fmfv{l=$\mu$}{v2c}
\fmfv{l=$\n$}{v1a}
\fmfv{l=$\n$}{v1b}
\end{fmfgraph*}
}
\notag}
\vspace{5mm}
\caption{$2\pi(p_0)\g^0\otimes\g^0\({1\o2\pi\ep}\)$
$\times\d^{\m0}\[-\d^{\n0}+\sum_j\d^{\n j}\]$}
\label{fig:4ptGamGamVert}
\end{subfigure}
\begin{subfigure}[b]{0.32\textwidth}
\eq{
\mff{
\scriptsize
\begin{fmfgraph*}(25,15)
\fmfpen{thin}
\fmfstraight
\fmfleft{i2,i1}
\fmfright{o2,o1}
\fmf{phantom}{i1,v1a}
\fmf{phantom}{i2,v2a}
\fmf{phantom,tension=1.55}{v1a,v1c,v1b}
\fmf{phantom,tension=1.55}{v2a,v2c,v2b}
\fmf{phantom}{v1b,o1}
\fmf{phantom}{v2b,o2}
\fmffreeze
\fmf{fermion}{i1,v1a}
\fmf{plain}{v1a,v1c}
\fmf{plain}{v1c,v1b}
\fmf{fermion}{v1b,o1}
\fmf{fermion}{i2,v2a}
\fmf{fermion}{v2a,v2c}
\fmf{fermion}{v2c,v2b}
\fmf{fermion}{v2b,o2}
\fmf{zigzag}{v1c,v2c}
\fmf{zigzag,right,tension=0}{v2a,v2b}
\fmfdot{v2a}
\fmfdot{v2b}
\fmfdot{v2c}
\fmfdot{v1c}
\fmfv{l=$\mu$,l.a=55}{v2c}
\fmfv{l=$\mu$}{v1c}
\fmfv{l=$\n$}{v2a}
\fmfv{l=$\n$}{v2b}
\end{fmfgraph*}
}
\notag}
\vspace{5mm}
\caption{$2\pi(p_0)\g^0\otimes\g^0\({1\o2\pi\ep}\)$
$\times\d^{\n0}\[-\d^{\m0}+\sum_j\d^{\m j}\]$}
\label{fig:4ptGamGamVert2}
\end{subfigure}
\begin{subfigure}[b]{0.32\textwidth}
\eq{
\mff{
\scriptsize
\begin{fmfgraph*}(25,15)
\fmfpen{thin}
\fmfstraight
\fmfleft{i2,i1}
\fmfright{o2,o1}
\fmf{phantom}{i1,v1a}
\fmf{phantom}{i2,v2a}
\fmf{phantom,tension=.6}{v1a,v1b}
\fmf{phantom,tension=.6}{v2a,v2b}
\fmf{phantom}{v1b,o1}
\fmf{phantom}{v2b,o2}
\fmffreeze
\fmf{fermion}{i1,v1a}
\fmf{fermion}{v1a,v1b}
\fmf{fermion}{v1b,o1}
\fmf{fermion}{i2,v2a}
\fmf{fermion}{v2a,v2b}
\fmf{fermion}{v2b,o2}
\fmf{zigzag}{v1a,v2a}
\fmf{photon}{v1b,v2b}
\fmfv{l=$\m$}{v1b}	
\fmfv{l=$\m$}{v2b}
\fmfv{l=$\n$}{v1a}	
\fmfv{l=$\n$}{v2a}
\fmfdot{v1a}
\fmfdot{v1b}
\fmfdot{v2a}
\fmfdot{v2b}
\end{fmfgraph*}}
\notag}\vspace{5mm}
\captionsetup{justification=centering}
\caption{Convergent
$\ph{2\pi(p_0)\g^0\otimes\g^0\({1\o2\pi\ep}\)}$}
\label{fig:4ptGaugePhotonNoX}
\end{subfigure}
\\\vspace{5mm}
\begin{subfigure}[b]{0.32\textwidth}
\eq{
\mff{
\scriptsize
\begin{fmfgraph*}(25,15)
\fmfpen{thin}
\fmfstraight
\fmfleft{i2,i1}
\fmfright{o2,o1}
\fmf{phantom}{i1,v1a}
\fmf{phantom}{i2,v2a}
\fmf{phantom,tension=1.55}{v1a,v1c,v1b}
\fmf{phantom,tension=1.55}{v2a,v2c,v2b}
\fmf{phantom}{v1b,o1}
\fmf{phantom}{v2b,o2}
\fmffreeze
\fmf{fermion}{i1,v1a}
\fmf{fermion}{v1a,v1c}
\fmf{fermion}{v1c,v1b}
\fmf{fermion}{v1b,o1}
\fmf{fermion}{i2,v2a}
\fmf{plain}{v2a,v2c}
\fmf{plain}{v2c,v2b}
\fmf{fermion}{v2b,o2}
\fmf{zigzag}{v1c,v2c}
\fmf{photon,left,tension=0}{v1a,v1b}
\fmfv{l=$\n$}{v1b}
\fmfv{l=$\n$}{v1a}
\fmfv{l=$\m$,l.a=-55}{v1c}
\fmfv{l=$\m$}{v2c}
\fmfdot{v1a}
\fmfdot{v1b}
\fmfdot{v1c}
\fmfdot{v2c}
\end{fmfgraph*}
}
\notag}\vspace{5mm}
\caption{{$2\pi\d(p_0)\g^\m\otimes\g^\m\(-{8\mathpzc{g}^2\o3\pi^2(2N)\ep}\)$}}
\label{fig:4ptsGammaPhotonArc}
\end{subfigure}
\begin{subfigure}[b]{0.32\textwidth}
\eq{
\mff{
\scriptsize
\begin{fmfgraph*}(25,15)
\fmfpen{thin}
\fmfstraight
\fmfleft{i2,i1}
\fmfright{o2,o1}
\fmf{phantom}{i1,v1a}
\fmf{phantom}{i2,v2a}
\fmf{phantom,tension=1.55}{v1a,v1c,v1b}
\fmf{phantom,tension=1.55}{v2a,v2c,v2b}
\fmf{phantom}{v1b,o1}
\fmf{phantom}{v2b,o2}
\fmffreeze
\fmf{fermion}{i1,v1a}
\fmf{fermion}{v2a,v2c}
\fmf{fermion}{v2c,v2b}
\fmf{fermion}{v1b,o1}
\fmf{fermion}{i2,v2a}
\fmf{plain}{v1a,v1c}
\fmf{plain}{v1c,v1b}
\fmf{fermion}{v2b,o2}
\fmf{zigzag}{v1c,v2c}
\fmf{photon,right,tension=0}{v2a,v2b}
\fmfv{l=$\n$}{v2b}
\fmfv{l=$\n$}{v2a}
\fmfv{l=$\m$}{v1c}
\fmfv{l=$\m$,l.a=55}{v2c}
\fmfdot{v2a}
\fmfdot{v2b}
\fmfdot{v2c}
\fmfdot{v1c}
\end{fmfgraph*}
}
\notag}\vspace{5mm}
\caption{{$2\pi\d(p_0)\g^\m\otimes\g^\m\(-{8\mathpzc{g}^2\o3\pi^2(2N)\ep}\)$}}
\label{fig:4ptsGammaPhotonArc2}
\end{subfigure}
\begin{subfigure}[b]{0.32\textwidth}
\eq{
\mff{
\scriptsize
\begin{fmfgraph*}(25,15)
\fmfpen{thin}
\fmfstraight
\fmfleft{i2,i1}
\fmfright{o2,o1}
\fmf{phantom}{i1,v1a}
\fmf{phantom}{i2,v2a}
\fmf{phantom,tension=.6}{v1a,v1b}
\fmf{phantom,tension=.6}{v2a,v2b}
\fmf{phantom}{v1b,o1}
\fmf{phantom}{v2b,o2}
\fmffreeze
\fmf{fermion}{i1,v1a}
\fmf{fermion}{v1a,v1b}
\fmf{fermion}{v1b,o1}
\fmf{fermion}{i2,v2a}
\fmf{fermion}{v2a,v2b}
\fmf{fermion}{v2b,o2}
\fmf{zigzag}{v1a,v2b}
\fmf{photon}{v1b,v2a}
\fmfv{l=$\m$}{v1b}	
\fmfv{l=$\m$}{v2a}
\fmfv{l=$\n$}{v1a}	
\fmfv{l=$\n$}{v2b}
\fmfdot{v1a}
\fmfdot{v1b}
\fmfdot{v2a}
\fmfdot{v2b}
\end{fmfgraph*}
}
\notag}\vspace{5mm}
\caption{Convergent}
\label{fig:4ptGaugePhotonX}
\end{subfigure}
\\\vspace{5mm}
\begin{subfigure}[b]{0.24\textwidth}
\eq{
\mff{
\scriptsize
\begin{fmfgraph*}(25,20)
\fmfpen{thin}
\fmfstraight
\fmfleft{i2,i1}
\fmfright{o2,o1}
\fmf{phantom}{i1,v1a}
\fmf{phantom}{i2,v2a}
\fmf{phantom,tension=.8}{v1a,v1c,v1b}
\fmf{phantom,tension=.8}{v2a,v2c,v2b}
\fmf{phantom}{v1b,o1}
\fmf{phantom}{v2b,o2}
\fmffreeze
\fmf{phantom,tension=0.7}{v1a,l1}
\fmf{phantom,tension=0.7}{v1b,r1}
\fmf{phantom,tension=0.7}{v1c,c1}
\fmf{phantom}{l1,l2,v2a}
\fmf{phantom}{r1,r2,v2b}
\fmf{phantom}{c1,c2,v2c}
\fmffreeze
\fmf{fermion}{i1,v1c}
\fmf{fermion}{v1c,o1}
\fmf{fermion}{i2,v2a}
\fmf{fermion}{v2a,v2b}
\fmf{fermion}{v2b,o2}
\fmf{zigzag}{v1c,c1}
\fmf{fermion}{c1,l2}
\fmf{fermion}{l2,r2}
\fmf{fermion}{r2,c1}
\fmf{photon}{l2,v2a}
\fmf{photon}{r2,v2b}
\fmfdot{v1c}
\fmfdot{c1}
\fmfdot{l2}
\fmfdot{r2}
\fmfdot{v2a}
\fmfdot{v2b}
\fmfv{l=$\m$}{v1c}
\fmfv{l=$\m$,l.a=0}{c1}
\fmfv{l=$\n$}{r2}
\fmfv{l=$\n$}{v2b}
\fmfv{l=$\s$}{l2}
\fmfv{l=$\s$}{v2a}
\end{fmfgraph*}
}
\notag}
\caption{Vanishes}
\label{fig:FermLoopGammaPhotonPhoton1}
\end{subfigure}
\begin{subfigure}[b]{0.24\textwidth}
\eq{
\mff{
\scriptsize
\begin{fmfgraph*}(25,20)
\fmfpen{thin}
\fmfstraight
\fmfleft{i2,i1}
\fmfright{o2,o1}
\fmf{phantom}{i1,v1a}
\fmf{phantom}{i2,v2a}
\fmf{phantom,tension=.8}{v1a,v1c,v1b}
\fmf{phantom,tension=.8}{v2a,v2c,v2b}
\fmf{phantom}{v1b,o1}
\fmf{phantom}{v2b,o2}
\fmffreeze
\fmf{phantom,tension=0.7}{v1a,l1}
\fmf{phantom,tension=0.7}{v1b,r1}
\fmf{phantom,tension=0.7}{v1c,c1}
\fmf{phantom}{l1,l2,v2a}
\fmf{phantom}{r1,r2,v2b}
\fmf{phantom}{c1,c2,v2c}
\fmffreeze
\fmf{fermion}{i1,v1c}
\fmf{fermion}{v1c,o1}
\fmf{fermion}{i2,v2a}
\fmf{fermion}{v2a,v2b}
\fmf{fermion}{v2b,o2}
\fmf{zigzag}{v1c,c1}
\fmf{fermion}{c1,r2}
\fmf{fermion}{r2,l2}
\fmf{fermion}{l2,c1}
\fmf{photon}{l2,v2a}
\fmf{photon}{r2,v2b}
\fmfdot{v1c}
\fmfdot{c1}
\fmfdot{l2}
\fmfdot{r2}
\fmfdot{v2a}
\fmfdot{v2b}
\fmfv{l=$\m$}{v1c}
\fmfv{l=$\m$,l.a=0}{c1}
\fmfv{l=$\n$}{r2}
\fmfv{l=$\n$}{v2b}
\fmfv{l=$\s$}{l2}
\fmfv{l=$\s$}{v2a}
\end{fmfgraph*}
}
\notag}
\caption{Vanishes}
\label{fig:FermLoopGammaPhotonPhoton2}
\end{subfigure}
\begin{subfigure}[b]{0.24\textwidth}
\eq{
\mff{
\scriptsize
\begin{fmfgraph*}(25,20)
\fmfpen{thin}
\fmfstraight
\fmfleft{i2,i1}
\fmfright{o2,o1}
\fmf{phantom}{i1,v1a}
\fmf{phantom}{i2,v2a}
\fmf{phantom,tension=.8}{v1a,v1c,v1b}
\fmf{phantom,tension=.8}{v2a,v2c,v2b}
\fmf{phantom}{v1b,o1}
\fmf{phantom}{v2b,o2}
\fmffreeze
\fmf{phantom,tension=0.7}{v1a,l1}
\fmf{phantom,tension=0.7}{v1b,r1}
\fmf{phantom,tension=0.7}{v1c,c1}
\fmf{phantom}{l1,l2,v2a}
\fmf{phantom}{r1,r2,v2b}
\fmf{phantom}{c1,c2,v2c}
\fmffreeze
\fmf{fermion}{i1,v1c}
\fmf{fermion}{v1c,o1}
\fmf{fermion}{i2,v2a}
\fmf{fermion}{v2a,v2b}
\fmf{fermion}{v2b,o2}
\fmf{zigzag}{v1c,c1}
\fmf{fermion}{c1,l2}
\fmf{fermion}{l2,r2}
\fmf{fermion}{r2,c1}
\fmf{photon}{l2,v2a}
\fmf{zigzag}{r2,v2b}
\fmfdot{v1c}
\fmfdot{c1}
\fmfdot{l2}
\fmfdot{r2}
\fmfdot{v2a}
\fmfdot{v2b}
\fmfv{l=$\m$}{v1c}
\fmfv{l=$\m$,l.a=0}{c1}
\fmfv{l=$\n$}{r2}
\fmfv{l=$\n$}{v2b}
\fmfv{l=$\s$}{l2}
\fmfv{l=$\s$}{v2a}
\end{fmfgraph*}
}
\notag}
\caption{Cancels Fig.~\ref{fig:FermLoopGammaPhotonDisorder2}}
\label{fig:FermLoopGammaPhotonDisorder1}
\end{subfigure}
\begin{subfigure}[b]{0.24\textwidth}
\eq{
\mff{
\scriptsize
\begin{fmfgraph*}(25,20)
\fmfpen{thin}
\fmfstraight
\fmfleft{i2,i1}
\fmfright{o2,o1}
\fmf{phantom}{i1,v1a}
\fmf{phantom}{i2,v2a}
\fmf{phantom,tension=.8}{v1a,v1c,v1b}
\fmf{phantom,tension=.8}{v2a,v2c,v2b}
\fmf{phantom}{v1b,o1}
\fmf{phantom}{v2b,o2}
\fmffreeze
\fmf{phantom,tension=0.7}{v1a,l1}
\fmf{phantom,tension=0.7}{v1b,r1}
\fmf{phantom,tension=0.7}{v1c,c1}
\fmf{phantom}{l1,l2,v2a}
\fmf{phantom}{r1,r2,v2b}
\fmf{phantom}{c1,c2,v2c}
\fmffreeze
\fmf{fermion}{i1,v1c}
\fmf{fermion}{v1c,o1}
\fmf{fermion}{i2,v2a}
\fmf{fermion}{v2a,v2b}
\fmf{fermion}{v2b,o2}
\fmf{zigzag}{v1c,c1}
\fmf{fermion}{c1,r2}
\fmf{fermion}{r2,l2}
\fmf{fermion}{l2,c1}
\fmf{photon}{l2,v2a}
\fmf{zigzag}{r2,v2b}
\fmfdot{v1c}
\fmfdot{c1}
\fmfdot{l2}
\fmfdot{r2}
\fmfdot{v2a}
\fmfdot{v2b}
\fmfv{l=$\m$}{v1c}
\fmfv{l=$\m$,l.a=0}{c1}
\fmfv{l=$\n$}{r2}
\fmfv{l=$\n$}{v2b}
\fmfv{l=$\s$}{l2}
\fmfv{l=$\s$}{v2a}
\end{fmfgraph*}
}
\notag}
\caption{Cancels Fig.~\ref{fig:FermLoopGammaPhotonDisorder1}}
\label{fig:FermLoopGammaPhotonDisorder2}
\end{subfigure}
\end{fmffile}

%% file: IntegralDiagramsMix.tex
{\centering
\begin{fmffile}{intdiagramsMix}
\fmfset{zigzag_width}{.7mm}
\fmfset{dot_size}{1.5thick}
\fmfset{arrow_len}{3mm}
\begin{subfigure}[b]{0.45\textwidth}
\eq{
\mff{
\scriptsize
\begin{fmfgraph*}(25,15)
\fmfpen{thin}
\fmfstraight
\fmfleft{i2,i1}
\fmfright{o2,o1}
\fmf{phantom}{i1,v1a}
\fmf{phantom}{i2,v2a}
\fmf{phantom,tension=.6}{v1a,v1b}
\fmf{phantom,tension=.6}{v2a,v2b}
\fmf{phantom}{v1b,o1}
\fmf{phantom}{v2b,o2}
\fmffreeze
\fmf{fermion}{i1,v1a}
\fmf{fermion}{v1a,v1b}
\fmf{fermion}{v1b,o1}
\fmf{fermion}{i2,v2a}
\fmf{fermion}{v2a,v2b}
\fmf{fermion}{v2b,o2}
\fmf{dbl_dots}{v1a,v2a}
\fmf{zigzag}{v1b,v2b}
\fmfdot{v1a}
\fmfdot{v1b}
\fmfdot{v2a}
\fmfdot{v2b}
\fmfv{l=$\mu$}{v1b}
\fmfv{l=$\m$}{v2b}
\end{fmfgraph*}}
\notag}\vspace{5mm}
\caption{
$\({1\o4\pi\ep}\)\begin{cases}
-\sum_j\g^j\otimes\g_j, & \m=0, \\
\id\otimes\id-\g^0\otimes\g^0, &\m=x,y
\end{cases}$}
\label{fig:4ptMixNoX}
\end{subfigure}
\begin{subfigure}[b]{0.45\textwidth}
\eq{
\mff{
\scriptsize
\begin{fmfgraph*}(25,15)
\fmfpen{thin}
\fmfstraight
\fmfleft{i2,i1}
\fmfright{o2,o1}
\fmf{phantom}{i1,v1a}
\fmf{phantom}{i2,v2a}
\fmf{phantom,tension=.6}{v1a,v1b}
\fmf{phantom,tension=.6}{v2a,v2b}
\fmf{phantom}{v1b,o1}
\fmf{phantom}{v2b,o2}
\fmffreeze
\fmf{fermion}{i1,v1a}
\fmf{fermion}{v1a,v1b}
\fmf{fermion}{v1b,o1}
\fmf{fermion}{i2,v2a}
\fmf{fermion}{v2a,v2b}
\fmf{fermion}{v2b,o2}
\fmf{dbl_dots}{v1a,v2b}
\fmf{zigzag}{v1b,v2a}
\fmfdot{v1a}
\fmfdot{v1b}
\fmfdot{v2a}
\fmfdot{v2b}
\fmfv{l=$\mu$}{v1b}
\fmfv{l=$\m$}{v2a}
\end{fmfgraph*}
}
\notag}\vspace{5mm}
\caption{
$\({1\o4\pi\ep}\)\begin{cases}
-\sum_j\g^j\otimes\g_j, & \m=0, \\
-\id\otimes\id-\g^0\otimes\g^0, &\m=x,y
\end{cases}$}
\label{fig:4ptMixX}
\end{subfigure}
\\\vspace{5mm}
\begin{subfigure}[b]{0.24\textwidth}
\eq{
\mff{
\scriptsize
\begin{fmfgraph*}(25,15)
\fmfpen{thin}
\fmfstraight
\fmfleft{i2,i1}
\fmfright{o2,o1}
\fmf{phantom}{i1,v1a}
\fmf{phantom}{i2,v2a}
\fmf{phantom,tension=1.55}{v1a,v1c,v1b}
\fmf{phantom,tension=1.55}{v2a,v2c,v2b}
\fmf{phantom}{v1b,o1}
\fmf{phantom}{v2b,o2}
\fmffreeze
\fmf{fermion}{i1,v1a}
\fmf{fermion}{v1a,v1c}
\fmf{fermion}{v1c,v1b}
\fmf{fermion}{v1b,o1}
\fmf{fermion}{i2,v2a}
\fmf{plain}{v2a,v2c}
\fmf{plain}{v2c,v2b}
\fmf{fermion}{v2b,o2}
\fmf{dbl_dots}{v1c,v2c}
\fmf{zigzag,left,tension=0}{v1a,v1b}
\fmfdot{v1a}
\fmfdot{v1b}
\fmfdot{v1c}
\fmfdot{v2c}
\fmfv{l=$\m$}{v1a}
\fmfv{l=$\m$}{v1b}
\end{fmfgraph*}
}
\notag}
\vspace{5mm}
\caption{${\id\otimes\id}\(-{1\o2\pi\ep}\)$}
\label{fig:4ptMixVert1a}
\end{subfigure}
\begin{subfigure}[b]{0.24\textwidth}
\eq{
\mff{
\scriptsize
\begin{fmfgraph*}(25,15)
\fmfpen{thin}
\fmfstraight
\fmfleft{i2,i1}
\fmfright{o2,o1}
\fmf{phantom}{i1,v1a}
\fmf{phantom}{i2,v2a}
\fmf{phantom,tension=1.55}{v1a,v1c,v1b}
\fmf{phantom,tension=1.55}{v2a,v2c,v2b}
\fmf{phantom}{v1b,o1}
\fmf{phantom}{v2b,o2}
\fmffreeze
\fmf{fermion}{i1,v1a}
\fmf{plain}{v1a,v1c}
\fmf{plain}{v1c,v1b}
\fmf{fermion}{v1b,o1}
\fmf{fermion}{i2,v2a}
\fmf{fermion}{v2a,v2c}
\fmf{fermion}{v2c,v2b}
\fmf{fermion}{v2b,o2}
\fmf{dbl_dots}{v1c,v2c}
\fmf{zigzag,right,tension=0}{v2a,v2b}
\fmfdot{v2a}
\fmfdot{v2b}
\fmfdot{v2c}
\fmfdot{v1c}
\fmfv{l=$\m$}{v2a}
\fmfv{l=$\m$}{v2b}
\end{fmfgraph*}
}
\notag}
\vspace{5mm}
\caption{${\id\otimes\id}\(-{1\o2\pi\ep}\)$}
\label{fig:4ptMixVert2a}
\end{subfigure}
\begin{subfigure}[b]{0.24\textwidth}
\eq{
\mff{
\scriptsize
\begin{fmfgraph*}(25,15)
\fmfset{zigzag_width}{.7mm}
\fmfpen{thin}
\fmfstraight
\fmfleft{i2,i1}
\fmfright{o2,o1}
\fmf{phantom}{i1,v1a}
\fmf{phantom}{i2,v2a}
\fmf{phantom,tension=1.55}{v1a,v1c,v1b}
\fmf{phantom,tension=1.55}{v2a,v2c,v2b}
\fmf{phantom}{v1b,o1}
\fmf{phantom}{v2b,o2}
\fmffreeze
\fmf{fermion}{i1,v1a}
\fmf{fermion}{v1a,v1c}
\fmf{fermion}{v1c,v1b}
\fmf{fermion}{v1b,o1}
\fmf{fermion}{i2,v2a}
\fmf{plain}{v2a,v2c}
\fmf{plain}{v2c,v2b}
\fmf{fermion}{v2b,o2}
\fmf{zigzag}{v1c,v2c}
\fmf{dbl_dots,left,tension=0}{v1a,v1b}
\fmfdot{v1a}
\fmfdot{v1b}
\fmfdot{v1c}
\fmfdot{v2c}
\fmfv{l=$\m$,l.a=-55}{v1c}
\fmfv{l=$\m$}{v2c}
\end{fmfgraph*}
}
\notag}
\vspace{5mm}
\caption{$\d^{\m0}\g^0\otimes\g^0\({1\o2\pi\ep}\)$}
\label{fig:4ptMixVert1b}
\end{subfigure}
\begin{subfigure}[b]{0.24\textwidth}
\eq{
\mff{
\scriptsize
\begin{fmfgraph*}(25,15)
\fmfpen{thin}
\fmfstraight
\fmfleft{i2,i1}
\fmfright{o2,o1}
\fmf{phantom}{i1,v1a}
\fmf{phantom}{i2,v2a}
\fmf{phantom,tension=1.55}{v1a,v1c,v1b}
\fmf{phantom,tension=1.55}{v2a,v2c,v2b}
\fmf{phantom}{v1b,o1}
\fmf{phantom}{v2b,o2}
\fmffreeze
\fmf{fermion}{i1,v1a}
\fmf{plain}{v1a,v1c}
\fmf{plain}{v1c,v1b}
\fmf{fermion}{v1b,o1}
\fmf{fermion}{i2,v2a}
\fmf{fermion}{v2a,v2c}
\fmf{fermion}{v2c,v2b}
\fmf{fermion}{v2b,o2}
\fmf{zigzag}{v1c,v2c}
\fmf{dbl_dots,right,tension=0}{v2a,v2b}
\fmfdot{v2a}
\fmfdot{v2b}
\fmfdot{v2c}
\fmfdot{v1c}
\fmfv{l=$\m$}{v1c}
\fmfv{l=$\m$,l.a=55}{v2c}
\end{fmfgraph*}
}
\notag}
\vspace{5mm}
\caption{$\d^{\m0}\g^0\otimes\g^0\({1\o2\pi\ep}\)$}
\label{fig:4ptMixVert2b}
\end{subfigure}
\\\vspace{5mm}
\begin{subfigure}[b]{0.24\textwidth}
\eq{
\mff{
\scriptsize
\begin{fmfgraph*}(25,20)
\fmfpen{thin}
\fmfstraight
\fmfleft{i2,i1}
\fmfright{o2,o1}
\fmf{phantom}{i1,v1a}
\fmf{phantom}{i2,v2a}
\fmf{phantom,tension=.8}{v1a,v1c,v1b}
\fmf{phantom,tension=.8}{v2a,v2c,v2b}
\fmf{phantom}{v1b,o1}
\fmf{phantom}{v2b,o2}
\fmffreeze
\fmf{phantom,tension=0.7}{v1a,l1}
\fmf{phantom,tension=0.7}{v1b,r1}
\fmf{phantom,tension=0.7}{v1c,c1}
\fmf{phantom}{l1,l2,v2a}
\fmf{phantom}{r1,r2,v2b}
\fmf{phantom}{c1,c2,v2c}
\fmffreeze
\fmf{fermion}{i1,v1c}
\fmf{fermion}{v1c,o1}
\fmf{fermion}{i2,v2a}
\fmf{fermion}{v2a,v2b}
\fmf{fermion}{v2b,o2}
\fmf{zigzag}{v1c,c1}
\fmf{fermion}{c1,l2}
\fmf{fermion}{l2,r2}
\fmf{fermion}{r2,c1}
\fmf{photon}{l2,v2a}
\fmf{dbl_dots}{r2,v2b}
\fmfdot{v1c}
\fmfdot{v2a}
\fmfdot{v2b}
\fmfdot{c1}
\fmfdot{r2}
\fmfdot{l2}
\fmfv{l=$\m$}{v1c}
\fmfv{l=$\m$,l.a=0}{c1}
\fmfv{l=$\n$}{l2}
\fmfv{l=$\n$}{v2a}
\end{fmfgraph*}
}
\notag}
\caption{Cancels Fig.~\ref{fig:FermLoopPotentialVertMassInt2}
$\ph{Cancels Fig.~\ref{fig:FermLoopPotentialVertMassInt1}}$}
\label{fig:FermLoopPotentialVertMassInt1}
\end{subfigure}
\begin{subfigure}[b]{0.24\textwidth}
\eq{
\mff{
\scriptsize
\begin{fmfgraph*}(25,20)
\fmfpen{thin}
\fmfstraight
\fmfleft{i2,i1}
\fmfright{o2,o1}
\fmf{phantom}{i1,v1a}
\fmf{phantom}{i2,v2a}
\fmf{phantom,tension=.8}{v1a,v1c,v1b}
\fmf{phantom,tension=.8}{v2a,v2c,v2b}
\fmf{phantom}{v1b,o1}
\fmf{phantom}{v2b,o2}
\fmffreeze
\fmf{phantom,tension=0.7}{v1a,l1}
\fmf{phantom,tension=0.7}{v1b,r1}
\fmf{phantom,tension=0.7}{v1c,c1}
\fmf{phantom}{l1,l2,v2a}
\fmf{phantom}{r1,r2,v2b}
\fmf{phantom}{c1,c2,v2c}
\fmffreeze
\fmf{fermion}{i1,v1c}
\fmf{fermion}{v1c,o1}
\fmf{fermion}{i2,v2a}
\fmf{fermion}{v2a,v2b}
\fmf{fermion}{v2b,o2}
\fmf{zigzag}{v1c,c1}
\fmf{fermion}{c1,r2}
\fmf{fermion}{r2,l2}
\fmf{fermion}{l2,c1}
\fmf{photon}{l2,v2a}
\fmf{dbl_dots}{r2,v2b}
\fmfdot{v1c}
\fmfdot{v2a}
\fmfdot{v2b}
\fmfdot{c1}
\fmfdot{r2}
\fmfdot{l2}
\fmfv{l=$\m$}{v1c}
\fmfv{l=$\m$,l.a=0}{c1}
\fmfv{l=$\n$}{l2}
\fmfv{l=$\n$}{v2a}
\end{fmfgraph*}
}
\notag}
\caption{Cancels Fig.~\ref{fig:FermLoopPotentialVertMassInt1}
$\ph{Cancels Fig.~\ref{fig:FermLoopPotentialVertMassInt1}}$}
\label{fig:FermLoopPotentialVertMassInt2}
\end{subfigure}
\begin{subfigure}[b]{0.24\textwidth}
\eq{
\mff{
\scriptsize
\begin{fmfgraph*}(25,20)
\fmfpen{thin}
\fmfstraight
\fmfleft{i2,i1}
\fmfright{o2,o1}
\fmf{phantom}{i1,v1a}
\fmf{phantom}{i2,v2a}
\fmf{phantom,tension=.8}{v1a,v1c,v1b}
\fmf{phantom,tension=.8}{v2a,v2c,v2b}
\fmf{phantom}{v1b,o1}
\fmf{phantom}{v2b,o2}
\fmffreeze
\fmf{phantom,tension=0.7}{v1a,l1}
\fmf{phantom,tension=0.7}{v1b,r1}
\fmf{phantom,tension=0.7}{v1c,c1}
\fmf{phantom}{l1,l2,v2a}
\fmf{phantom}{r1,r2,v2b}
\fmf{phantom}{c1,c2,v2c}
\fmffreeze
\fmf{fermion}{i1,v1c}
\fmf{fermion}{v1c,o1}
\fmf{fermion}{i2,v2a}
\fmf{fermion}{v2a,v2b}
\fmf{fermion}{v2b,o2}
\fmf{dbl_dots}{v1c,c1}
\fmf{fermion}{c1,l2}
\fmf{fermion}{l2,r2}
\fmf{fermion}{r2,c1}
\fmf{photon}{l2,v2a}
\fmf{zigzag}{r2,v2b}
\fmfdot{v1c}
\fmfdot{v2a}
\fmfdot{v2b}
\fmfdot{c1}
\fmfdot{r2}
\fmfdot{l2}
\fmfv{l=$\m$}{v2b}
\fmfv{l=$\m$}{r2}
\fmfv{l=$\n$}{l2}
\fmfv{l=$\n$}{v2a}
\end{fmfgraph*}
}
\notag}
\caption{$\tr\[\O_{fl}\]{\mathpzc{g}^2\o2N}\({1\o4\pi\ep}\)$
$\times\(2-\d_{\m j}\)\id\otimes\id$}
\label{fig:FermLoopMassVerPotentialInt1}
\end{subfigure}
\begin{subfigure}[b]{0.24\textwidth}
\eq{
\mff{
\scriptsize
\begin{fmfgraph*}(25,20)
\fmfpen{thin}
\fmfstraight
\fmfleft{i2,i1}
\fmfright{o2,o1}
\fmf{phantom}{i1,v1a}
\fmf{phantom}{i2,v2a}
\fmf{phantom,tension=.8}{v1a,v1c,v1b}
\fmf{phantom,tension=.8}{v2a,v2c,v2b}
\fmf{phantom}{v1b,o1}
\fmf{phantom}{v2b,o2}
\fmffreeze
\fmf{phantom,tension=0.7}{v1a,l1}
\fmf{phantom,tension=0.7}{v1b,r1}
\fmf{phantom,tension=0.7}{v1c,c1}
\fmf{phantom}{l1,l2,v2a}
\fmf{phantom}{r1,r2,v2b}
\fmf{phantom}{c1,c2,v2c}
\fmffreeze
\fmf{fermion}{i1,v1c}
\fmf{fermion}{v1c,o1}
\fmf{fermion}{i2,v2a}
\fmf{fermion}{v2a,v2b}
\fmf{fermion}{v2b,o2}
\fmf{dbl_dots}{v1c,c1}
\fmf{fermion}{c1,r2}
\fmf{fermion}{r2,l2}
\fmf{fermion}{l2,c1}
\fmf{photon}{l2,v2a}
\fmf{zigzag}{r2,v2b}
\fmfdot{v1c}
\fmfdot{v2a}
\fmfdot{v2b}
\fmfdot{c1}
\fmfdot{l2}
\fmfdot{r2}
\fmfv{l=$\m$}{v2b}
\fmfv{l=$\m$}{r2}
\fmfv{l=$\n$}{l2}
\fmfv{l=$\n$}{v2a}
\end{fmfgraph*}
}
\notag}
\caption{$\tr\[\O_{fl}\]{\mathpzc{g}^2\o2N}\({1\o4\pi\ep}\)$
$\times\(2-\d_{\m j}\)\id\otimes\id$}
\label{fig:FermLoopMassVerPotentialInt2}
\end{subfigure}
\end{fmffile}}

%% file: InternalFermionLoops.tex
\centering
\begin{fmffile}{InternalFermionLoop3Gauge}
\fmfset{zigzag_width}{.7mm}
\fmfset{dot_size}{1.5thick}
\fmfset{arrow_len}{3mm}
\begin{subfigure}[b]{0.24\textwidth}
\centering
\eq{
\mfff{
\scriptsize
\begin{fmfgraph*}(20,17)
\fmfpen{thin}
\fmfstraight
\fmfleft{i1,i2,i3}
\fmfright{o1,o2,o3}
\fmf{phantom}{i2,v2a}
\fmf{phantom,tension=0.6}{v2a,v2b}
\fmf{phantom}{v2b,o2}
\fmf{phantom}{i1,v1a}
\fmf{phantom,tension=0.6}{v1a,v1b}
\fmf{phantom}{v1b,o1}
\fmf{phantom}{i3,v3a}
\fmf{phantom,tension=0.6}{v3a,v3b}
\fmf{phantom}{v3b,o3}
\fmffreeze
\fmf{dbl_dots}{i2,v2a}
\fmf{fermion,label=$q$,label.side=right}{v2a,v1b}
\fmf{fermion,label=$q-k$,label.side=right}{v1b,v3b}
\fmf{fermion,label=$q$,label.side=right}{v3b,v2a}
\fmf{photon}{v1b,o1}
\fmf{photon}{v3b,o3}
\fmfdot{v2a}
\fmfdot{v1b}
\fmfdot{v3b}
\fmfv{label=$\mu$}{v1b}
\fmfv{label=$\n$}{v3b}
\end{fmfgraph*}
}
\notag}
\caption{}
\label{fig:IntFermionLoopMass2Phot1}
\end{subfigure}
\begin{subfigure}[b]{0.24\textwidth}
\centering
\eq{
\mfff{
\scriptsize
\begin{fmfgraph*}(20,17)
\fmfpen{thin}
\fmfstraight
\fmfleft{i1,i2,i3}
\fmfright{o1,o2,o3}
\fmf{phantom}{i2,v2a}
\fmf{phantom,tension=0.6}{v2a,v2b}
\fmf{phantom}{v2b,o2}
\fmf{phantom}{i1,v1a}
\fmf{phantom,tension=0.6}{v1a,v1b}
\fmf{phantom}{v1b,o1}
\fmf{phantom}{i3,v3a}
\fmf{phantom,tension=0.6}{v3a,v3b}
\fmf{phantom}{v3b,o3}
\fmffreeze
\fmf{dbl_dots}{i2,v2a}
\fmf{fermion,label=$q$,label.side=left}{v1b,v2a}
\fmf{fermion,label=$q+k$,label.side=left}{v3b,v1b}
\fmf{fermion,label=$q$,label.side=left}{v2a,v3b}
\fmf{photon}{v1b,o1}
\fmf{photon}{v3b,o3}
\fmfdot{v2a}
\fmfdot{v1b}
\fmfdot{v3b}
\fmfv{label=$\mu$}{v1b}
\fmfv{label=$\n$}{v3b}
\end{fmfgraph*}
}
\notag}
\caption{}
\label{fig:IntFermionLoopMass2Phot2}
\end{subfigure}
\begin{subfigure}[b]{0.24\textwidth}
\centering
\eq{
\mfff{
\scriptsize
\begin{fmfgraph*}(20,17)
\fmfpen{thin}
\fmfstraight
\fmfleft{i1,i2,i3}
\fmfright{o1,o2,o3}
\fmf{phantom}{i2,v2a}
\fmf{phantom,tension=0.6}{v2a,v2b}
\fmf{phantom}{v2b,o2}
\fmf{phantom}{i1,v1a}
\fmf{phantom,tension=0.6}{v1a,v1b}
\fmf{phantom}{v1b,o1}
\fmf{phantom}{i3,v3a}
\fmf{phantom,tension=0.6}{v3a,v3b}
\fmf{phantom}{v3b,o3}
\fmffreeze
\fmf{dbl_dots}{i2,v2a}
\fmf{fermion,label=$q$,label.side=right}{v2a,v1b}
\fmf{fermion,label=$q-k$,label.side=right}{v1b,v3b}
\fmf{fermion,label=$q$,label.side=right}{v3b,v2a}
\fmf{dbl_dots}{v1b,o1}
\fmf{photon}{v3b,o3}
\fmfdot{v2a}
\fmfdot{v1b}
\fmfdot{v3b}
\fmfv{label=$\m$}{v3b}
\end{fmfgraph*}
}
\notag}
\caption{}
\label{fig:IntFermionLoopMass1Mass1Phot1}
\end{subfigure}
\begin{subfigure}[b]{0.24\textwidth}
\centering
\eq{
\mfff{
\scriptsize
\begin{fmfgraph*}(20,17)
\fmfpen{thin}
\fmfstraight
\fmfleft{i1,i2,i3}
\fmfright{o1,o2,o3}
\fmf{phantom}{i2,v2a}
\fmf{phantom,tension=0.6}{v2a,v2b}
\fmf{phantom}{v2b,o2}
\fmf{phantom}{i1,v1a}
\fmf{phantom,tension=0.6}{v1a,v1b}
\fmf{phantom}{v1b,o1}
\fmf{phantom}{i3,v3a}
\fmf{phantom,tension=0.6}{v3a,v3b}
\fmf{phantom}{v3b,o3}
\fmffreeze
\fmf{dbl_dots}{i2,v2a}
\fmf{fermion,label=$q$,label.side=left}{v1b,v2a}
\fmf{fermion,label=$q+k$,label.side=left}{v3b,v1b}
\fmf{fermion,label=$q$,label.side=left}{v2a,v3b}
\fmf{dbl_dots}{v1b,o1}
\fmf{photon}{v3b,o3}
\fmfdot{v2a}
\fmfdot{v1b}
\fmfdot{v3b}
\fmfv{label=$\m$}{v3b}
\end{fmfgraph*}
}
\notag}
\caption{}
\label{fig:IntFermionLoopMass1Mass1Phot2}
\end{subfigure}
\\\vspace{3mm}
\begin{subfigure}[b]{0.24\textwidth}
\centering
\eq{
\mfff{
\scriptsize
\begin{fmfgraph*}(20,17)
\fmfpen{thin}
\fmfstraight
\fmfleft{i1,i2,i3}
\fmfright{o1,o2,o3}
\fmf{phantom}{i2,v2a}
\fmf{phantom,tension=0.6}{v2a,v2b}
\fmf{phantom}{v2b,o2}
\fmf{phantom}{i1,v1a}
\fmf{phantom,tension=0.6}{v1a,v1b}
\fmf{phantom}{v1b,o1}
\fmf{phantom}{i3,v3a}
\fmf{phantom,tension=0.6}{v3a,v3b}
\fmf{phantom}{v3b,o3}
\fmffreeze
\fmf{zigzag}{i2,v2a}
\fmf{fermion,label=$q$,label.side=right}{v2a,v1b}
\fmf{fermion,label=$q-k$,label.side=right}{v1b,v3b}
\fmf{fermion,label=$q$,label.side=right}{v3b,v2a}
\fmf{photon}{v1b,o1}
\fmf{zigzag}{v3b,o3}
\fmfdot{v2a}
\fmfdot{v1b}
\fmfdot{v3b}
\fmfv{label=$\mu$,l.a=90}{v2a}
\fmfv{label=$\nu$}{v3b}
\fmfv{label=$\s$}{v3b}
\end{fmfgraph*}
}
\notag}
\caption{}
\label{fig:IntFermionLoop3Gauge1}
\end{subfigure}
\begin{subfigure}[b]{0.24\textwidth}
\centering
\eq{
\mfff{
\scriptsize
\begin{fmfgraph*}(20,17)
\fmfpen{thin}
\fmfstraight
\fmfleft{i1,i2,i3}
\fmfright{o1,o2,o3}
\fmf{phantom}{i2,v2a}
\fmf{phantom,tension=0.6}{v2a,v2b}
\fmf{phantom}{v2b,o2}
\fmf{phantom}{i1,v1a}
\fmf{phantom,tension=0.6}{v1a,v1b}
\fmf{phantom}{v1b,o1}
\fmf{phantom}{i3,v3a}
\fmf{phantom,tension=0.6}{v3a,v3b}
\fmf{phantom}{v3b,o3}
\fmffreeze
\fmf{zigzag}{i2,v2a}
\fmf{fermion,label=$q$,label.side=left}{v1b,v2a}
\fmf{fermion,label=$q+k$,label.side=left}{v3b,v1b}
\fmf{fermion,label=$q$,label.side=left}{v2a,v3b}
\fmf{photon}{v1b,o1}
\fmf{zigzag}{v3b,o3}
\fmfdot{v2a}
\fmfdot{v1b}
\fmfdot{v3b}
\fmfv{label=$\mu$,l.a=90}{v2a}
\fmfv{label=$\nu$}{v1b}
\fmfv{label=$\s$}{v3b}
\end{fmfgraph*}
}
\notag}
\caption{}
\label{fig:IntFermionLoop3Gauge2}
\end{subfigure}
\begin{subfigure}[b]{0.24\textwidth}
\centering
\eq{
\mfff{
\scriptsize
\begin{fmfgraph*}(20,17)
\fmfpen{thin}
\fmfstraight
\fmfleft{i1,i2,i3}
\fmfright{o1,o2,o3}
\fmf{phantom}{i2,v2a}
\fmf{phantom,tension=0.6}{v2a,v2b}
\fmf{phantom}{v2b,o2}
\fmf{phantom}{i1,v1a}
\fmf{phantom,tension=0.6}{v1a,v1b}
\fmf{phantom}{v1b,o1}
\fmf{phantom}{i3,v3a}
\fmf{phantom,tension=0.6}{v3a,v3b}
\fmf{phantom}{v3b,o3}
\fmffreeze
\fmf{zigzag}{i2,v2a}
\fmf{fermion,label=$q$,label.side=right}{v2a,v1b}
\fmf{fermion,label=$q-k$,label.side=right}{v1b,v3b}
\fmf{fermion,label=$q$,label.side=right}{v3b,v2a}
\fmf{dbl_dots}{v1b,o1}
\fmf{photon}{v3b,o3}
\fmfdot{v2a}
\fmfdot{v1b}
\fmfdot{v3b}
\fmfv{label=$\mu$,l.a=90}{v2a}
\fmfv{label=$\nu$}{v3b}
\end{fmfgraph*}
}
\notag
}
\caption{}
\label{fig:IntFermLoopGaugeMassPhot1}
\end{subfigure}
\\\vspace{3mm}
\begin{subfigure}[b]{0.24\textwidth}
\centering
\eq{
\mfff{
\scriptsize
\begin{fmfgraph*}(20,17)
\fmfpen{thin}
\fmfstraight
\fmfleft{i1,i2,i3}
\fmfright{o1,o2,o3}
\fmf{phantom}{i2,v2a}
\fmf{phantom,tension=0.6}{v2a,v2b}
\fmf{phantom}{v2b,o2}
\fmf{phantom}{i1,v1a}
\fmf{phantom,tension=0.6}{v1a,v1b}
\fmf{phantom}{v1b,o1}
\fmf{phantom}{i3,v3a}
\fmf{phantom,tension=0.6}{v3a,v3b}
\fmf{phantom}{v3b,o3}
\fmffreeze
\fmf{zigzag}{i2,v2a}
\fmf{fermion,label=$q$,label.side=left}{v1b,v2a}
\fmf{fermion,label=$q+k$,label.side=left}{v3b,v1b}
\fmf{fermion,label=$q$,label.side=left}{v2a,v3b}
\fmf{photon}{v1b,o1}
\fmf{dbl_dots}{v3b,o3}
\fmfdot{v2a}
\fmfdot{v1b}
\fmfdot{v3b}
\fmfv{label=$\mu$,l.a=90}{v2a}
\fmfv{label=$\nu$}{v1b}
\end{fmfgraph*}
}
\notag}
\caption{}
\label{fig:IntFermLoopGaugeMassPhot2}
\end{subfigure}
\begin{subfigure}[b]{0.24\textwidth}
\centering
\eq{
\mfff{
\scriptsize
\begin{fmfgraph*}(20,17)
\fmfpen{thin}
\fmfstraight
\fmfleft{i1,i2,i3}
\fmfright{o1,o2,o3}
\fmf{phantom}{i2,v2a}
\fmf{phantom,tension=0.6}{v2a,v2b}
\fmf{phantom}{v2b,o2}
\fmf{phantom}{i1,v1a}
\fmf{phantom,tension=0.6}{v1a,v1b}
\fmf{phantom}{v1b,o1}
\fmf{phantom}{i3,v3a}
\fmf{phantom,tension=0.6}{v3a,v3b}
\fmf{phantom}{v3b,o3}
\fmffreeze
\fmf{dbl_dots}{i2,v2a}
\fmf{fermion,label=$q$,label.side=right}{v2a,v1b}
\fmf{fermion,label=$q-k$,label.side=right}{v1b,v3b}
\fmf{fermion,label=$q$,label.side=right}{v3b,v2a}
\fmf{photon}{v1b,o1}
\fmf{zigzag}{v3b,o3}
\fmfdot{v2a}
\fmfdot{v1b}
\fmfdot{v3b}
\fmfv{label=$\mu$}{v1b}
\fmfv{label=$\n$}{v3b}
\end{fmfgraph*}
}
\notag}
\caption{}
\label{fig:IntFermLoopMassGaugePho1}
\end{subfigure}
\begin{subfigure}[b]{0.24\textwidth}
\centering
\eq{
\mfff{
\scriptsize
\begin{fmfgraph*}(20,17)
\fmfpen{thin}
\fmfstraight
\fmfleft{i1,i2,i3}
\fmfright{o1,o2,o3}
\fmf{phantom}{i2,v2a}
\fmf{phantom,tension=0.6}{v2a,v2b}
\fmf{phantom}{v2b,o2}
\fmf{phantom}{i1,v1a}
\fmf{phantom,tension=0.6}{v1a,v1b}
\fmf{phantom}{v1b,o1}
\fmf{phantom}{i3,v3a}
\fmf{phantom,tension=0.6}{v3a,v3b}
\fmf{phantom}{v3b,o3}
\fmffreeze
\fmf{dbl_dots}{i2,v2a}
\fmf{fermion,label=$q$,label.side=left}{v1b,v2a}
\fmf{fermion,label=$q+k$,label.side=left}{v3b,v1b}
\fmf{fermion,label=$q$,label.side=left}{v2a,v3b}
\fmf{zigzag}{v1b,o1}
\fmf{photon}{v3b,o3}
\fmfdot{v2a}
\fmfdot{v1b}
\fmfdot{v3b}
\fmfv{label=$\mu$}{v1b}
\fmfv{label=$\n$}{v3b}
\end{fmfgraph*}
}
\notag}
\caption{}
\label{fig:IntFermLoopMassGaugePho2}
\end{subfigure}
\end{fmffile}
\vspace{5mm}

%% file: DiagramTable.tex
%
%
%
%
%
%
%

{\scriptsize
\centering
\begin{fmffile}{DiagramTable1}
\fmfset{dot_size}{1.5thick}
\fmfset{arrow_len}{3mm}
\begin{table}\scriptsize\centering
\begin{tabular}{| c | c |  m{0.35\textwidth} || c |  c | m{0.35\textwidth} |}
\hline
\normalsize
diagram & \normalsize$n_d$ &\normalsize divergence & \normalsize diagram & \normalsize $n_d$ & \normalsize divergence
\\\hline\hline
\centering
\begin{minipage}{.12\textwidth}\vspace{8mm}\centering
\centering
\scriptsize
\begin{fmfgraph*}(18,8)
\fmfpen{thin}
\fmfstraight
\fmfleft{i2,i1}
\fmfright{o2,o1}
\fmf{phantom}{i1,v1a}
\fmf{phantom}{i2,v2a}
\fmf{phantom,tension=.6}{v1a,v1b}
\fmf{phantom,tension=.6}{v2a,v2b}
\fmf{phantom}{v1b,o1}
\fmf{phantom}{v2b,o2}
\fmffreeze
\fmf{fermion}{i1,v1a}
\fmf{fermion}{v1a,v1b}
\fmf{fermion}{v1b,o1}
\fmf{fermion}{i2,v2a}
\fmf{fermion}{v2a,v2b}
\fmf{fermion}{v2b,o2}
\fmf{dbl_dashes}{v1a,v2a}
\fmf{dbl_dashes}{v1b,v2b}
\fmfdot{v1a}
\fmfdot{v2a}
\fmfdot{v1b}
\fmfdot{v2b}
\fmfv{l=$a$}{v1a}
\fmfv{l=$a$}{v2a}
\fmfv{l=$b$}{v1b}
\fmfv{l=$b$}{v2b}
\end{fmfgraph*}
\vspace{8mm}\end{minipage}
&
1
&
{$\sum_{ab}\(g_{t,a}g_{t,b}\o4\pi\ep\)\bigg\{{-\d^{ab}}\[\g^j\otimes\g_j\]\[\id\otimes\id\]$
$+\sum_c\abs{\ep^{abc}}\[\g^j\otimes\g_j\]\[\s_c\otimes\s_c\]\bigg\}$}
& 
\begin{minipage}{.12\textwidth}\vspace{8mm}\centering
\centering
\scriptsize
\begin{fmfgraph*}(18,8)
\fmfpen{thin}
\fmfstraight
\fmfleft{i2,i1}
\fmfright{o2,o1}
\fmf{phantom}{i1,v1a}
\fmf{phantom}{i2,v2a}
\fmf{phantom,tension=.6}{v1a,v1b}
\fmf{phantom,tension=.6}{v2a,v2b}
\fmf{phantom}{v1b,o1}
\fmf{phantom}{v2b,o2}
\fmffreeze
\fmf{fermion}{i1,v1a}
\fmf{fermion}{v1a,v1b}
\fmf{fermion}{v1b,o1}
\fmf{fermion}{i2,v2a}
\fmf{fermion}{v2a,v2b}
\fmf{fermion}{v2b,o2}
\fmf{dbl_dashes}{v1a,v2b}
\fmf{dbl_dashes}{v1b,v2a}
\fmfdot{v1a}
\fmfdot{v2a}
\fmfdot{v1b}
\fmfdot{v2b}
\fmfv{l=$a$}{v1a}
\fmfv{l=$a$}{v2b}
\fmfv{l=$b$}{v1b}
\fmfv{l=$b$}{v2a}
\end{fmfgraph*}
\vspace{8mm}\end{minipage}
&
1
&
$\sum_{ab}\(g_{t,a}g_{t,b}\o4\pi\ep\)\bigg\{\d^{ab}\[\g^j\otimes\g_j\]\[\id\otimes\id\]$
$+\sum_c\abs{\ep^{abc}}\[\g^j\otimes\g_j\]\[\s_c\otimes\s_c\]\bigg\}$
\\\hline
\begin{minipage}{.12\textwidth}
\vspace{8mm}
\centering
\scriptsize
\begin{fmfgraph*}(18,8)
\fmfpen{thin}
\fmfstraight
\fmfleft{i2,i1}
\fmfright{o2,o1}
\fmf{phantom}{i1,v1a}
\fmf{phantom}{i2,v2a}
\fmf{phantom,tension=1.55}{v1a,v1c,v1b}
\fmf{phantom,tension=1.55}{v2a,v2c,v2b}
\fmf{phantom}{v1b,o1}
\fmf{phantom}{v2b,o2}
\fmffreeze
\fmf{fermion}{i1,v1a}
\fmf{fermion}{v1a,v1c}
\fmf{fermion}{v1c,v1b}
\fmf{fermion}{v1b,o1}
\fmf{fermion}{i2,v2a}
\fmf{plain}{v2a,v2c}
\fmf{plain}{v2c,v2b}
\fmf{fermion}{v2b,o2}
\fmf{dbl_dashes}{v1c,v2c}
\fmf{dbl_dashes,left,tension=0}{v1a,v1b}
\fmfdot{v1a}
\fmfdot{v1c}
\fmfdot{v1b}
\fmfdot{v2c}
\fmfv{l=$b$}{v1a}
\fmfv{l=$b$}{v1b}
\fmfv{l=$a$,l.a=-55}{v1c}
\fmfv{l=$a$}{v2c}
\end{fmfgraph*}
\vspace{8mm}\end{minipage}
&
2
&
$\sum_a{g_{t,a}\o\pi\ep}\(2g_{t,a}-\sum_bg_{t,b}\)\[\id\otimes\id\]\[\s^a\otimes\s_a\]$
&
\begin{minipage}{.12\textwidth}\vspace{8mm}\centering
\centering
\scriptsize
\begin{fmfgraph*}(18,8)
\fmfpen{thin}
\fmfstraight
\fmfleft{i2,i1}
\fmfright{o2,o1}
\fmf{phantom}{i1,v1a}
\fmf{phantom}{i2,v2a}
\fmf{phantom,tension=1.55}{v1a,v1c,v1b}
\fmf{phantom,tension=1.55}{v2a,v2c,v2b}
\fmf{phantom}{v1b,o1}
\fmf{phantom}{v2b,o2}
\fmffreeze
\fmf{fermion}{i1,v1a}
\fmf{fermion}{v1a,v1c}
\fmf{fermion}{v1c,v1b}
\fmf{fermion}{v1b,o1}
\fmf{fermion}{i2,v2a}
\fmf{plain}{v2a,v2c}
\fmf{plain}{v2c,v2b}
\fmf{fermion}{v2b,o2}
\fmf{dbl_dashes}{v1c,v2c}
\fmf{photon,left,tension=0}{v1a,v1b}
\fmfv{l=$\m$}{v1b}
\fmfv{l=$\m$}{v1a}
\fmfv{l=$a$,l.a=-55}{v1c}
\fmfv{l=$a$}{v2c}
\fmfdot{v1a}
\fmfdot{v1c}
\fmfdot{v1b}
\fmfdot{v2c}
\end{fmfgraph*}
\vspace{8mm}\end{minipage}
&
2
&
$\sum_a\(-{48g_{t,a}\mathpzc{g}^2\o\pi^2(4N)\ep}\)\[\id\otimes\id\]\[\s^a\otimes\s_a\]$
\\\hline
\centering
\begin{minipage}{.12\textwidth}\vspace{8mm}\centering
\scriptsize
\begin{fmfgraph*}(18,8)
\fmfpen{thin}
\fmfstraight
\fmfleft{i2,i1}
\fmfright{o2,o1}
\fmf{phantom}{i1,v1a}
\fmf{phantom}{i2,v2a}
\fmf{phantom,tension=.6}{v1a,v1b}
\fmf{phantom,tension=.6}{v2a,v2b}
\fmf{phantom}{v1b,o1}
\fmf{phantom}{v2b,o2}
\fmffreeze
\fmf{fermion}{i1,v1a}
\fmf{fermion}{v1a,v1b}
\fmf{fermion}{v1b,o1}
\fmf{fermion}{i2,v2a}
\fmf{fermion}{v2a,v2b}
\fmf{fermion}{v2b,o2}
\fmf{dbl_dashes}{v1a,v2a}
\fmf{dashes}{v1b,v2b}
\fmfv{l=$a$}{v1a}
\fmfv{l=$a$}{v2a}
\fmfdot{v1a}
\fmfdot{v2a}
\fmfdot{v1b}
\fmfdot{v2b}
\end{fmfgraph*}
\vspace{8mm}\end{minipage}
&
2
&
$\sum_a\(-{g_sg_{t,a}\o2\pi\ep}\)\[\g^j\otimes\g_j\]\[\s^a\otimes\s_a\]$
&
\begin{minipage}{.12\textwidth}\vspace{8mm}\centering
\scriptsize
\begin{fmfgraph*}(18,8)
\fmfpen{thin}
\fmfstraight
\fmfleft{i2,i1}
\fmfright{o2,o1}
\fmf{phantom}{i1,v1a}
\fmf{phantom}{i2,v2a}
\fmf{phantom,tension=.6}{v1a,v1b}
\fmf{phantom,tension=.6}{v2a,v2b}
\fmf{phantom}{v1b,o1}
\fmf{phantom}{v2b,o2}
\fmffreeze
\fmf{fermion}{i1,v1a}
\fmf{fermion}{v1a,v1b}
\fmf{fermion}{v1b,o1}
\fmf{fermion}{i2,v2a}
\fmf{fermion}{v2a,v2b}
\fmf{fermion}{v2b,o2}
\fmf{dbl_dashes}{v1a,v2b}
\fmf{dashes}{v1b,v2a}
\fmfv{l=$a$}{v1a}
\fmfv{l=$a$}{v2b}
\fmfdot{v1a}
\fmfdot{v2a}
\fmfdot{v1b}
\fmfdot{v2b}
\end{fmfgraph*}
\vspace{8mm}\end{minipage}
&
2
&
$\sum_a\({g_sg_{t,a}\o2\pi\ep}\)\[\g^j\otimes\g_j\]\[\s^a\otimes\s_a\]$
\\\hline
\begin{minipage}{.12\textwidth}\vspace{8mm}\centering
\scriptsize
\begin{fmfgraph*}(18,8)
\fmfpen{thin}
\fmfstraight
\fmfleft{i2,i1}
\fmfright{o2,o1}
\fmf{phantom}{i1,v1a}
\fmf{phantom}{i2,v2a}
\fmf{phantom,tension=1.55}{v1a,v1c,v1b}
\fmf{phantom,tension=1.55}{v2a,v2c,v2b}
\fmf{phantom}{v1b,o1}
\fmf{phantom}{v2b,o2}
\fmffreeze
\fmf{fermion}{i1,v1a}
\fmf{fermion}{v1a,v1c}
\fmf{fermion}{v1c,v1b}
\fmf{fermion}{v1b,o1}
\fmf{fermion}{i2,v2a}
\fmf{plain}{v2a,v2c}
\fmf{plain}{v2c,v2b}
\fmf{fermion}{v2b,o2}
\fmf{dashes}{v1c,v2c}
\fmf{dbl_dashes,left,tension=0}{v1a,v1b}
\fmfv{l=$a$}{v1b}
\fmfv{l=$a$}{v1a}
\fmfdot{v1a}
\fmfdot{v1c}
\fmfdot{v1b}
\fmfdot{v2c}
\end{fmfgraph*}
\vspace{8mm}\end{minipage}
&
2
&
$\sum_a\(g_sg_{t,a}\o\pi\ep\)\[\id\otimes\id\]\[\id\otimes\id\]$
&
\begin{minipage}{.12\textwidth}\vspace{8mm}\centering
\scriptsize
\begin{fmfgraph*}(18,8)
\fmfpen{thin}
\fmfstraight
\fmfleft{i2,i1}
\fmfright{o2,o1}
\fmf{phantom}{i1,v1a}
\fmf{phantom}{i2,v2a}
\fmf{phantom,tension=1.55}{v1a,v1c,v1b}
\fmf{phantom,tension=1.55}{v2a,v2c,v2b}
\fmf{phantom}{v1b,o1}
\fmf{phantom}{v2b,o2}
\fmffreeze
\fmf{fermion}{i1,v1a}
\fmf{fermion}{v1a,v1c}
\fmf{fermion}{v1c,v1b}
\fmf{fermion}{v1b,o1}
\fmf{fermion}{i2,v2a}
\fmf{plain}{v2a,v2c}
\fmf{plain}{v2c,v2b}
\fmf{fermion}{v2b,o2}
\fmf{dbl_dashes}{v1c,v2c}
\fmf{dashes,left,tension=0}{v1a,v1b}
\fmfdot{v1a}
\fmfdot{v1c}
\fmfdot{v1b}
\fmfdot{v2c}
\fmfv{l=$a$,l.a=-55}{v1c}
\fmfv{l=$a$}{v2c}
\end{fmfgraph*}
\vspace{8mm}\end{minipage}
&
2
&
$\sum_a\(g_sg_{t,a}\o\pi\ep\)\[\id\otimes\id\]\[\s^a\otimes\s_a\]$
\\\hline
\centering
\begin{minipage}{.12\textwidth}\vspace{8mm}\centering
\scriptsize
\begin{fmfgraph*}(18,8)
\fmfpen{thin}
\fmfstraight
\fmfleft{i2,i1}
\fmfright{o2,o1}
\fmf{phantom}{i1,v1a}
\fmf{phantom}{i2,v2a}
\fmf{phantom,tension=.6}{v1a,v1b}
\fmf{phantom,tension=.6}{v2a,v2b}
\fmf{phantom}{v1b,o1}
\fmf{phantom}{v2b,o2}
\fmffreeze
\fmf{fermion}{i1,v1a}
\fmf{fermion}{v1a,v1b}
\fmf{fermion}{v1b,o1}
\fmf{fermion}{i2,v2a}
\fmf{fermion}{v2a,v2b}
\fmf{fermion}{v2b,o2}
\fmf{dbl_plain}{v1a,v2a}
\fmf{dbl_plain}{v1b,v2b}
\fmfv{l=$b,,j$}{v1a}
\fmfv{l=$b,,j$}{v2a}
\fmfv{l=$a,,i$}{v1b}
\fmfv{l=$a,,i$}{v2b}
\fmfdot{v1a}
\fmfdot{v2a}
\fmfdot{v1b}
\fmfdot{v2b}
\end{fmfgraph*}
\vspace{8mm}\end{minipage}
&
1
&
{$\sum_{ab}\(g_{\A,a}g_{\A,b}\o\pi\ep\)\sum_c\abs{\ep^{abc}}\[\g^j\otimes\g_j\]\[\s_c\otimes\s_c\]$
$+\sum_a\(-{g_{\A,a}^2\o\pi\ep}\)\[\g^j\otimes\g_j\]\[\id\otimes\id\]$}
&
\begin{minipage}{.12\textwidth}\vspace{8mm}\centering
\scriptsize
\begin{fmfgraph*}(18,8)
\fmfpen{thin}
\fmfstraight
\fmfleft{i2,i1}
\fmfright{o2,o1}
\fmf{phantom}{i1,v1a}
\fmf{phantom}{i2,v2a}
\fmf{phantom,tension=.6}{v1a,v1b}
\fmf{phantom,tension=.6}{v2a,v2b}
\fmf{phantom}{v1b,o1}
\fmf{phantom}{v2b,o2}
\fmffreeze
\fmf{fermion}{i1,v1a}
\fmf{fermion}{v1a,v1b}
\fmf{fermion}{v1b,o1}
\fmf{fermion}{i2,v2a}
\fmf{fermion}{v2a,v2b}
\fmf{fermion}{v2b,o2}
\fmf{dbl_plain}{v1a,v2b}
\fmf{dbl_plain}{v1b,v2a}
\fmfv{l=$b,,j$}{v1a}
\fmfv{l=$b,,j$}{v2b}
\fmfv{l=$a,,i$}{v1b}
\fmfv{l=$a,,i$}{v2a}
\fmfdot{v1a}
\fmfdot{v2a}
\fmfdot{v1b}
\fmfdot{v2b}
\end{fmfgraph*}
\vspace{8mm}\end{minipage}
&
1
&
{$\sum_{ab}\(g_{\A,a}g_{\A,b}\o\pi\ep\)\sum_c\abs{\ep^{abc}}\[\g^j\otimes\g_j\]\[\s_c\otimes\s_c\]$
$+\sum_a\({g_{\A,a}^2\o\pi\ep}\)\[\g^j\otimes\g_j\]\[\id\otimes\id\]$}
\\\hline
\begin{minipage}{.12\textwidth}\vspace{8mm}\centering
\scriptsize
\begin{fmfgraph*}(18,8)
\scriptsize
\fmfpen{thin}
\fmfstraight
\fmfleft{i2,i1}
\fmfright{o2,o1}
\fmf{phantom}{i1,v1a}
\fmf{phantom}{i2,v2a}
\fmf{phantom,tension=1.55}{v1a,v1c,v1b}
\fmf{phantom,tension=1.55}{v2a,v2c,v2b}
\fmf{phantom}{v1b,o1}
\fmf{phantom}{v2b,o2}
\fmffreeze
\fmf{fermion}{i1,v1a}
\fmf{fermion}{v1a,v1c}
\fmf{fermion}{v1c,v1b}
\fmf{fermion}{v1b,o1}
\fmf{fermion}{i2,v2a}
\fmf{plain}{v2a,v2c}
\fmf{plain}{v2c,v2b}
\fmf{fermion}{v2b,o2}
\fmf{dbl_plain}{v1c,v2c}
\fmf{photon,left,tension=0}{v1a,v1b}
\fmfv{l=$\m$}{v1b}
\fmfv{l=$\m$}{v1a}
\fmfv{l=$a$,l.a=-55}{v1c}
\fmfv{l=$a$}{v2c}
\fmfdot{v1a}
\fmfdot{v1c}
\fmfdot{v1b}
\fmfdot{v2c}
\fmfv{l=$a,,j$,l.a=-55}{v1c}
\fmfv{l=$a,,j$}{v2c}
\end{fmfgraph*}
\vspace{8mm}\end{minipage}
&
1
&
{$\sum_a\({16g_{\A,a}\mathpzc{g}^2\o3\pi^2(2N)\ep}\)\[\g^j\otimes\g_j\]\[\s^a\otimes\s_a\]$}
&
\begin{minipage}{.12\textwidth}\vspace{8mm}\centering
\scriptsize
\begin{fmfgraph*}(18,8)
\scriptsize
\fmfpen{thin}
\fmfstraight
\fmfleft{i2,i1}
\fmfright{o2,o1}
\fmf{phantom}{i1,v1a}
\fmf{phantom}{i2,v2a}
\fmf{phantom,tension=1.55}{v1a,v1c,v1b}
\fmf{phantom,tension=1.55}{v2a,v2c,v2b}
\fmf{phantom}{v1b,o1}
\fmf{phantom}{v2b,o2}
\fmffreeze
\fmf{fermion}{i1,v1a}
\fmf{fermion}{v1a,v1c}
\fmf{fermion}{v1c,v1b}
\fmf{fermion}{v1b,o1}
\fmf{fermion}{i2,v2a}
\fmf{plain}{v2a,v2c}
\fmf{plain}{v2c,v2b}
\fmf{fermion}{v2b,o2}
\fmf{dbl_wiggly}{v1c,v2c}
\fmf{dbl_wiggly,left,tension=0}{v1a,v1b}
\fmfv{l=$b,,0$}{v1b}
\fmfv{l=$b,,0$}{v1a}
\fmfv{l=$a$,l.a=-55}{v1c}
\fmfv{l=$a$}{v2c}
\fmfdot{v1a}
\fmfdot{v1c}
\fmfdot{v1b}
\fmfdot{v2c}
\fmfv{l=$a,,0$,l.a=-55}{v1c}
\fmfv{l=$a,,0$}{v2c}
\end{fmfgraph*}
\vspace{8mm}\end{minipage}
&
2
&
{$\sum_a\(-{g_{v,a}\o\pi\ep}\)\(2g_{v,a}-\sum_bg_{v,b}\)$
$\times\[\g^0\otimes\g^0\]\[\s^a\otimes\s_a\]$}
\\\hline
\begin{minipage}{.12\textwidth}\vspace{8mm}\centering
\scriptsize
\begin{fmfgraph*}(18,8)
\scriptsize
\fmfpen{thin}
\fmfstraight
\fmfleft{i2,i1}
\fmfright{o2,o1}
\fmf{phantom}{i1,v1a}
\fmf{phantom}{i2,v2a}
\fmf{phantom,tension=.6}{v1a,v1b}
\fmf{phantom,tension=.6}{v2a,v2b}
\fmf{phantom}{v1b,o1}
\fmf{phantom}{v2b,o2}
\fmffreeze
\fmf{fermion}{i1,v1a}
\fmf{fermion}{v1a,v1b}
\fmf{fermion}{v1b,o1}
\fmf{fermion}{i2,v2a}
\fmf{fermion}{v2a,v2b}
\fmf{fermion}{v2b,o2}
\fmf{dbl_wiggly}{v1a,v2b}
\fmf{dbl_wiggly}{v1b,v2a}
\fmfv{l=$b,,0$}{v1a}
\fmfv{l=$b,,0$}{v2b}
\fmfv{l=$a,,0$}{v1b}
\fmfv{l=$a,,0$}{v2a}
\fmfdot{v1a}
\fmfdot{v2a}
\fmfdot{v1b}
\fmfdot{v2b}
\end{fmfgraph*}
\vspace{8mm}\end{minipage}
&
1
&
{$
\sum_{ab}{g_{v,a}g_{v,b}\o4\pi\ep}\sum_c\abs{\ep^{abc}}\[\g^j\otimes\g_j\]\[\s_c\otimes\s_c\]$
$+
\sum_a\({g_{v,a}^2\o4\pi\ep}\)\[\g^j\otimes\g_j\]\[\id\otimes\id\]$}
&
\begin{minipage}{.12\textwidth}\vspace{8mm}\centering
\scriptsize
\begin{fmfgraph*}(18,8)
\scriptsize
\fmfpen{thin}
\fmfstraight
\fmfleft{i2,i1}
\fmfright{o2,o1}
\fmf{phantom}{i1,v1a}
\fmf{phantom}{i2,v2a}
\fmf{phantom,tension=.6}{v1a,v1b}
\fmf{phantom,tension=.6}{v2a,v2b}
\fmf{phantom}{v1b,o1}
\fmf{phantom}{v2b,o2}
\fmffreeze
\fmf{fermion}{i1,v1a}
\fmf{fermion}{v1a,v1b}
\fmf{fermion}{v1b,o1}
\fmf{fermion}{i2,v2a}
\fmf{fermion}{v2a,v2b}
\fmf{fermion}{v2b,o2}
\fmf{dbl_wiggly}{v1a,v2a}
\fmf{dbl_wiggly}{v1b,v2b}
\fmfv{l=$b,,0$}{v1a}
\fmfv{l=$b,,0$}{v2a}
\fmfv{l=$a,,0$}{v1b}
\fmfv{l=$a,,0$}{v2b}
\fmfdot{v1a}
\fmfdot{v2a}
\fmfdot{v1b}
\fmfdot{v2b}
\end{fmfgraph*}
\vspace{8mm}\end{minipage}
&
1
&
{$
\sum_{ab}{g_{v,a}g_{v,b}\o4\pi\ep}\sum_c\abs{\ep^{abc}}\[\g^j\otimes\g_j\]\[\s_c\otimes\s_c\]$
$+
\sum_a\(-{g_{v,a}^2\o4\pi\ep}\)\[\g^j\otimes\g_j\]\[\id\otimes\id\]$}
\\
\hline
\begin{minipage}{.12\textwidth}\vspace{8mm}\centering
\scriptsize
\begin{fmfgraph*}(18,8)
\scriptsize
\fmfpen{thin}
\fmfstraight
\fmfleft{i2,i1}
\fmfright{o2,o1}
\fmf{phantom}{i1,v1a}
\fmf{phantom}{i2,v2a}
\fmf{phantom,tension=1.55}{v1a,v1c,v1b}
\fmf{phantom,tension=1.55}{v2a,v2c,v2b}
\fmf{phantom}{v1b,o1}
\fmf{phantom}{v2b,o2}
\fmffreeze
\fmf{fermion}{i1,v1a}
\fmf{fermion}{v1a,v1c}
\fmf{fermion}{v1c,v1b}
\fmf{fermion}{v1b,o1}
\fmf{fermion}{i2,v2a}
\fmf{plain}{v2a,v2c}
\fmf{plain}{v2c,v2b}
\fmf{fermion}{v2b,o2}
\fmf{dbl_wiggly}{v1c,v2c}
\fmf{photon,left,tension=0}{v1a,v1b}
\fmfv{l=$\m$}{v1b}
\fmfv{l=$\m$}{v1a}
\fmfv{l=$a$,l.a=-55}{v1c}
\fmfv{l=$a$}{v2c}
\fmfdot{v1a}
\fmfdot{v1c}
\fmfdot{v1b}
\fmfdot{v2c}
\fmfv{l=$a,,0$,l.a=-55}{v1c}
\fmfv{l=$a,,0$}{v2c}
\end{fmfgraph*}
\vspace{8mm}\end{minipage}
&
2
&
{$\sum_a\(-{16g_{v,a}\mathpzc{g}^2\o3\pi^2(2N)\ep}\)\[\g^0\otimes\g^0\]\[\s^a\otimes\s_a\]$}
&
\begin{minipage}{.12\textwidth}\vspace{8mm}\centering
\scriptsize
\begin{fmfgraph*}(18,8)
\fmfpen{thin}
\fmfstraight
\fmfleft{i2,i1}
\fmfright{o2,o1}
\fmf{phantom}{i1,v1a}
\fmf{phantom}{i2,v2a}
\fmf{phantom,tension=1.55}{v1a,v1c,v1b}
\fmf{phantom,tension=1.55}{v2a,v2c,v2b}
\fmf{phantom}{v1b,o1}
\fmf{phantom}{v2b,o2}
\fmffreeze
\fmf{fermion}{i1,v1a}
\fmf{fermion}{v1a,v1c}
\fmf{fermion}{v1c,v1b}
\fmf{fermion}{v1b,o1}
\fmf{fermion}{i2,v2a}
\fmf{plain}{v2a,v2c}
\fmf{plain}{v2c,v2b}
\fmf{fermion}{v2b,o2}
\fmf{dbl_wiggly}{v1c,v2c}
\fmf{dbl_plain,left,tension=0}{v1a,v1b}
\fmfv{l=$b,,j$}{v1b}
\fmfv{l=$b,,j$}{v1a}
\fmfdot{v1a}
\fmfdot{v1c}
\fmfdot{v1b}
\fmfdot{v2c}
\fmfv{l=$a,,0$,l.a=-55}{v1c}
\fmfv{l=$a,,0$}{v2c}
\end{fmfgraph*}
\vspace{8mm}\end{minipage}
&
2
&
{$\sum_a\(-{2g_{v,a}\o\pi\ep}\)\(2g_{\A,a}-\sum_bg_{\A,b}\)$
$\times\[\g^0\otimes\g^0\]\[\s^a\otimes\s_a\]$}
\\\hline
\end{tabular}
\caption{Feynman diagrams which determine the bilinear counter terms.}
\label{tab:DiagramTab1}
\end{table}
\begin{table}
\centering\scriptsize
\begin{tabular}{| c | c |  m{0.35\textwidth} || c |  c | m{0.35\textwidth} |}
\hline
\normalsize
diagram & \normalsize$n_d$ &\normalsize divergence & \normalsize diagram & \normalsize $n_d$ & \normalsize divergence
\\\hline\hline
\begin{minipage}{.12\textwidth}\vspace{8mm}\centering
\scriptsize
\begin{fmfgraph*}(18,8)
\fmfpen{thin}
\fmfstraight
\fmfleft{i2,i1}
\fmfright{o2,o1}
\fmf{phantom}{i1,v1a}
\fmf{phantom}{i2,v2a}
\fmf{phantom,tension=.6}{v1a,v1b}
\fmf{phantom,tension=.6}{v2a,v2b}
\fmf{phantom}{v1b,o1}
\fmf{phantom}{v2b,o2}
\fmffreeze
\fmf{fermion}{i1,v1a}
\fmf{fermion}{v1a,v1b}
\fmf{fermion}{v1b,o1}
\fmf{fermion}{i2,v2a}
\fmf{fermion}{v2a,v2b}
\fmf{fermion}{v2b,o2}
\fmf{dbl_plain}{v1a,v2b}
\fmf{dbl_wiggly}{v1b,v2a}
\fmfv{l=$b,,j$}{v1a}
\fmfv{l=$b,,j$}{v2b}
\fmfv{l=$a,,0$}{v1b}
\fmfv{l=$a,,0$}{v2a}
\fmfdot{v1a}
\fmfdot{v2a}
\fmfdot{v1b}
\fmfdot{v2b}
\end{fmfgraph*}
\vspace{8mm}\end{minipage}
&
2
&{
$\sum_a{g_{v,a}g_{\A,a}\o\pi\ep}\bigg\{{-\[\id\otimes\id\]}\[\id\otimes\id\]$
$-\[\g^0\otimes\g^0\]\[\id\otimes\id\]\bigg\}$
$+\sum_{ab}{g_{v,a}g_{\A,b}\o\pi\ep}\sum_c\abs{\ep^{abc}}\bigg\{$
$-\[\id\otimes\id\]\[\s_c\otimes\s_c\]$
$-\[\g^0\otimes\g^0\]\[\s_c\otimes\s_c\]\bigg\}
$}
&
\begin{minipage}{.12\textwidth}\vspace{8mm}\centering
\scriptsize
\begin{fmfgraph*}(18,8)
\fmfpen{thin}
\fmfstraight
\fmfleft{i2,i1}
\fmfright{o2,o1}
\fmf{phantom}{i1,v1a}
\fmf{phantom}{i2,v2a}
\fmf{phantom,tension=.6}{v1a,v1b}
\fmf{phantom,tension=.6}{v2a,v2b}
\fmf{phantom}{v1b,o1}
\fmf{phantom}{v2b,o2}
\fmffreeze
\fmf{fermion}{i1,v1a}
\fmf{fermion}{v1a,v1b}
\fmf{fermion}{v1b,o1}
\fmf{fermion}{i2,v2a}
\fmf{fermion}{v2a,v2b}
\fmf{fermion}{v2b,o2}
\fmf{dbl_plain}{v1a,v2a}
\fmf{dbl_wiggly}{v1b,v2b}
\fmfv{l=$b,,j$}{v1a}
\fmfv{l=$b,,j$}{v2a}
\fmfv{l=$a,,0$}{v1b}
\fmfv{l=$a,,0$}{v2b}
\fmfdot{v1a}
\fmfdot{v2a}
\fmfdot{v1b}
\fmfdot{v2b}
\end{fmfgraph*}\vspace{8mm}\end{minipage}
&
2
&
$\sum_a{g_{v,a}g_{\A,a}\o\pi\ep}\bigg\{{-\[\id\otimes\id\]}\[\id\otimes\id\]$
$+\[\g^0\otimes\g^0\]\[\id\otimes\id\]\bigg\}$
$+\sum_{ab}{g_{v,a}g_{\A,b}\o\pi\ep}\sum_c\abs{\ep^{abc}}\bigg\{
\[\id\otimes\id\]\[\s_c\otimes\s_c\]$
$-\[\g^0\otimes\g^0\]\[\s_c\otimes\s_c\]\bigg\}$
\\\hline
\begin{minipage}{.12\textwidth}\vspace{8mm}\centering
\scriptsize
\begin{fmfgraph*}(18,8)
\fmfpen{thin}
\fmfstraight
\fmfleft{i2,i1}
\fmfright{o2,o1}
\fmf{phantom}{i1,v1a}
\fmf{phantom}{i2,v2a}
\fmf{phantom,tension=.6}{v1a,v1b}
\fmf{phantom,tension=.6}{v2a,v2b}
\fmf{phantom}{v1b,o1}
\fmf{phantom}{v2b,o2}
\fmffreeze
\fmf{fermion}{i1,v1a}
\fmf{fermion}{v1a,v1b}
\fmf{fermion}{v1b,o1}
\fmf{fermion}{i2,v2a}
\fmf{fermion}{v2a,v2b}
\fmf{fermion}{v2b,o2}
\fmf{dashes}{v1a,v2a}
\fmf{dbl_wiggly}{v1b,v2b}
\fmfv{l=$a,,0$}{v1b}
\fmfv{l=$a,,0$}{v2b}
\fmfdot{v1a}
\fmfdot{v2a}
\fmfdot{v1b}
\fmfdot{v2b}
\end{fmfgraph*}
\vspace{8mm}\end{minipage}
&
2
&
{
$\sum_a\({ g_sg_{v,a}\o2\pi\ep}\)\[\g^j\otimes\g_j\]\[\s^a\otimes\s_a\]
$}
&
\begin{minipage}{.12\textwidth}\vspace{8mm}\centering
\scriptsize
\begin{fmfgraph*}(18,8)
\fmfpen{thin}
\fmfstraight
\fmfleft{i2,i1}
\fmfright{o2,o1}
\fmf{phantom}{i1,v1a}
\fmf{phantom}{i2,v2a}
\fmf{phantom,tension=.6}{v1a,v1b}
\fmf{phantom,tension=.6}{v2a,v2b}
\fmf{phantom}{v1b,o1}
\fmf{phantom}{v2b,o2}
\fmffreeze
\fmf{fermion}{i1,v1a}
\fmf{fermion}{v1a,v1b}
\fmf{fermion}{v1b,o1}
\fmf{fermion}{i2,v2a}
\fmf{fermion}{v2a,v2b}
\fmf{fermion}{v2b,o2}
\fmf{dashes}{v1a,v2b}
\fmf{dbl_wiggly}{v1b,v2a}
\fmfv{l=$a,,0$}{v1b}
\fmfv{l=$a,,0$}{v2a}
\fmfdot{v1a}
\fmfdot{v2a}
\fmfdot{v1b}
\fmfdot{v2b}
\end{fmfgraph*}
\vspace{8mm}\end{minipage}
&
2
&{
$\sum_a\({ g_sg_{v,a}\o2\pi\ep}\)\[\g^j\otimes\g_j\]\[\s^a\otimes\s_a\]
$}
\\\hline
\begin{minipage}{.12\textwidth}\vspace{8mm}\centering
\scriptsize\begin{fmfgraph*}(18,8)
\fmfpen{thin}
\fmfstraight
\fmfleft{i2,i1}
\fmfright{o2,o1}
\fmf{phantom}{i1,v1a}
\fmf{phantom}{i2,v2a}
\fmf{phantom,tension=1.55}{v1a,v1c,v1b}
\fmf{phantom,tension=1.55}{v2a,v2c,v2b}
\fmf{phantom}{v1b,o1}
\fmf{phantom}{v2b,o2}
\fmffreeze
\fmf{fermion}{i1,v1a}
\fmf{fermion}{v1a,v1c}
\fmf{fermion}{v1c,v1b}
\fmf{fermion}{v1b,o1}
\fmf{fermion}{i2,v2a}
\fmf{plain}{v2a,v2c}
\fmf{plain}{v2c,v2b}
\fmf{fermion}{v2b,o2}
\fmf{dbl_wiggly}{v1c,v2c}
\fmf{dashes,left,tension=0}{v1a,v1b}
\fmfdot{v1a}
\fmfdot{v1c}
\fmfdot{v1b}
\fmfdot{v2c}
\fmfv{l=$a,,0$,l.a=-55}{v1c}
\fmfv{l=$a,,0$}{v2c}
\end{fmfgraph*}
\vspace{8mm}\end{minipage}
&
2
&
{$\sum_a\(-{ g_sg_{v,a}\o\pi\ep}\)\[\g^0\otimes\g^0\]\[\s^a\otimes\s_a\]$}
&
\begin{minipage}{.12\textwidth}\vspace{8mm}\centering
\scriptsize
\begin{fmfgraph*}(18,8)\fmfpen{thin}
\fmfstraight
\fmfleft{i2,i1}
\fmfright{o2,o1}
\fmf{phantom}{i1,v1a}
\fmf{phantom}{i2,v2a}
\fmf{phantom,tension=1.55}{v1a,v1c,v1b}
\fmf{phantom,tension=1.55}{v2a,v2c,v2b}
\fmf{phantom}{v1b,o1}
\fmf{phantom}{v2b,o2}
\fmffreeze
\fmf{fermion}{i1,v1a}
\fmf{fermion}{v1a,v1c}
\fmf{fermion}{v1c,v1b}
\fmf{fermion}{v1b,o1}
\fmf{fermion}{i2,v2a}
\fmf{plain}{v2a,v2c}
\fmf{plain}{v2c,v2b}
\fmf{fermion}{v2b,o2}
\fmf{dashes}{v1c,v2c}
\fmf{dbl_wiggly,left,tension=0}{v1a,v1b}
\fmfv{l=$a,,0$}{v1b}
\fmfv{l=$a,,0$}{v1a}
\fmfdot{v1a}
\fmfdot{v1c}
\fmfdot{v1b}
\fmfdot{v2c}
\end{fmfgraph*}
\vspace{8mm}\end{minipage}
&
2
&
{$\[\id\otimes\id\]\[\id\otimes\id\]\({ g_s\o\pi\ep}\sum_ag_{v,a}\)$}
\\\hline
\begin{minipage}{.12\textwidth}\vspace{8mm}\centering
\scriptsize
\begin{fmfgraph*}(18,8)
\fmfpen{thin}
\fmfstraight
\fmfleft{i2,i1}
\fmfright{o2,o1}
\fmf{phantom}{i1,v1a}
\fmf{phantom}{i2,v2a}
\fmf{phantom,tension=.6}{v1a,v1b}
\fmf{phantom,tension=.6}{v2a,v2b}
\fmf{phantom}{v1b,o1}
\fmf{phantom}{v2b,o2}
\fmffreeze
\fmf{fermion}{i1,v1a}
\fmf{fermion}{v1a,v1b}
\fmf{fermion}{v1b,o1}
\fmf{fermion}{i2,v2a}
\fmf{fermion}{v2a,v2b}
\fmf{fermion}{v2b,o2}
\fmf{dashes}{v1a,v2a}
\fmf{dbl_plain}{v1b,v2b}
\fmfv{l=$a,,j$}{v1b}
\fmfv{l=$a,,j$}{v2b}
\fmfdot{v1a}
\fmfdot{v2a}
\fmfdot{v1b}
\fmfdot{v2b}
\end{fmfgraph*}\vspace{8mm}\end{minipage}
&
2
&{
$\sum_a\( g_sg_{\A,a}\o\pi\ep\)\[\id\otimes\id\]\[\s^a\otimes\s_a\]$
$+\sum_a\(-{ g_sg_{\A,a}\o\pi\ep}\)\[\g^0\otimes\g^0\]\[\s^a\otimes\s_a\]
$}
&
\begin{minipage}{.12\textwidth}\vspace{8mm}\centering
\scriptsize
\begin{fmfgraph*}(18,8)
\fmfpen{thin}
\fmfstraight
\fmfleft{i2,i1}
\fmfright{o2,o1}
\fmf{phantom}{i1,v1a}
\fmf{phantom}{i2,v2a}
\fmf{phantom,tension=.6}{v1a,v1b}
\fmf{phantom,tension=.6}{v2a,v2b}
\fmf{phantom}{v1b,o1}
\fmf{phantom}{v2b,o2}
\fmffreeze
\fmf{fermion}{i1,v1a}
\fmf{fermion}{v1a,v1b}
\fmf{fermion}{v1b,o1}
\fmf{fermion}{i2,v2a}
\fmf{fermion}{v2a,v2b}
\fmf{fermion}{v2b,o2}
\fmf{dashes}{v1a,v2b}
\fmf{dbl_plain}{v1b,v2a}
\fmfv{l=$a,,j$}{v1b}
\fmfv{l=$a,,j$}{v2a}
\fmfdot{v1a}
\fmfdot{v2a}
\fmfdot{v1b}
\fmfdot{v2b}
\end{fmfgraph*}
\vspace{8mm}\end{minipage}
&
2
&
{
$\sum_a\(-{ g_sg_{\A,a}\o\pi\ep}\)\[\id\otimes\id\]\[\s^a\otimes\s_a\]$
$+\sum_a\(-{ g_sg_{\A,a}\o\pi\ep}\)\[\g^0\otimes\g^0\]\[\s^a\otimes\s_a\]
$}
\\\hline
\begin{minipage}{.12\textwidth}\vspace{8mm}\centering
\scriptsize
\begin{fmfgraph*}(18,8)
\fmfpen{thin}
\fmfstraight
\fmfleft{i2,i1}
\fmfright{o2,o1}
\fmf{phantom}{i1,v1a}
\fmf{phantom}{i2,v2a}
\fmf{phantom,tension=1.55}{v1a,v1c,v1b}
\fmf{phantom,tension=1.55}{v2a,v2c,v2b}
\fmf{phantom}{v1b,o1}
\fmf{phantom}{v2b,o2}
\fmffreeze
\fmf{fermion}{i1,v1a}
\fmf{fermion}{v1a,v1c}
\fmf{fermion}{v1c,v1b}
\fmf{fermion}{v1b,o1}
\fmf{fermion}{i2,v2a}
\fmf{plain}{v2a,v2c}
\fmf{plain}{v2c,v2b}
\fmf{fermion}{v2b,o2}
\fmf{dashes}{v1c,v2c}
\fmf{dbl_plain,left,tension=0}{v1a,v1b}
\fmfv{l=$a,,j$}{v1b}
\fmfv{l=$a,,j$}{v1a}
\fmfdot{v1a}
\fmfdot{v1c}
\fmfdot{v1b}
\fmfdot{v2c}
\end{fmfgraph*}
\vspace{8mm}\end{minipage}
&
2
&
{$\(-{2 g_s\o\pi\ep}\sum_ag_{\A,a}\)\[\id\otimes\id\]\[\id\otimes\id\]$}
&
\begin{minipage}{.12\textwidth}\vspace{8mm}\centering
\scriptsize
\begin{fmfgraph*}(18,8)
\fmfpen{thin}
\fmfstraight
\fmfleft{i2,i1}
\fmfright{o2,o1}
\fmf{phantom}{i1,v1a}
\fmf{phantom}{i2,v2a}
\fmf{phantom,tension=.6}{v1a,v1b}
\fmf{phantom,tension=.6}{v2a,v2b}
\fmf{phantom}{v1b,o1}
\fmf{phantom}{v2b,o2}
\fmffreeze
\fmf{fermion}{i1,v1a}
\fmf{fermion}{v1a,v1b}
\fmf{fermion}{v1b,o1}
\fmf{fermion}{i2,v2a}
\fmf{fermion}{v2a,v2b}
\fmf{fermion}{v2b,o2}
\fmf{dbl_dashes}{v1a,v2a}
\fmf{dbl_wiggly}{v1b,v2b}
\fmfv{l=$a,,0$}{v1b}
\fmfv{l=$a,,0$}{v2b}
\fmfv{l=$b$}{v1a}
\fmfv{l=$b$}{v2a}
\fmfdot{v1a}
\fmfdot{v2a}
\fmfdot{v1b}
\fmfdot{v2b}
\end{fmfgraph*}
\vspace{8mm}\end{minipage}
&
2
&
{
$
\sum_{a}\(g_{v,a}g_{t,a}\o2\pi\ep\)\[\g^j\otimes\g_j\]\[\id\otimes\id\]$
$-\sum_{ab}\({g_{v,a}g_{t,b}\o2\pi\ep}\)\sum_c\abs{\ep^{abc}}\[\g^j\otimes\g_j\]\[\s_c\otimes\s_c\]
$}
\\\hline
\begin{minipage}{.12\textwidth}\vspace{8mm}\centering
\scriptsize
\begin{fmfgraph*}(18,8)
\fmfpen{thin}
\fmfstraight
\fmfleft{i2,i1}
\fmfright{o2,o1}
\fmf{phantom}{i1,v1a}
\fmf{phantom}{i2,v2a}
\fmf{phantom,tension=.6}{v1a,v1b}
\fmf{phantom,tension=.6}{v2a,v2b}
\fmf{phantom}{v1b,o1}
\fmf{phantom}{v2b,o2}
\fmffreeze
\fmf{fermion}{i1,v1a}
\fmf{fermion}{v1a,v1b}
\fmf{fermion}{v1b,o1}
\fmf{fermion}{i2,v2a}
\fmf{fermion}{v2a,v2b}
\fmf{fermion}{v2b,o2}
\fmf{dbl_dashes}{v1a,v2b}
\fmf{dbl_wiggly}{v1b,v2a}
\fmfv{l=$a,,0$}{v1b}
\fmfv{l=$a,,0$}{v2a}
\fmfv{l=$b$}{v1a}
\fmfv{l=$b$}{v2b}
\fmfdot{v1a}
\fmfdot{v2a}
\fmfdot{v1b}
\fmfdot{v2b}
\end{fmfgraph*}
\vspace{8mm}\end{minipage}
&
2
&
{
$\sum_{a}\(g_{v,a}g_{t,a}\o2\pi\ep\)\[\g^j\otimes\g_j\]\[\id\otimes\id\]$
$+\sum_{ab}\({g_{v,a}g_{t,b}\o2\pi\ep}\)\sum_c\abs{\ep^{abc}}\[\g^j\otimes\g_j\]\[\s_c\otimes\s_c\]
$}
&
\begin{minipage}{.12\textwidth}\vspace{8mm}\centering
\scriptsize
\begin{fmfgraph*}(18,8)
\fmfpen{thin}
\fmfstraight
\fmfleft{i2,i1}
\fmfright{o2,o1}
\fmf{phantom}{i1,v1a}
\fmf{phantom}{i2,v2a}
\fmf{phantom,tension=1.55}{v1a,v1c,v1b}
\fmf{phantom,tension=1.55}{v2a,v2c,v2b}
\fmf{phantom}{v1b,o1}
\fmf{phantom}{v2b,o2}
\fmffreeze
\fmf{fermion}{i1,v1a}
\fmf{fermion}{v1a,v1c}
\fmf{fermion}{v1c,v1b}
\fmf{fermion}{v1b,o1}
\fmf{fermion}{i2,v2a}
\fmf{plain}{v2a,v2c}
\fmf{plain}{v2c,v2b}
\fmf{fermion}{v2b,o2}
\fmf{dbl_wiggly}{v1c,v2c}
\fmf{dbl_dashes,left,tension=0}{v1a,v1b}
\fmfdot{v1a}
\fmfdot{v1c}
\fmfdot{v1b}
\fmfdot{v2c}
\fmfv{l=$a,,0$,l.a=-55}{v1c}
\fmfv{l=$a,,0$}{v2c}
\fmfv{l=$b$}{v1a}
\fmfv{l=$b$}{v1b}
\end{fmfgraph*}
\vspace{8mm}\end{minipage}
&
2
&
{$\sum_a\(-{g_{v,a}\o\pi\ep}\)\(2g_{t,a}-\sum_bg_{t,b}\)$
$\times\[\g^0\otimes\g^0\]\[\s^a\otimes\s_a\]$}
\\\hline
\begin{minipage}{.12\textwidth}\vspace{8mm}\centering
\scriptsize
\begin{fmfgraph*}(18,8)
\fmfpen{thin}
\fmfstraight
\fmfleft{i2,i1}
\fmfright{o2,o1}
\fmf{phantom}{i1,v1a}
\fmf{phantom}{i2,v2a}
\fmf{phantom,tension=1.55}{v1a,v1c,v1b}
\fmf{phantom,tension=1.55}{v2a,v2c,v2b}
\fmf{phantom}{v1b,o1}
\fmf{phantom}{v2b,o2}
\fmffreeze
\fmf{fermion}{i1,v1a}
\fmf{fermion}{v1a,v1c}
\fmf{fermion}{v1c,v1b}
\fmf{fermion}{v1b,o1}
\fmf{fermion}{i2,v2a}
\fmf{plain}{v2a,v2c}
\fmf{plain}{v2c,v2b}
\fmf{fermion}{v2b,o2}
\fmf{dbl_dashes}{v1c,v2c}
\fmf{dbl_wiggly,left,tension=0}{v1a,v1b}
\fmfv{l=$a,,0$}{v1b}
\fmfv{l=$a,,0$}{v1a}
\fmfv{l=$b$,l.a=-55}{v1c}
\fmfv{l=$b$}{v2c}
\fmfdot{v1a}
\fmfdot{v1c}
\fmfdot{v1b}
\fmfdot{v2c}
\end{fmfgraph*}
\vspace{8mm}\end{minipage}
&
2
&
{$\sum_a\({g_{t,a}\o\pi\ep}\)\(2g_{v,a}-\sum_bg_{v,b}\)\[\id\otimes\id\]\[\s^a\otimes\s_a\]$}
&
\begin{minipage}{.1\textwidth}\vspace{8mm}\centering
\begin{fmfgraph*}(18,8)
\fmfpen{thin}
\fmfstraight
\fmfleft{i2,i1}
\fmfright{o2,o1}
\fmf{phantom}{i1,v1a}
\fmf{phantom}{i2,v2a}
\fmf{phantom,tension=.6}{v1a,v1b}
\fmf{phantom,tension=.6}{v2a,v2b}
\fmf{phantom}{v1b,o1}
\fmf{phantom}{v2b,o2}
\fmffreeze
\fmf{fermion}{i1,v1a}
\fmf{fermion}{v1a,v1b}
\fmf{fermion}{v1b,o1}
\fmf{fermion}{i2,v2a}
\fmf{fermion}{v2a,v2b}
\fmf{fermion}{v2b,o2}
\fmf{dbl_dashes}{v1a,v2a}
\fmf{dbl_plain}{v1b,v2b}
\fmfv{l=$a,,j$}{v1b}
\fmfv{l=$a,,j$}{v2b}
\fmfv{l=$b$}{v1a}
\fmfv{l=$b$}{v2a}
\fmfdot{v1a}
\fmfdot{v2a}
\fmfdot{v1b}
\fmfdot{v2b}
\end{fmfgraph*}
\vspace{8mm}\end{minipage}
&
2
&{
$\sum_a{g_{t,a}g_{\A,a}\o\pi\ep}\bigg\{
\[\id\otimes\id\]\[\id\otimes\id\]$
${-\[\g^0\otimes\g^0\]\[\id\otimes\id\]\bigg\}}$
$+\sum_{ab}{g_{t,a}g_{\A,b}\o\pi\ep}\sum_c\abs{\ep^{abc}}\bigg\{
{-\[\id\otimes\id\]}\[\s^c\otimes\s_c\]$
$+\[\g^0\otimes\g^0\]\[\s^a\otimes\s_a\]\bigg\}
$}
\\\hline
\end{tabular}
\caption{Feynman diagrams which determine the bilinear counter terms.}
\label{tab:DiagramTab2}
\end{table}
\begin{table}
\scriptsize\centering
\begin{tabular}{| c | c |  m{0.35\textwidth} || c |  c | m{0.35\textwidth} |}
\hline
\normalsize
diagram & \normalsize$n_d$ &\normalsize divergence & \normalsize diagram & \normalsize $n_d$ & \normalsize divergence
\\\hline\hline
\begin{minipage}{.12\textwidth}\vspace{8mm}\centering
\scriptsize
\begin{fmfgraph*}(18,8)
\fmfpen{thin}
\fmfstraight
\fmfleft{i2,i1}
\fmfright{o2,o1}
\fmf{phantom}{i1,v1a}
\fmf{phantom}{i2,v2a}
\fmf{phantom,tension=.6}{v1a,v1b}
\fmf{phantom,tension=.6}{v2a,v2b}
\fmf{phantom}{v1b,o1}
\fmf{phantom}{v2b,o2}
\fmffreeze
\fmf{fermion}{i1,v1a}
\fmf{fermion}{v1a,v1b}
\fmf{fermion}{v1b,o1}
\fmf{fermion}{i2,v2a}
\fmf{fermion}{v2a,v2b}
\fmf{fermion}{v2b,o2}
\fmf{dbl_dashes}{v1a,v2b}
\fmf{dbl_plain}{v1b,v2a}
\fmfv{l=$a,,j$}{v1b}
\fmfv{l=$a,,j$}{v2a}
\fmfv{l=$b$}{v1a}
\fmfv{l=$b$}{v2b}
\fmfdot{v1a}
\fmfdot{v2a}
\fmfdot{v1b}
\fmfdot{v2b}
\end{fmfgraph*}
\vspace{8mm}\end{minipage}
&
2
&
{
$\sum_a{g_{t,a}g_{\A,a}\o\pi\ep}\bigg\{
{-\[\id\otimes\id\]\[\id\otimes\id\]}$
${-\[\g^0\otimes\g^0\]\[\id\otimes\id\]}\bigg\}$ 
$+\sum_{ab} {g_{t,a}g_{\A,b}\o\pi\ep}\sum_c\abs{\ep^{abc}}
\bigg\{ {-\[\id\otimes\id\]}\[\s^c\otimes\s_c\]$ 
${-\[\g^0\otimes\g^0\]}\[\s^a\otimes\s_a\]\bigg\}
$}
&
\begin{minipage}{.12\textwidth}\vspace{8mm}\centering
\scriptsize
\begin{fmfgraph*}(18,8)
\fmfpen{thin}
\fmfstraight
\fmfleft{i2,i1}
\fmfright{o2,o1}
\fmf{phantom}{i1,v1a}
\fmf{phantom}{i2,v2a}
\fmf{phantom,tension=1.55}{v1a,v1c,v1b}
\fmf{phantom,tension=1.55}{v2a,v2c,v2b}
\fmf{phantom}{v1b,o1}
\fmf{phantom}{v2b,o2}
\fmffreeze
\fmf{fermion}{i1,v1a}
\fmf{fermion}{v1a,v1c}
\fmf{fermion}{v1c,v1b}
\fmf{fermion}{v1b,o1}
\fmf{fermion}{i2,v2a}
\fmf{plain}{v2a,v2c}
\fmf{plain}{v2c,v2b}
\fmf{fermion}{v2b,o2}
\fmf{dbl_dashes}{v1c,v2c}
\fmf{dbl_plain,left,tension=0}{v1a,v1b}
\fmfv{l=$a,,j$}{v1b}
\fmfv{l=$a,,j$}{v1a}
\fmfv{l=$b$,l.a=-55}{v1c}
\fmfv{l=$b$}{v2c}
\fmfdot{v1a}
\fmfdot{v1c}
\fmfdot{v1b}
\fmfdot{v2c}
\end{fmfgraph*}
\vspace{8mm}\end{minipage}
&
2
&
{$\sum_a\(-{2g_{t,a}\o\pi\ep}\)\(2g_{\A,a}-\sum_bg_{\A,b}\)$
$\times\[\id\otimes\id\]\[\s^a\otimes\s_a\]$}
\\\hline
\begin{minipage}{.12\textwidth}\vspace{8mm}\centering
\scriptsize
\begin{fmfgraph*}(18,12)
\fmfpen{thin}
\fmfstraight
\fmfleft{i2,i1}
\fmfright{o2,o1}
\fmf{phantom}{i1,v1a}
\fmf{phantom}{i2,v2a}
\fmf{phantom,tension=.8}{v1a,v1c,v1b}
\fmf{phantom,tension=.8}{v2a,v2c,v2b}
\fmf{phantom}{v1b,o1}
\fmf{phantom}{v2b,o2}
\fmffreeze
\fmf{phantom,tension=0.7}{v1a,l1}
\fmf{phantom,tension=0.7}{v1b,r1}
\fmf{phantom,tension=0.7}{v1c,c1}
\fmf{phantom}{l1,l2,v2a}
\fmf{phantom}{r1,r2,v2b}
\fmf{phantom}{c1,c2,v2c}
\fmffreeze
\fmf{fermion}{i1,v1c}
\fmf{fermion}{v1c,o1}
\fmf{fermion}{i2,v2a}
\fmf{fermion}{v2a,v2b}
\fmf{fermion}{v2b,o2}
\fmf{dbl_dashes}{v1c,c1}
\fmf{fermion}{c1,l2}
\fmf{fermion}{l2,r2}
\fmf{fermion}{r2,c1}
\fmf{photon}{l2,v2a}
\fmf{dbl_plain}{r2,v2b}
\fmfdot{v1c}
\fmfdot{v2a}
\fmfdot{v2b}
\fmfdot{c1}
\fmfdot{r2}
\fmfdot{l2}
\fmfv{l=$b$}{v1c}
\fmfv{l=$b$,l.a=0}{c1}
\fmfv{l=$a,,j$}{v2b}
\fmfv{l=$a,,j$}{r2}
\fmfv{l=$\m$}{l2}
\fmfv{l=$\m$}{v2a}
\end{fmfgraph*}
\vspace{8mm}\end{minipage}
&
8
&
{$\sum_a{4g_{t,a}g_{\A,a}\mathpzc{g}^2\o\pi\ep}\[\id\otimes\id\]\[\s^a\otimes\s_a\]$}
&
\begin{minipage}{.12\textwidth}\vspace{8mm}\centering
\scriptsize
\begin{fmfgraph*}(18,12)
\fmfpen{thin}
\fmfstraight
\fmfleft{i2,i1}
\fmfright{o2,o1}
\fmf{phantom}{i1,v1a}
\fmf{phantom}{i2,v2a}
\fmf{phantom,tension=.8}{v1a,v1c,v1b}
\fmf{phantom,tension=.8}{v2a,v2c,v2b}
\fmf{phantom}{v1b,o1}
\fmf{phantom}{v2b,o2}
\fmffreeze
\fmf{phantom,tension=0.7}{v1a,l1}
\fmf{phantom,tension=0.7}{v1b,r1}
\fmf{phantom,tension=0.7}{v1c,c1}
\fmf{phantom}{l1,l2,v2a}
\fmf{phantom}{r1,r2,v2b}
\fmf{phantom}{c1,c2,v2c}
\fmffreeze
\fmf{fermion}{i1,v1c}
\fmf{fermion}{v1c,o1}
\fmf{fermion}{i2,v2a}
\fmf{fermion}{v2a,v2b}
\fmf{fermion}{v2b,o2}
\fmf{dbl_dashes}{v1c,c1}
\fmf{fermion}{c1,l2}
\fmf{fermion}{l2,r2}
\fmf{fermion}{r2,c1}
\fmf{photon}{l2,v2a}
\fmf{dbl_wiggly}{r2,v2b}
\fmfdot{v1c}
\fmfdot{v2a}
\fmfdot{v2b}
\fmfdot{c1}
\fmfdot{r2}
\fmfdot{l2}
\fmfv{l=$b$}{v1c}
\fmfv{l=$b$,l.a=0}{c1}
\fmfv{l=$a,,0$}{v2b}
\fmfv{l=$a,,0$}{r2}
\fmfv{l=$\m$}{l2}
\fmfv{l=$\m$}{v2a}
\end{fmfgraph*}
\vspace{8mm}\end{minipage}
&
8
&
{$\sum_a\(-{4g_{t,a}g_{v,a}\mathpzc{g}^2\o\pi\ep}\)\[\id\otimes\id\]\[\s^a\otimes\s_a\]$}
\\\hline
\end{tabular}
\caption{Feynman diagrams which determine the bilinear counter terms.}
\label{tab:DiagramTab3}
\end{table}
\end{fmffile}
}

%% file: FluxDisorder2.tex
%
%
%
%
%
%

The renormalization of $g_\E$ and $g_\B$ result from terms in the photon self-energy which are proportional to $2\pi\d(p_0)$.
It follows that the usual $1/2N$ corrections to the photon propagator, like shown in Fig.~\ref{fig:photoSelfE}, do not renormalize the flux disorder.

In order to renormalize $g_\E$ and $g_\B$ we must have a disorder line going through the middle.
This would allow a diagram like that shown in Fig.~\ref{fig:FluxTwo Loop}. 
The trace over fermion flavours means that the only disorder we could place between the two loops is the singlet mass-like disorder, with coupling $g_s$.
This diagram is $\O(2Ng_s)\sim\O(1)$ and so thankfully it vanishes:
\eq{
\text{Fig.~\ref{fig:FluxTwo Loop}}&=
2\pi\d(p_0)2N\m^{-2\ep}\mathpzc{g}^2 g_s
\underbrace{\int{d^Dq\o(2\pi)^D}\tr\[{iq_\a\g^\a\o q^2}{i(q+p)_\b\g^\b\o(q+p)^2}i\g^\m\]}_{\mathcal{I}^\m(p)}
\underbrace{\int{d^Dk\o(2\pi)^D}\tr\[{ik_\s\g^\s\o k^2}{i(k+p)_\r\g^\r\o(k+p)^2}i\g^\n\]}_{\mathcal{I}^\n(p)}
}
where
\eq{
\mathcal{I}^\m(p)&=-i\tr\[\g^\a\g^\b\g^\m\]\int{d^Dq\o(2\pi)^D}\int_0^1dx{q_\a q_\b-x(1-x)p_\a p_b\o\[q^2+x(1-x)p^2\]^2}
 =0.
}
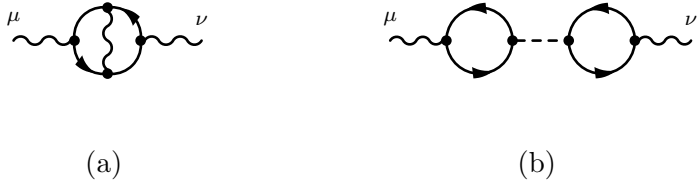
\begin{figure}
\input{FluxRenormVanish}
\caption{Diagrams which enter into the photon self-energy at leading order. (a) will not renormalize the disorder and (b) vanishes.}
\end{figure}

\begin{figure}
\input{3LoopFluxRenorm.tex}
\caption{Diagrams which renormalizes the flux disorder at $\O(g_\xi,g_\xi/2N)$.  
Depending on whether the internal indices are $(\s,\r)=(0,0)$ or $(i,j)$, the coupling constant are $-g_\B$ or $g_\E$ respectively.}
\label{fig:3LoopRenorm}
\end{figure}
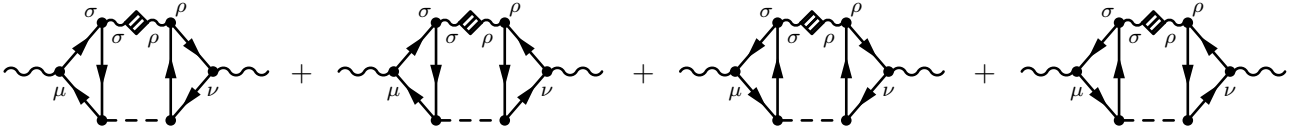

We next consider the situation with two internal disorder lines. These lines must go between the two bubbles  otherwise they will be cancelled by a vertex or a field strength renormalization and will not lead to a renormalization of the flux disorder.
Furthermore, one of the internal lines must correspond to a flux disorder interaction since otherwise the divergence will be cancelled by one of the bilinear disorder counter terms we determined in the previous two sections.  
This leaves the diagrams with one internal disorder line coupling to the topological current and one to the mass since all other bilinear disorder types will vanish upon tracing over the flavour indices.
These diagrams are shown in Fig.~\ref{fig:3LoopRenorm}.
Depending on whether the internal indices $(\s,\r)$ are $(0,0)$ or $(i,j)$ the diagrams are proportional to $-\mathpzc{g}^4g_sg_\b$ or $\mathpzc{g}^4g_sg_\E$ respectively. They therefore contribute at the same order as the diagrams in the previous two sections.
We note that diagrams which two internal flux disorder lines appear at a order in $g_\xi$ and $1/2N$.

Ignoring coupling constants for the moment, for any give $\m,\n, \s,$ and $\r$, it's easy to check that the four diagrams being added in Fig.~\ref{fig:3LoopRenorm} all have the same value. 
Therefore, their sum is equal to
\eq{
\text{Fig.~\ref{fig:3LoopRenorm}}&=
4(-1)^2(16)^2\int {d^Dk\o(2\pi)^D}\int{d^Dq\o(2\pi)^D}{d^D\ell\o(2\pi)^D}2\pi\d(k_0)2\pi\d(-k_0+p_0)
\(\d_{\s\r}-{k_\s k_\r\o k^2}\)
\nt
&\quad\times
\tr\[i\g^\m{iq_\a\g^\a\o q^2}i\g^\s{i(q+k)_\b\o(q+k)^2}{i(q+p)_\lam\g^\lam\o(q+p)^2}\]
\tr\[{i(\ell+p)_{\lam'}\g^{\lam'}\o(q+p)^2}{i(\ell+k)_{\b'}\o(\ell+k)^2}i\g^\r{i\ell_{\a'}\g^{\a'}\o \ell^2}i\g^\n\]
}
Noting that
\eq{
\tr\[{i(\ell+p)_{\lam'}\g^{\lam'}\o(q+p)^2}{i(\ell+k)_{\b'}\o(\ell+k)^2}i\g^\r{i\ell_{\a'}\g^{\a'}\o \ell^2}i\g^\n\]&=-
\tr\[i\g^\n{i\ell_{\a'}\g^{\a'}\o \ell^2}i\g^\r{i(\ell+k)_{\b'}\o(\ell+k)^2}{i(\ell+p)_{\lam'}\g^{\lam'}\o(q+p)^2}\],
}
we define a function
\eq{
\mathcal{F}^{\m\s}(k,p)&=\int {d^Dq\o(2\pi)^D}\tr\[\g^\m\g^\a\g^\s\g^\b\g^\lam\]{q_\a(q+k)_\b(q+p)_\lam\o q^2(q+k)^2(q+p)^2}\cdot
}
It follows that 
\eq{
\text{Fig.~\ref{fig:3LoopRenorm}}&=4(16)^2\cdot2\pi\d(p_0)\int {d^dk\o(2\pi)^d}\mathcal{F}^{\m\s}(k,p)\mathcal{F}^{\n\r}(k,p)\(\d_{\s\r}-{k_\s k_\r\o k^2}\)\cdot
}
By dimensional analysis and gauge invariance, we know that any divergence arising from the sum of these diagrams must take the form
\eq{
\text{Fig.~\ref{fig:3LoopRenorm}}&=C^{\m\n,\s\r}\times2\pi\d(p_0)p^2\(\d_{\m\n}-{p_\m p_\n\o p^2}\)+\mathrm{finite}
}
where $C^{\m\n,\s\r}$ is a constant proportional to $1/\ep$.
It follows that our problem can be significantly simplified by differentiating twice with respect to $p$, setting it to zero, and using a cuttoff $\m_\mathrm{IR}$ to regulate the IR divergence.
That is
\eq{\label{eqn:divFluxConst}
C^{\m\n,\s\r}&={4(16)^2}\int {d^dk\o(2\pi)^d}\(\d_{\s\r}-{k_\s k_\r\o k^2}\){\ptl\o \ptl p^2}\[\mathcal{F}^{\m\s}(k,0)\mathcal{F}^{\n\r}(k,0)\],
}
up to finite pieces.
Noting that we should only differentiate with respect to $p^2=p_x^2+p_y^2$, since $p_0=0$, we have
\eq{
{\ptl^2\o \ptl p^2}\[\mathcal{F}^{\m\s}(k,0)\mathcal{F}^{\n\r}(k,0)\]&={1\o 2d}\sum_{j}\[{\ptl_j\ptl^j}\mathcal{F}^{\m\s}\mathcal{F}^{\n\r}+\mathcal{F}^{\m\s}\ptl^j\ptl_j\mathcal{F}^{\n\r}+2\ptl_j \mathcal{F}^{\m\s}\ptl^j \mathcal{F}^{\n\r}\].
}
where $\ptl_j=\ptl/\ptl p_j$.

We start by finding $\mathcal{F}^{\m\s}(k,0)$:
\eq{
\mathcal{F}^{\m\s}(k,0)&=\tr\[\g^\m\g^\a\g^\s\g^\b\g^\lam\]\int{d^Dq\o(2\pi)^D}{q_\a(q+k)_\b q_\lam\o(q^2)^2(q+k)^2}
\nt
&=\tr\[\g^\m\g^\a\g^\s\g^\b\g^\lam\]\int{d^Dq\o(2\pi)^D}\int_0^1dx{2(1-x)\o\[q^2+x(1-x)k^2\]^3}
\nt
&\quad\times
\({q^2\o D}\[-x\d_{\a\b}k_\lam-x\d_{\b\lam k_\a}+(1-x)\d_{\a\lam}k_\b\]+x^2(1-x)k_\a k_\b k_\lam\)
\nt
&={\tr\[\g^\m\g^\a\g^\s\g^\b\g^\lam\]\o128\abs{k}}\(3\d_{\a\lam}k_\b-\d_{\a\b}k_\lam-\d_{\b\lam}k_\a+{k_\a k_\b k_\lam\o k^2}\)
\nt
&=0.
}
Here, we have set $D=3$ since the integral is finite; we will continue to do so below.
So the first two terms in the derivative of $\mathcal{F}^{\m\s}\mathcal{F}^{\n\r}$ vanish, leaving only the third. 
We are left to find
\eq{
\ptl_j \mathcal{F}^{\m\s}(k,0)&=\d_{j\eta}\tr\[\g^\m\g^\a\g^\s\g^\b\g^\lam\]\int{d^Dq\o(2\pi)}{q_\a(q+k)_\b\o(q^2)^2(q+k)^2}\(\d_{\eta\lam}-{2q_\lam q_\eta\o q^2}\)\cdot
}
We separate this into two terms:
\eq{
\[\ptl_j \mathcal{F}^{\m\s}(k,0)\]_A&=\d_{j\eta}\tr\[\g^\m\g^\a\g^\s\g^\b\g^\eta\]\int{d^Dq\o(2\pi)}{q_\a(q+k)_\b\o(q^2)^2(q+k)^2},
\nt
\[\ptl_j \mathcal{F}^{\m\s}(k,0)\]_B&=-2\d_{j\eta}\,\tr\[\g^\m\g^\a\g^\s\g^\b\g^\lam\]\int{d^Dq\o(2\pi)}{q_\a(q+k)_\b q_\lam q_\eta\o(q^2)^3(q+k)^2}\cdot
}
The ``$A$" contribution is
\eq{
\[\ptl_j \mathcal{F}^{\m\s}(k,0)\]_A&=\d_{j\eta}\tr\[\g^\m\g^\a\g^\s\g^\b\g^\eta\]\int {d^Dq\o(2\pi)^D}{2(1-x)\o[q^2+x(1-x)k^2]^3}\({q^2\o D}\d_{\a\b}-x(1-x)k_\a k_\b\)
\nt
&=\d_{j\eta}\,{\tr\[\g^\m\g^\a\g^\s\g^\b\g^\eta\]\o32\abs{k}}\(\d_{\a\b}-{k_\a k_\b \o k^2}\)
\nt
&=\d_{j\eta}\,{2i\o 32\abs{k}}\(-\ep^{\m\s\eta}+{1\o k^2}\[\ep^{\s\eta\a}k_\a k^\m+\ep^{\m\s\a}k_\a k^\eta+\ep^{\m\eta\a}k_\a k^\s\]\)\cdot
}
The ``$B$" part is slightly more complicated,
\eq{
\[\ptl_j \mathcal{F}^{\m\s}(k,0)\]_B&=-6\,\d_{j\eta}\,\,\tr\[\g^\m\g^\a\g^\s\g^\b\g^\lam\]\int {d^Dq\o(2\pi)^D}\int_0^1dx{(1-x)^2\o\[q^2+x(1-x)k^2\]^4}\bigg(
q_\a q_\b q_\lam q_\eta
\nt
&\quad
 +{q^2\o D}\bigg[x^2\(\d_{\a\b} k_\lam k_\eta +\d_{\b\lam}k_\a k_\eta +\d_{\b\eta} k_\a k_\lam \)
 -x(1-x)\(\d_{\a\lam}k_\b k_\eta-\d_{\a\eta}k_\b k_\lam-\d_{\lam\eta}k_\a k_\b\)\bigg]
 \nt
 &\quad
-x^3(1-x)k_\a k_\b k_\lam k_\eta
\bigg),
}
and so we further separate this into three pieces:
\eq{
\[\ptl_j \mathcal{F}^{\m\s}(k,0)\]_{B}^n&=-6\,\d_{j\eta}\,\,\tr\[\g^\m\g^\a\g^\s\g^\b\g^\lam\]\int {d^Dq\o(2\pi)^D}\int_0^1dx{(1-x)^2\o\[q^2+x(1-x)k^2\]^4}f^n_{\a\b\lam\eta}(q,k),
}
where
\eq{
f^1_{\a\b\lam\eta}(q,k)&=q_\a q_\b q_\lam q_\eta,
\nt
f^2_{\a\b\lam\eta}(q,k)&={q^2\o D}\bigg[x^2\(\d_{\a\b} k_\lam k_\eta +\d_{\b\lam}k_\a k_\eta +\d_{\b\eta} k_\a k_\lam \)
 -x(1-x)\(\d_{\a\lam}k_\b k_\eta-\d_{\a\eta}k_\b k_\lam-\d_{\lam\eta}k_\a k_\b\)\bigg],
 \nt
f^3_{\a\b\lam\eta}(q,k)&=-x^3(1-x)k_\a k_\b k_\lam k_\eta.
}
For the first part of $\[\ptl_j \mathcal{F}^{\m\s}(k,0)\]_{B}$ we replace the four $q$'s with
\eq{
q_\a q_\b q_\lam q_\eta\rightarrow {(q^2)^2\o D(D+2)}\(\d_{\a\b}\d_{\lam \eta}+\d_{\a\lam}\d_{\b\eta}+\d_{\a\eta}\d_{\b\lam}\)
}
which gives
\eq{
\[\ptl_j \mathcal{F}^{\m\s}(k,0)\]_{B}^1&=-\d_{j\eta}\,{3\o 256\abs{k}}\tr\[\g^\m\g^\a\g^\s\g^\b\g^\lam\]\(\d_{\a\b}\d_{\lam \eta}+\d_{\a\lam}\d_{\b\eta}+\d_{\a\eta}\d_{\b\lam}\)
\nt
&=\d_{j\eta}\,{2i\o 256\abs{k}}\cdot 15\ep^{\m\s\eta}.
}
The second piece evaluates to
\eq{
\[\ptl_j \mathcal{F}^{\m\s}(k,0)\]_{B}^2&=\d_{j\eta}\,\tr\[\g^\m\g^\a\g^\s\g^\b\g^\lam\]\Bigg(-{1\o 256\abs{k}^3}\[\d_{\a\b} k_\lam k_\eta +\d_{\b\lam}k_\a k_\eta +\d_{\b\eta} k_\a k_\lam \]
\nt
&\quad
+{3\o 256\abs{k}^3}\[\d_{\a\lam}k_\b k_\eta-\d_{\a\eta}k_\b k_\lam-\d_{\lam\eta}k_\a k_\b\]
\Bigg)
\nt
&=-\d_{j\eta}\,{2i\o256\abs{k}^3}\(
\ep^{\m\s\a}k_\a k^\eta+4\ep^{\s\eta \a}k_\a k^\m+4\ep^{\m\eta\a}k_\a k^\s+3\ep^{\m\s\eta}k^2\).
}
Finally, the third part is
\eq{
\[\ptl_j \mathcal{F}^{\m\s}(k,0)\]_{B}^3&=\d_{j\eta}\,{3\o 256\abs{k}^5}\tr\[\g^\m\g^\a\g^\s\g^\b\g^\lam\]k_\a k_\b k_\lam k_\eta
=-{2i \o 256\abs{k}^3}\cdot 3\ep^{\m\s\a}k_\a k^\eta.
}
Adding the three contributions, we find
\eq{
\[\ptl_j \mathcal{F}^{\m\s}(k,0)\]_{B}&=\d_{j\eta}\,{2i\o64\abs{k}}\(3\ep^{\m\s\eta}-{1\o k^2}\[\ep^{\m\s\a}k_\a k^\eta+\ep^{\s\eta\a}k_\a k^\m+\ep^{\m\eta \a}k_\a k^\s\]\),
}
and, upon including $\[\ptl_j \mathcal{F}^{\m\s}(k,0)\]_{A}$, we obtain
\eq{
\ptl_j \mathcal{F}^{\m\s}(k,0)&=\d_{j\eta}\,{i\o32\abs{k}}\(\ep^{\m\s\eta}+{1\o k^2}\[\ep^{\m\s\a}k_\a k^\eta+\ep^{\s\eta\a}k_\a k^\m+\ep^{\m\eta \a}k_\a k^\s\]\)\cdot
}

We can now extract the divergence.
When we only consider the magnetic disorder, the internal indices in Eq.~\eqref{eqn:divFluxConst} are fixed at $(\s,\r)=(0,0)$. 
In this case, we have
\eq{
C^{\m\n,00}&={4(16)^2}\int{d^dk\o(2\pi)^d}{1\o k^2}\,{2\o 2d}\sum_{j=x,y}\ptl^j \mathcal{F}^{\m\s}(k,0)\ptl_j \mathcal{F}^{\n\r}(k,0)
=-{\d^{\m i}\d^{\n j}\d_{ij}}{2(16)^2}{4\o 4(16)^2}\int{d^dk\o(2\pi)^d}{1\o k^2}
\nt
&=\d^{\m i}\d^{\n j}\d_{ij}\({1\o \pi\ep}\)\cdot
}
When we have $\(\s,\r\)=(i,j)$ we find
\eq{
\sum_{i,j=x,y}C^{\m\n,ij}&=4(16)^2\int{d^dk\o(2\pi)^d}{1\o k^2}\,{2\o2d}\sum_{\ell,i,j=x,y}\ptl^\ell \mathcal{F}^{\m i}(k,0)\ptl_\ell \mathcal{F}^{\n j}(k,0)\(\d_{ij}-{k_i k_j\o k^2}\)
\nt
&
=\d^{\m 0}\d^{\n0}\(1\o \pi\ep\)\cdot
}
Multiplying by the corresponding coupling constants, we obtain the counter terms cited in Eq.~\eqref{eqn:FluxCounterTerms}:
\eq{
\d_\E&=\mathpzc{g}^4 g_s g_\B \(1\o \pi\ep\),
&
\d_\B&=\mathpzc{g}^4g_s g_\E \(1\o \pi\ep\)\cdot
}
%

%% file: FluxRenormVanish.tex
\centering
\begin{fmffile}{FluxRenormVanish}
\fmfset{dot_size}{1.5thick}
\fmfset{arrow_len}{3mm}
\begin{subfigure}[b]{0.32\textwidth}
\scriptsize
\mff{
\begin{fmfgraph*}(25,20)
\fmfpen{thin}
\fmfstraight
\fmfleft{i1}
\fmfright{o1}
\fmfv{label=$\m$,l.a=90}{i1}
\fmfv{label=$\n$,l.a=90}{o1}
\fmf{photon}{i1,v1}
\fmf{phantom,tension=.922}{v1,v2}
\fmf{photon}{v2,o1}
\fmfdot{v1}
\fmfdot{v2}
\fmffreeze
\fmfpoly{phantom,smooth}{v1,vU,v2,vB}
\fmffreeze
\fmf{photon}{vU,vB}
\fmf{fermion,right=.5}{v1,vU}
\fmf{plain,right=.5}{vU,v2}
\fmf{fermion,right=.5}{v2,vB}
\fmf{plain,right=.5}{vB,v1}
\fmfdot{vU}
\fmfdot{vB}
\end{fmfgraph*}
}
\caption{}
\label{fig:photoSelfE}
\end{subfigure}
\begin{subfigure}[b]{0.32\textwidth}
\scriptsize
\mff{
\begin{fmfgraph*}(40,20)
\fmfpen{thin}
\fmfstraight
\fmfleft{i1}
\fmfright{o1}
\fmfv{label=$\m$,l.a=90}{i1}
\fmfv{label=$\n$,l.a=90}{o1}
\fmf{photon}{i1,v1}
\fmf{phantom,tension=0.85}{v1,v2}
\fmf{dashes}{v2,v3}
\fmf{phantom,tension=0.85}{v3,v4}
\fmf{photon}{v4,o1}
\fmfdot{v1}
\fmfdot{v2}
\fmfdot{v3}
\fmfdot{v4}
\fmffreeze
\fmf{fermion,right}{v1,v2}
\fmf{fermion,right}{v2,v1}
\fmf{fermion,right}{v3,v4}
\fmf{fermion,right}{v4,v3}
\end{fmfgraph*}
}
\caption{}
\label{fig:FluxTwo Loop}
\end{subfigure}
\end{fmffile}

%% file: 3LoopFluxRenorm.tex
\begin{fmffile}{3LoopFluxRenorm}
\fmfset{dot_size}{1.5thick}
\fmfset{arrow_len}{3mm}
\centering
\eq{
\mf{
\scriptsize
\begin{fmfgraph*}(35,13)
\fmfpen{thin}
\fmfstraight
\fmfleft{i1,i2,i3}
\fmfright{o1,o2,o3}
\fmf{phantom}{i1,v1a}
\fmf{phantom,tension=1.3}{v1a,v1b}
\fmf{phantom,tension=0.8}{v1b,v1c}
\fmf{phantom,tension=1.3}{v1c,v1d}
\fmf{phantom}{v1d,o1}
\fmf{phantom}{i3,v3a}
\fmf{phantom,tension=1.3}{v3a,v3b}
\fmf{phantom,tension=0.8}{v3b,v3c}
\fmf{phantom,tension=1.3}{v3c,v3d}
\fmf{phantom}{v3d,o3}
\fmf{photon}{i2,v2a}
\fmf{phantom,tension=1.3}{v2a,v2b}
\fmf{phantom,tension=0.8}{v2b,v2c}
\fmf{phantom,tension=1.3}{v2c,v2d}
\fmf{photon}{v2d,o2}
\fmffreeze
\fmf{fermion}{v2a,v3b}
\fmf{fermion}{v3b,v1b}
\fmf{fermion}{v1b,v2a}
\fmf{fermion}{v2d,v1c,v3c,v2d}
\fmf{photon}{v3b,vC,v3c}
\fmfv{decor.shape=diamond,decor.filled=shaded,decor.size=3mm}{vC}
\fmf{dashes}{v1b,v1c}
\fmfdot{v2a,v3b,v1b,v3c,v1c,v2d}
\fmf{phantom}{v3b,vC2,v3c}
\fmfv{l=$\s$,l.a=-135}{vC}
\fmfv{l=$\r$,l.a=-45}{vC2}
\fmfv{l=$\s$,l.d=1mm}{v3b}
\fmfv{l=$\r$,l.d=1mm}{v3c}
\fmfv{l=$\m$,l.a=-90}{v2a}
\fmfv{l=$\n$,l.a=-90}{v2d}
\end{fmfgraph*}}
\;\;
+
\mf{
\scriptsize
\begin{fmfgraph*}(35,13)
\fmfpen{thin}
\fmfstraight
\fmfleft{i1,i2,i3}
\fmfright{o1,o2,o3}
\fmf{phantom}{i1,v1a}
\fmf{phantom,tension=1.3}{v1a,v1b}
\fmf{phantom,tension=0.8}{v1b,v1c}
\fmf{phantom,tension=1.3}{v1c,v1d}
\fmf{phantom}{v1d,o1}
\fmf{phantom}{i3,v3a}
\fmf{phantom,tension=1.3}{v3a,v3b}
\fmf{phantom,tension=0.8}{v3b,v3c}
\fmf{phantom,tension=1.3}{v3c,v3d}
\fmf{phantom}{v3d,o3}
\fmf{photon}{i2,v2a}
\fmf{phantom,tension=1.3}{v2a,v2b}
\fmf{phantom,tension=0.8}{v2b,v2c}
\fmf{phantom,tension=1.3}{v2c,v2d}
\fmf{photon}{v2d,o2}
\fmffreeze
\fmf{fermion}{v2a,v3b,v1b,v2a}
\fmf{fermion}{v2d,v3c,v1c,v2d}
\fmf{photon}{v3b,vC,v3c}
\fmfv{decor.shape=diamond,decor.filled=shaded,decor.size=3mm}{vC}
\fmf{dashes}{v1b,v1c}
\fmfdot{v2a,v3b,v1b,v3c,v1c,v2d}
\fmf{phantom}{v3b,vC2,v3c}
\fmfv{l=$\s$,l.a=-135}{vC}
\fmfv{l=$\r$,l.a=-45}{vC2}
\fmfv{l=$\s$,l.d=1mm}{v3b}
\fmfv{l=$\r$,l.d=1mm}{v3c}
\fmfv{l=$\m$,l.a=-90}{v2a}
\fmfv{l=$\n$,l.a=-90}{v2d}
\end{fmfgraph*}
}
\;\;
+
\mf{
\scriptsize
\begin{fmfgraph*}(35,13)
\fmfpen{thin}
\fmfstraight
\fmfleft{i1,i2,i3}
\fmfright{o1,o2,o3}
\fmf{phantom}{i1,v1a}
\fmf{phantom,tension=1.3}{v1a,v1b}
\fmf{phantom,tension=0.8}{v1b,v1c}
\fmf{phantom,tension=1.3}{v1c,v1d}
\fmf{phantom}{v1d,o1}
\fmf{phantom}{i3,v3a}
\fmf{phantom,tension=1.3}{v3a,v3b}
\fmf{phantom,tension=0.8}{v3b,v3c}
\fmf{phantom,tension=1.3}{v3c,v3d}
\fmf{phantom}{v3d,o3}
\fmf{photon}{i2,v2a}
\fmf{phantom,tension=1.3}{v2a,v2b}
\fmf{phantom,tension=0.8}{v2b,v2c}
\fmf{phantom,tension=1.3}{v2c,v2d}
\fmf{photon}{v2d,o2}
\fmffreeze
\fmf{fermion}{v2a,v1b,v3b,v2a}
\fmf{fermion}{v2d,v1c,v3c,v2d}
\fmf{photon}{v3b,vC,v3c}
\fmfv{decor.shape=diamond,decor.filled=shaded,decor.size=3mm}{vC}
\fmf{dashes}{v1b,v1c}
\fmfdot{v2a,v3b,v1b,v3c,v1c,v2d}
\fmf{phantom}{v3b,vC2,v3c}
\fmfv{l=$\s$,l.a=-135}{vC}
\fmfv{l=$\r$,l.a=-45}{vC2}
\fmfv{l=$\s$,l.d=1mm}{v3b}
\fmfv{l=$\r$,l.d=1mm}{v3c}
\fmfv{l=$\m$,l.a=-90}{v2a}
\fmfv{l=$\n$,l.a=-90}{v2d}
\end{fmfgraph*}
}
\;\;
+
\mf{
\scriptsize
\begin{fmfgraph*}(35,13)
\fmfpen{thin}
\fmfstraight
\fmfleft{i1,i2,i3}
\fmfright{o1,o2,o3}
\fmf{phantom}{i1,v1a}
\fmf{phantom,tension=1.3}{v1a,v1b}
\fmf{phantom,tension=0.8}{v1b,v1c}
\fmf{phantom,tension=1.3}{v1c,v1d}
\fmf{phantom}{v1d,o1}
\fmf{phantom}{i3,v3a}
\fmf{phantom,tension=1.3}{v3a,v3b}
\fmf{phantom,tension=0.8}{v3b,v3c}
\fmf{phantom,tension=1.3}{v3c,v3d}
\fmf{phantom}{v3d,o3}
\fmf{photon}{i2,v2a}
\fmf{phantom,tension=1.3}{v2a,v2b}
\fmf{phantom,tension=0.8}{v2b,v2c}
\fmf{phantom,tension=1.3}{v2c,v2d}
\fmf{photon}{v2d,o2}
\fmffreeze
\fmf{fermion}{v2a,v1b,v3b,v2a}
\fmf{fermion}{v2d,v3c,v1c,v2d}
\fmf{photon}{v3b,vC,v3c}
\fmfv{decor.shape=diamond,decor.filled=shaded,decor.size=3mm}{vC}
\fmf{dashes}{v1b,v1c}
\fmfdot{v2a,v3b,v1b,v3c,v1c,v2d}
\fmf{phantom}{v3b,vC2,v3c}
\fmfv{l=$\s$,l.a=-135}{vC}
\fmfv{l=$\r$,l.a=-45}{vC2}
\fmfv{l=$\s$,l.d=1mm}{v3b}
\fmfv{l=$\r$,l.d=1mm}{v3c}
\fmfv{l=$\m$,l.a=-90}{v2a}
\fmfv{l=$\n$,l.a=-90}{v2d}
\end{fmfgraph*}
}
\notag}
\end{fmffile}

%% file: appendixConductivity.tex
%
%
%
%
%
%


\section{Current-current correlators}\label{app:CurrentCurrentCorr}

In this appendix we review our calculation of the Feynman diagrams shown in Figs.~\ref{fig:currentLoopBare} to ~\ref{fig:currentLoopSelfE2}. 
Since no divergences are present in these diagrams, no counter-terms will be necessary.

\subsection{Bare loop}\label{app:PolarizationLoop}

The leading term is shown in Fig.~\ref{fig:currentLoopBare}. It is simply
\eq{
\text{Fig.~\ref{fig:currentLoopBare}}
&=(-1)\tr\[T^rT^s\]\tr\[\s^a\s^b\]\int{d^Dq\o(2\pi)^D}\tr\[i\g^\m{iq_\a\g^\a\o q^2}i\g^\n{i(q+p)_\b\g^\b\o(q+p)^2}\]
\nt
&=-\d^{rs}\d^{ab}\,{{\abs{p}\o16}\(\d^{\m\n}-{p^\m p^\n\o p^2}\)}
}
where we used  $\tr\[T^rT^s\]={\d^{rs}\o2}$. Setting $\vp=0$ and $\m=\n=x$,we have
\eq{
\text{Fig.~\ref{fig:currentLoopBare}}&=-{\abs{p_0}\o16}\cdot
}

\subsection{Vertex diagrams}

\begin{figure}
\input{currentCurrentSubDiagrams}
\caption{Subdiagrams which contribute to the flavour conductivity.}
\end{figure}
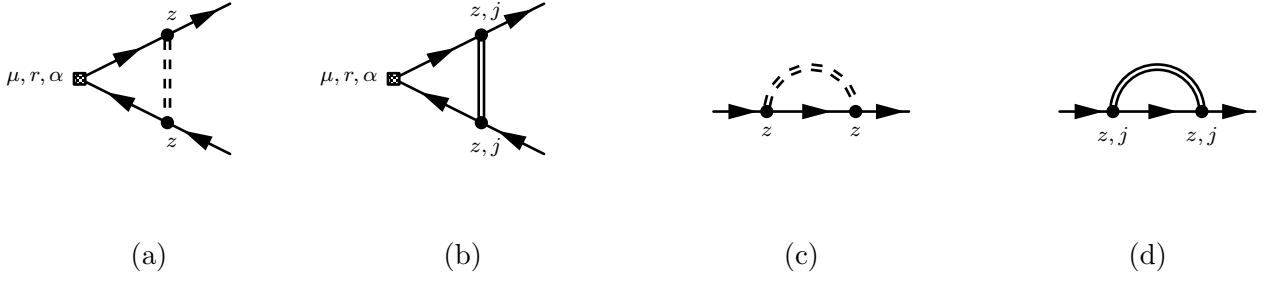

\subsubsection{Contribution proportional to $g_{t,z}$}

We begin by calculating the 1-loop vertex contribution shown in Fig.~\ref{fig:VertsubDiagram}:
\eq{\label{eqn:currentLoopVert}
\text{Fig.~\ref{fig:VertsubDiagram}}&=
g_{t,z}T^r\otimes\s^z\s^a\s^z\int{d^Dk\o(2\pi)^d}2\pi\d(k_0){i(q+p)_\a\g^\a\o(q+p)^2}{i(q+k+p)_\b\g^\b\o(q+k+p)^2}i\g^\m{i(q+k)_\s\g^\s\o q^2}{iq_\r\g^\r\o q^2}
\nt
&=
ig_{t,z}T^r\otimes\s^z\s^a\s^z\otimes\g^a\g^\r\g^\m\g^\s\g^\r\,{(q+p)_
a q_\r\o(q+p)^2 q^2}I_{\b\s}(q_0,p_0)
}
where
\eq{\label{eqn:VertIntDef}
I_{\b\s}(q_0,p_0)&=
\d_{\b0}\d_{\s0}\underbrace{\int {d^dk\o(2\pi)^d}{q_0(q_0+p_0)\o\[(q_0+p_0)^2+\vk^2\]\[q_0^2+\vk^2\]}}_{I_0(q_0,p_0)}
+
{\d_{\b j}\d_{\s i}\d^{ij}\o d}\underbrace{\int {d^dk\o(2\pi)^d}{\vk^2\o\[(q_0+p_0)^2+\vk^2\]\[q_0^2+\vk^2\]}}_{I_d(q_0,p_0)}
\cdot
}
The full diagram in Fig.~\ref{fig:currentLoopVert} is then
\eq{\label{eqn:fullVertLoop}
\text{Fig.~\ref{fig:currentLoopVert}}&=-1\times2\times g_{t,z}\tr[T^rT^s]\tr[\s^z\s^a\s^z\s^b]\tr\[\g^\a\g^\b\g^x\g^\s\g^\r\g^x\](i)^2\int{d^3q\o(2\pi)^3}{(q+p)_
a q_\r\o(q+p)^2 q^2}I_{\b\s}(q_0,p_0)
\nt
&=-2\eta_{a}\d^{rs}\d^{ab}\,\int{d^3q\o(2\pi)^3}\tr\[\g^\r\g^x\g^\a\g^x\]\(I_{\mathrm{V},0}(q_0,p_0)+{(d-2)\o2}I_{\mathrm{V},d}(q_0,p_0)\){(q+p)_
a q_\r\o(q+p)^2 q^2}
\nt
&=4\eta_{a}\d^{rs}\d^{ab}\,\int{d^3q\o(2\pi)^3}\(I_{\mathrm{V},0}(q_0,p_0)+{(d-2)\o2}I_{\mathrm{V},d}(q_0,p_0)\){q_0(q_0+p_0)\o\[(q_0+p_0)^2+\vq^2\]\[q_0^2+\vq^2\]}
}
where $\eta_z=+1$ and $\eta_{x,y}=-1$.

We perform the integral over $\vk$ in $I_{\mathrm{V},0}$ and $I_{\mathrm{V},d}$ and analytically continuing to $d=2+\ep$ spatial dimensions:
\eq{
I^\mathrm{V}_\mathrm{tot}(q_0,p_0)&=I_{\mathrm{V},0}(q_0,p_0)+{(d-2)\o2}I_{\mathrm{V},d}(q_0,p_0)
=-{1\o4\pi}\Bigg\{1+{q_0(q_0+p_0)\o p_0(p_0+2q_0)}\log\[q_0^2\o(q_0+p_0)^2\]\Bigg\}\cdot
}
Performing the $\vq$ integral, we have
\eq{
\int{d^2q\o(2\pi)^2}{q_0(q_0+p_0)\o\[(q_0+p_0)^2+\vq^2\]\[q_0^2+\vq^2\]}
&=
-{1\o4\pi}{q_0(q_0+p_0)\o p_0(p_0+2q_0)}\log\[q_0^2\o(q_0+p_0)^2\].
}
Plugging these into Eq.~\ref{eqn:fullVertLoop} and integrating over $q_0$ we find,
\eq{
\text{Fig.~\ref{fig:currentLoopVert}}&=\eta_a\d^{rs}\d^{ab}\cdot g_{t,z}{\abs{p_0}\o96\pi}\cdot
}

\subsubsection{Contribution proportional to $g_{\A,z}$}
The diagram in Fig.~\ref{fig:currentLoopVert2} vanishes. We can see this by noting that
\eq{\label{eqn:currentLoopVert2}
\text{Fig.~\ref{fig:VertsubDiagram2}}&=
g_{\A,z}T^r\otimes\s^z\s^a\s^z\int{d^Dk\o(2\pi)^d}2\pi\d(k_0){i(q+p)_\a\g^\a\o(q+p)^2}i\g^j{i(q+k+p)_\b\g^\b\o(q+k+p)^2}i\g^\m{i(q+k)_\s\g^\s\o q^2}\i\g_j{iq_\r\g^\r\o q^2}
\nt
&=
-ig_{\A,z}T^r\otimes\s^z\s^a\s^z\otimes\g^\a\g^j\g^\r\g^\m\g^\s\g_j\g^\r\,{(q+p)_
\a q_\r\o(q+p)^2 q^2}I_{\b\s}(q_0,p_0)
}
where $I_{\b\s}(q_0,p_0)$ is defined in Eq.~\ref{eqn:VertIntDef}. The full diagram is therefore
\eq{\label{eqn:fullVertLoop2}
\text{Fig.~\ref{fig:currentLoopVert2}}&=(-1)^2\times2\times g_{\A,z}\tr[T^rT^s]\tr[\s^z\s^a\s^z\s^b]\tr\[\g^\a\g^j\g^\b\g^x\g^\s\g_j\g^\r\g^x\](i)^2\int{d^3q\o(2\pi)^3}{(q+p)_
a q_\r\o(q+p)^2 q^2}I_{\b\s}(q_0,p_0)
\nt
&=2\eta_{a}g_{\A,z}\d^{rs}\d^{ab}\,\int{d^3q\o(2\pi)^3}\tr\[\g^\r\g^x\g^\a\g^x\](2-d)\(I_{\mathrm{V},0}(q_0,p_0)+{(d-2)\o2}I_{\mathrm{V},d}(q_0,p_0)\){(q+p)_
a q_\r\o(q+p)^2 q^2}
\nt
&=(d-2)g_{\A,z}\times\({1\o g_{t,z}}\text{Fig.~\ref{fig:currentLoopVert}}\)\cdot
}
Noting that Fig.~\ref{fig:currentLoopVert} has no epsilon pole, when $\ep\rightarrow0$, this diagram vanishes: $\text{Fig.~\ref{fig:currentLoopVert2}}=0$.

\subsection{Self-energy diagram}

\subsubsection{Contribution proportional to $g_{t,z}$}

The self-energy subdiagram is
\eq{\label{eqn:SelfEsubDiagram}
\text{Fig.~\ref{fig:SelfEsubDiagram}}&=
g_{t,z}\s^z\s^z\otimes\,\int{d^Dk\o(2\pi)^D}2\pi\d^(k_0){i(q+k)_\b\g^\b\o(q+k)^2}
=g_{t,z}i\g^0\int{d^dk\o(2\pi)^d}{q_0\o q_0^2+\vk^2}
\nt
&=g_{t,z}i\g^0q_0\underbrace{\[-{1\o2\pi\ep}+{1\o4\pi}\log\[4\pi e^{-\g_E}\]-{1\o4\pi}\log q_0^2\]}_{I_\Sigma(q_0)}\cdot
}
The full diagram is therefore
\eq{\label{eqn:currentLoopSelfE}
\text{Fig.~\ref{fig:currentLoopSelfE}}&=
-1\times2\times g_{t,z}\tr[T^rT^s]\tr\[\s^a\s^b\]\int {d^3q\o(2\pi)^3}\tr\[{iq_\a\g^\a\o q^2}i\g^0{iq_\b\g^\b\o q^2}i\g^x{i(q+p)_\r\g^\r\o (q+p)^2}i\g^x\]q_0I_\Sigma(q_0)
\nt
&=4g_{t,z}\d^{rs}\d^{ab}\int {d^3q\o(2\pi)^3}{q_0(q_0+p_0)(\vq^2-q_0^2)\o \[q_0^2+\vq^2\]^2\[(q_0+p_0)^2+\vq^2\]}I_\Sigma(q_0)
\nt
&=-g_{t,z}\d^{rs}\d^{ab}{2\o \pi}\int {dq_0\o2\pi}{q_0(q_0+p_0)I_\Sigma(q_0)\o p_0(p_0+2q_0)}\(1+{q_0^2+(q_0+p_0)^2\o2p_0(p_0+2q_0)}\log\[q_0^2\o(q_0+p_0)^2\]\)
}
We see that the constant (and divergent) portion of $I_\Sigma(p_0)$ integrate to zero since it is odd. The term proportional to the log on the other hand, can be rewritten and solved:
\eq{
\text{Fig.~\ref{fig:currentLoopSelfE}}&=
g_{t,z}\d^{rs}\d^{ab}{1\o4\pi^2}\int {dq_0\o2\pi}\(1+{q_0^2+(q_0+p_0)^2\o2p_0(p_0+2q_0)}\log\[q_0^2\o(q_0+p_0)^2\]\){q_0(q_0+p_0)\o p_0(p_0+2q_0)}\log\[q_0^2\o(q_0+p_0)^2\]
\nt
&=\d^{rs}\d^{ab}\cdot g_{t,z}{\abs{p_0}\o 96\pi}
}

\subsubsection{Contribution proportional to $g_{\A,z}$}
This diagram is nearly identical to the previous one:
\eq{
\text{Fig.~\ref{fig:currentLoopSelfE2}}
&=
-1\times2\times g_{\A,z}\tr[T^rT^s]\tr\[\s^a\s^b\]\int {d^3q\o(2\pi)^3}\tr\[{iq_\a\g^\a\o q^2}i\g^ji\g^0i\g_j{iq_\b\g^\b\o q^2}i\g^x{i(q+p)_\r\g^\r\o (q+p)^2}i\g^x\]q_0I_\Sigma(q_0)
\nt
&=2\times4g_{\A,z}\d^{rs}\d^{ab}\int {d^3q\o(2\pi)^3}{q_0(q_0+p_0)(\vq^2-q_0^2)\o \[q_0^2+\vq^2\]^2\[(q_0+p_0)^2+\vq^2\]}I_\Sigma(q_0)
\nt
&=\d^{rs}\d^{ab}\cdot 2g_{\A,z}{\abs{p_0}\o 96\pi}\cdot
}
where $I_\Sigma(q_0)$ is given in Eq.~\ref{eqn:SelfEsubDiagram}.


%% file: currentCurrentSubDiagrams.tex
\centering
\begin{fmffile}{CurrentCurrentSugDiagrams}
\begin{subfigure}[b]{0.23\textwidth}
\centering
\eq{
\mfff{
\scriptsize
\begin{fmfgraph*}(20,20)
\fmfpen{thin}
\fmfstraight
\fmfleft{i1}
\fmfright{o2,o1}
\fmf{phantom,tension=1.4}{v1,o1}
\fmf{phantom}{i1,v1}
\fmf{phantom,tension=1.4}{v2,o2}
\fmf{phantom}{i1,v2}
\fmffreeze
\fmf{fermion}{i1,v1}
\fmf{fermion}{v1,o1}
\fmf{fermion}{v2,i1}
\fmf{fermion}{o2,v2}
\fmf{dbl_dashes}{v1,v2}
\fmfdot{v1}
\fmfdot{v2}
\fmfv{decor.size=1.5mm,decor.shape=square,decor.filled=gray50}{i1}
\fmfv{label=$z$}{v1}
\fmfv{label=$z$}{v2}
\fmfv{label=$\mu,,r,,\a$}{i1}
\end{fmfgraph*}
}
\notag}
\vspace{4mm}
\caption{}
\label{fig:VertsubDiagram}
\end{subfigure}
\begin{subfigure}[b]{0.23\textwidth}
\centering
\eq{
\mfff{
\scriptsize
\begin{fmfgraph*}(20,20)
\fmfpen{thin}
\fmfstraight
\fmfleft{i1}
\fmfright{o2,o1}
\fmf{phantom,tension=1.4}{v1,o1}
\fmf{phantom}{i1,v1}
\fmf{phantom,tension=1.4}{v2,o2}
\fmf{phantom}{i1,v2}
\fmffreeze
\fmf{fermion}{i1,v1}
\fmf{fermion}{v1,o1}
\fmf{fermion}{v2,i1}
\fmf{fermion}{o2,v2}
\fmf{dbl_plain}{v1,v2}
\fmfdot{v1}
\fmfdot{v2}
\fmfv{decor.size=1.5mm,decor.shape=square,decor.filled=gray50}{i1}
\fmfv{label=$z,,j$}{v1}
\fmfv{label=$z,,j$}{v2}
\fmfv{label=$\mu,,r,,\a$}{i1}
\end{fmfgraph*}
}
\notag}
\vspace{4mm}
\caption{}
\label{fig:VertsubDiagram2}
\end{subfigure}
\hspace{2mm}
\begin{subfigure}[b]{0.23\textwidth}
\centering
\eq{
\mff{
\scriptsize
\begin{fmfgraph*}(26,20)
\fmfpen{thin}
\fmfstraight
\fmfleft{i1}
\fmfright{o1}
\fmf{fermion}{i1,v1a}
\fmf{fermion,tension=0.6}{v1a,v1b}
\fmf{fermion}{v1b,o1}
\fmf{dbl_dashes,left,tension=0}{v1a,v1b}
\fmfdot{v1a}
\fmfdot{v1b}
\fmfv{l=$z$,l.a=-90}{v1a}
\fmfv{l=$z$,l.a=-90}{v1b}
\end{fmfgraph*}}
\notag}
\caption{}
\label{fig:SelfEsubDiagram}
\end{subfigure}
\hspace{3mm}
\begin{subfigure}[b]{0.23\textwidth}
\centering
\eq{
\mff{
\scriptsize
\begin{fmfgraph*}(26,20)
\fmfpen{thin}
\fmfstraight
\fmfleft{i1}
\fmfright{o1}
\fmf{fermion}{i1,v1a}
\fmf{fermion,tension=0.6}{v1a,v1b}
\fmf{fermion}{v1b,o1}
\fmf{dbl_plain,left,tension=0}{v1a,v1b}
\fmfdot{v1a}
\fmfdot{v1b}
\fmfv{l=$z,,j$,l.a=-90}{v1a}
\fmfv{l=$z,,j$,l.a=-90}{v1b}
\end{fmfgraph*}}
\notag}
\caption{}
\label{fig:SelfEsubDiagram2}
\end{subfigure}
\end{fmffile}
\vspace{5mm}